\documentclass[a4paper]{book}
\RequirePackage[english]{babel}
\RequirePackage{amsbsy}
\RequirePackage{amssymb}
\RequirePackage{amsmath}
\RequirePackage{mathrsfs}
\RequirePackage{bm}
\newtheorem{dfn}{Definition}
\newtheorem{pre}{Principle}
\newtheorem{axm}{Axiom}
\newtheorem{thm}{Theorem}
\newtheorem{dt}[thm]{Definition-Theorem}
\begin{document}
%%%%%%%%%%%%%%%%%%%%%%%%%%%%%%%%%%%%%%%%%%%%%%%%%%%%%%%%%%%%%%%%%%%%%%%%%%%%%%%%%%%%%%%%%%%%%%%%%%%
\title{\LARGE{\bf{HIGHER-ORDER THEORIES\\ OF GRAVITATION}}}
\author{Luca Fabbri}
\date{}
\maketitle
%%%%%%%%%%%%%%%%%%%%%%%%%%%%%%%%%%%%%%%%%%%%%%%%%%%%%%%%%%%%%%%%%%%%%%%%%%%%%%%%%%%%%%%%%%%%%%%%%%%
\thispagestyle{empty}
\begin{flushright}
``Who happened to follow this path\\
is led to the conviction\\
that the human mind\\
can \textit{work out}\\
physical theories,\\
that can \textit{foretell}\\
the behaviour of the physical world\\
under still unexplored circumstances.''\\
\vspace{1cm}
Silvio Bergia
\end{flushright}
%%%%%%%%%%%%%%%%%%%%%%%%%%%%%%%%%%%%%%%%%%%%%%%%%%%%%%%%%%%%%%%%%%%%%%%%%%%%%%%%%%%%%%%%%%%%%%%%%%%
\part{Prologue}
\chapter{Phenomenology,\\ Theoretical Physics,\\ Geometry}
Ever since Physics has been asked to be the Mathematical description of Nature, Physicists have observed a gradual, although not always linear, growth of its complexity.

Born as the Observation of Nature, Physics had further become the Modelization of Nature, in which the collected data were interpreted in the frame of a Model, suitable to be mathematized; this model, providing the description of natural phenomena, is the Phenomenology.

But this is not the only way in which we can approach the description of the Universe; another way consists in starting from what we \emph{think} to be the fundamental principles leading natural phenomena, and developing all the possible consequences, till when we are able to make predictions about something we can measure; the development of a theory for which the principles are not asked to nature, but \emph{thought} to be true \emph{a priori} is Theoretical Physics.

While observation of nature is always intertwined with phenomenological concepts, Theoretical Physics does not depend on them; it is a superstructure at first disconnected from the real universe, in which we find ourselves free to create new theories, whose status of truth is assigned only at the moment of the last, final check: this feature is what allow us to call Theoretical Physics, according to how philosophers of science (see for example Russo in \cite{r}) define it, a high-order science.

In Theoretical Physics one has the possibility to invent theories by using principles of any sort; nevertheless, one of the most elegant ways is to use Principles of Symmetry: a principle of symmetry is essentially a principle that states that our description of nature does not change, changing some of its features.

Thus, in the following, we will consider Theoretical Physics as the mathematical formulation of the principles of symmetry that we think to be the fundamental character of nature.

When we talk about the mathematical formulation we are commonly lead to talk (and this was indeed its very first meaning) about Geometry. 

Geometry is the description of the character a generic space has, and in physics, it deals with the features possessed by the physical space itself; it is then our goal to see which features a space should have in order to represent the real physical one.

To get started, the first thing we have to do is to fix the number of dimensions. 

Since ancient Greeks, and for two millennia, the space was thought to be $3$-dimensional: this lasted even through the I Scientific Revolution, in the XVII century; it was only after the II Scientific Revolution, at the beginning of XIX century, that physicists recognized Time to be the fourth dimension of an extended spacetime. 

This unification is a formal consequence of the so-called Lorentz transformations (with respect to which physical laws were recognized to be invariant, symmetric), in which space and time are able to mix between each other, losing their own individual identity, and becoming parts of a unique spacetime.

From this moment on, the space was thought to be $4$-dimensional, and in a $4$-dimensional space, Einstein has been able to describe gravity, in the framework of his General Relativity.

Nevertheless, many think that all the other physical interactions can be placed into an enlarged theory of general relativity, in which the enlargement is obtained by an extension of the number of dimensions: these are the so-called Multidimensional Theories.

In these theories the space is considered \emph{a priori} a generic $n$-dimensional space, and then the number of dimensions is fixed by using phenomenological considerations. 

The first time this idea has been used is found in the works of Kaluza, followed by those of Klein, who supposed the universe to have just one dimension more than the $4$ we were used to consider: the $5$-dimensional Kaluza-Klein Universe described Electromagnetism as a $4$-dimensional effect of the $5$-dimensional, enlarged gravity, providing the first attempt of Electro-gravitational unification. 

But, even without considering the technical problems such a theory had, $5$ dimensions were certainly not enough to create the room necessary to include the nuclear interactions, and even more enlarged geometries were needed.

According to Witten's observation that $11$ is the only dimension for which a space is big enough to contain $U(1)\times SU(2)\times SU(3)$ \emph{and} small enough to allow Supersymmetry (\cite{wi}), the $11$-dimensional space was immediately thought to be the space we live in (an alternative reason to consider an $11$-dimensional universe comes from String Theory, in which anomalies cancel only in $10$ dimensions, and it has been noticed that a particular strong coupling limit of type IIA theory (called M-theory) develops an additional dimension, giving, once again, the $11$-dimensional space): for this reason, $11$-dimensional spaces are quite an attractive choice for the background of Kaluza-Klein's theories (for a general introduction to Kaluza-Klein theory, see the original works of Kaluza and Klein; for generalized, multidimensional theories see the reviews of Toms and Duff, in \cite{kk1}, and also the more comprehensive reference \cite{kk2}).

Nevertheless, the Universe looks $4$-dimensional.

The process those $11$-dimensional theories need to justify the quadridimensionality of the universe consists in taking the $11$-dimensional space, and breaking (spontaneously!) its symmetry, in order to split it into two subspaces, one of which being the $4$-dimensional Minkowskian space, the other being the space of the remaining $11-4=7$ dimensions, which must be somehow hidden from observations.

The most common way to hide this space is to consider its dimensions compactified to a typical length so small that they are invisible, because inaccessible at low energy scales. 

The mechanism for the compactification of the $7$-dimensional sub-space can indeed be accomplished through the parallelization of the $7$-sphere. 

The parallelization of a space is a process for which, given a metric and a completely antisymmetric Cartan tensor, it is possible to find a way in which they cancel each other inside the expression of the Riemann tensor, so that Riemann tensor vanishes; for $7$-dimensional spaces, it is actually possible to get the vanishing of Riemann tensor through this process, and so $S^{7}$ can indeed be parallelized, as shown by Englert in \cite{en}.

In the framework of $11$-dimensional Kaluza-Klein theories, and after the parallelization of the $7$-sphere, the space is structured as a direct product of the Minkowskian $(1+3)$-dimensional spacetime plus the compactified $7$-dimensional internal sub-space, in which the completely antisymmetric Cartan tensor is a completely antisymmetric potential for a correspondent completely antisymmetric strength: this strength is the supersymmetric field we need to induce the spontaneous compactification mechanism for $11$-dimensional Kaluza-Klein theory of Supergravity (other interpretations are assigned in the case of Superstring theories, as showed by Agricola, Friedrich, Nagy and Puhle \cite{a-f-n-p}, of Strominger \cite{st} and Gauntlett, Martelli and Waldram \cite{g-m-w}).

For whom is concerned by the supersymmetric field of the $11$-dimensional theories, and finds arbitrariness in the choice of the number of dimensions, $4$-dimensional spacetime is then the most natural space in which physical theories can take place.

Anyway, the number of dimensions of a space is not something that can be proven to be equal to $4$, and that's why, if we think it actually is the real number of dimensions of our Universe, we have to fix it as a matter of principle.

So, in all we are going to say hereafter, we will consider the Physical Universe to be $4$-dimensional.

But fixing the number of the dimensions is not enough. 

There is, in fact, also the intrinsic structure of the spacetime that has to be determined, both from the point of view of the geometry, and, moreover, from the point of view of the dynamics that physical fields will have in it. 

The latter point is what we will drive us to the variational formulation of the fundamental field equations; the former, to the construction of the geometrical background.

And this is the point we are going to start from.
%%%%%%%%%%%%%%%%%%%%%%%%%%%%%%%%%%%%%%%%%%%%%%%%%%%%%%%%%%%%%%%%%%%%%%%%%%%%%%%%%%%%%%%%%%%%%%%%%%%
%%%%%%%%%%%%%%%%%%%%%%%%%%%%%%%%%%%%%%%%%%%%%%%%%%%%%%%%%%%%%%%%%%%%%%%%%%%%%%%%%%%%%%%%%%%%%%%%%%%
%%%%%%%%%%%%%%%%%%%%%%%%%%%%%%%%%%%%%%%%%%%%%%%%%%%%%%%%%%%%%%%%%%%%%%%%%%%%%%%%%%%%%%%%%%%%%%%%%%%
%%%%%%%%%%%%%%%%%%%%%%%%%%%%%%%%%%%%%%%%%%%%%%%%%%%%%%%%%%%%%%%%%%%%%%%%%%%%%%%%%%%%%%%%%%%%%%%%%%%
\part{Fundamentals}
\chapter{Covariant Theories as\\ Theories of Absolute Relativity}
Historically, the Foundations of the Theory of Relativity have to be searched in the generalization of the \emph{a posteriori} called special theory of relativity. 

\emph{A posteriori} because, chronologically speaking, special theory of relativity had, at the time in which it was born, nothing special, since, before its generalization, it was the only theory of relativity to be known. 

Its fundamental idea is the conception of a unified structure of the spacetime, that is, a structure in which time and space can mix between each other; in order to have this mixing in such a way that time and space do lose their own individual character we have to consider them completely indistinguishable, thus they must be dimensionally homogeneous, and so we will choose to measure them in Natural Units for which $c=1$, i.e. the speed of light has a normalized value.

The fact that space and time can mix is a mathematical consequence of the inclusion of the transformation between two systems of reference moving with a relative velocity among the whole set of the possible transformations that leave a physical situation unchanged.

Providing this geometrical description of the spacetime, special relativity was able to achieve many conceptual improvements, but it had a fundamental problem it was not able to solve: it was not able to include among the set of the spacetime symmetries the transformation between two systems of reference moving with a relative acceleration.

Nevertheless, Einstein thought that any good geometry should have been able to describe also accelerated systems\footnote{Moreover, if we think that accelerated systems are related to each other via a non linear transformation of coordinates involving time, then we see how they are a particular case of a general non linear transformation of coordinates; since a non linear transformation of coordinates is the transformation that perform a change between Cartesian and general curvilinear coordinates, even if it does not involve time, then it is already in a $3$-dimensional geometry that the passage from Cartesian to curvilinear coordinates can not be achieved, if the geometry does not admit non-linear transformations between systems.}.

In the extension Einstein got in order to describe any sort of accelerated frames, transformations between different spacetimes, which were linear before, were not linear any longer; also, the possibility to describe reference systems is given by using a metric, and the metric of the spacetime, which was constant in special relativity, was not constant any more (of course, fixing a point, it could have always been possible to choose a non-linear transformation such that a non-constant metric could have been transformed into a constant one, giving to the spacetime written through those coordinates the appearance it would have had in special relativity; but this choice would have been point-dependent, the final metric would have described a locally inertial reference system, and the spacetime written in terms of those coordinates would have been like the one Einstein had in special relativity only in a small neighborhood of the given point): the fact that a general curvilinear system of reference is point dependent produces the consequence that the derivatives of local fields are non-zero. The extra terms Einstein got in the theory after this generalization were enough to produce a drastic distortion in the large scale structure of the geometry, leading it from something he knew very well to something different at all. 

Around 1912, Einstein arrived till this very point. To him, the foundations of this new theory were clear, but he had no idea whatsoever about the mathematical way in which he could have had accomplished its formalization: desperate, he wrote to his friend: $<<$Grossmann, you've got to help me, or I'll go crazy!$>>$...

\

Some years before, during the second half of the XIX century, mathematicians saw the birth of non-Euclidean Geometries. 

There are basically two ways in which those geometries can be thought to be non-Euclidean: the first, the most known one, touches the contents, and corresponds to the fact that the fifth axiom of Euclid, in those geometries, was non considered; the second, methodological, corresponds, instead, to our way of approaching geometrical entities. 

For the latter, the revolution came before, by hand of Ren\'e Descartes, who took the synthetic Euclidean geometry and put a frame on it, allowing the possibility to describe the geometry in an analytic way: geometrical entities were not imagined, but written in terms of equations -- they were not in the brain of the mathematicians, but on their sheets of paper. After that, mathematicians had tools of calculus so powerful that it was easier to consider geometrical objects in three dimensions, and it became even possible to consider them in four dimensions, and more; in this new, more fertile, environment, mathematicians were endowed with so a deeper view of the concepts they were working on that some of them tried to criticize even the axiomatic foundations of geometry themselves: removing the fifth postulate of Euclid, and performing calculations in such generalized spaces without meeting any contradiction, Lobacevskij and Beltrami, among others, have been able to give examples of spaces in which the parallel straight lines of Euclidean geometry were able to get further or closer, as if the space were curved. 

However, the possibility to describe curved spaces in any number of dimensions, had a fundamental problem, namely, the frame Descartes introduced carried with it the peculiarity of something put by hand, without any conceptual reason, which was not something elegant for the taste of mathematicians. It is due to Karl Friedrich Gauss and Bernhard Riemann the development of methods able to restore, in this powerful analytic environment, the syntheticity possessed by an intrinsic approach.

Finally, the opera was completed by an article in which a theory of differential calculus independent on the frame, and therefore called absolute, was developed: the bases of Absolute Differential Calculus were thrown by the Italian mathematicians Gregorio Ricci-Curbastro and his pupil Tullio Levi-Civita in a paper written in French, \textit{M\'ethodes de calcul differentiel absolu et leurs applications}, published in the German review ``Mathematische Annalen'', 1900, way before that the first insight of relativity was shown.

\

Marcel Grossmann was a mathematician, and he was aware of this work on the Absolute Differential Calculus, and he introduced his friend Einstein to it; the physical theory got the mathematical tool it needed, and during 1914, Einstein and Grossmann published an article called ``\textit{Entwurf}'', containing the draft of the generalization of the theory of special relativity that deals with all the possible and most general systems of reference.

\

The Theory of Relativity, as we said, can be seen as a geometry of curved spacetimes, in which reference systems, described through non-constant metrics, are linked via non-linear transformations; the geometry of curved spacetimes, moreover, has a very intriguing feature, because it can manifest itself, in the effects on the motion of a test body, as a gravitational field!

This is amazingly powerful, if we think that in this way Einstein has been able to make a double take: he explained gravity as an effect of the curvature of the spacetime, reducing the gravitational phenomenology to the theoretical concept of spacetime curvature, by ``creating'' the most general relativistic environment any physical theory should find place in.

Actually, the way Einstein followed to write it in the final form is \emph{not} the one exposed here; in his path, he proceeded lead by intuition, luck, ability, and, as we saw, desperation: it was a path full of ambiguities, tricks, traps and problems; but after ten years of troubles, this is the final product he gave us. 

Philosophically, what moved Einstein was that we didn't have a relativistic theory for physical fields; in particular we didn't have a relativistic theory of gravity: we had a theory of gravity, yes, which nowadays we know to be an approximation of the Einsteinian one, but, at that time, the small discrepancies measured could have possibly been thought as due to some other, not fundamental, effects. 

It is worth noticing that this need is a theoretical one: empiricism didn't determine, and not even influenced, Einstein in his ideas. The final form of the theory of relativity is a pure product of a human mind.

Theory of relativity has its fundamental statement in the principle of relativity, which states that physical equations have to be written in a form independent on the system of reference (whatever it is) used to describe them.

It is at least weird that a physical theory named Relativity is mathematized by a geometry built up on a differential calculus called Absolute; this is the reason for which, in the following, we would like to reverse this tendency, calling it Theory of Absolute Relativity.
%%%%%%%%%%%%%%%%%%%%%%%%%%%%%%%%%%%%%%%%%%%%%%%%%%%%%%%%%%%%%%%%%%%%%%%%%%%%%%%%%%%%%%%%%%%%%%%%%%%
\section{The Principle of Covariance\\ as Principle of Absolute Relativity}
If we decide to call Relativity with the fair name of Absolute Relativity, then we should call its principle Principle of Absolute Relativity. 

Or we can choose the more neutral and technical name of Principle of Covariance; it states that:
\begin{pre}[Covariance]
Physical quantities and their properties have to be expressed in a form that must be independent on the system of reference used to describe them.
\end{pre}

And this is the principle that we are now going to mathematize.
%%%%%%%%%%%%%%%%%%%%%%%%%%%%%%%%%%%%%%%%%%%%%%%%%%%%%%%%%%%%%%%%%%%%%%%%%%%%%%%%%%%%%%%%%%%%%%%%%%%
\subsection{Tensors}
Let be given the following convention on the notation, for which for any upper index equal to a lower one, we will perform the sum over all the possible values of the indices, omitting the sign of summation $\sum$, unless otherwise specified.

Now, let be given the following 
\begin{dfn}
Let be given a $4$-dimensional space $A$, in which (at least) two systems of coordinates $x'^{\mu}$ and $x^{\nu}$ are related by the equation $x'=x'(x)$, and let be $S$
\begin{eqnarray}\nonumber
S=\{z \vert \left(J(z)=\mathrm{det}\left(\frac{\partial x'}{\partial x}\right)=0\right)\}
\end{eqnarray}
the sub-domain of the space in which the transformation is singular, which we will require to be a set of points that is zero almost everywhere in $A$; let us consider two sets of fields $T^{\alpha_{1}...\alpha_{i}}_{\rho_{1}...\rho_{j}}$ and $T'^{\alpha'_{1}...\alpha'_{i}}_{\rho'_{1}...\rho'_{j}}$, which are $4^{(i+j)}$ functions of the coordinates defined in the domain $A \setminus S$: then, if they verify the transformation relation
\begin{eqnarray}
\begin{tabular}{|c|}
\hline\\
$T'^{\alpha'_{1}...\alpha'_{m}}_{\rho'_{1}...\rho'_{n}}(x')\!=\!\mathrm{sign}(J)\!\!
\left(\frac{\partial x^{\rho_{1}}}{\partial x^{\rho'_{1}}}
... \frac{\partial x^{\rho_{n}}}{\partial x^{\rho'_{n}}}\right)\!\!
\left(\frac{\partial x^{\alpha'_{1}}}{\partial x^{\alpha_{1}}}
... \frac{\partial x^{\alpha'_{m}}}{\partial x^{\alpha_{m}}}
T^{\alpha_{1}...\alpha_{m}}_{\rho_{1}...\rho_{n}}(x)\right)_{x=x(x')}$\\ \\
\hline
\end{tabular}
\end{eqnarray}
they are said to be the components of a pseudo-tensorial field, while if they verify the transformation relation 
\begin{eqnarray}
\begin{tabular}{|c|}
\hline\\
$T'^{\alpha'_{1}...\alpha'_{m}}_{\rho'_{1}...\rho'_{n}}(x')\!=\!
\left(\frac{\partial x^{\rho_{1}}}{\partial x^{\rho'_{1}}}
... \frac{\partial x^{\rho_{n}}}{\partial x^{\rho'_{n}}}\right)\!\!
\left(\frac{\partial x^{\alpha'_{1}}}{\partial x^{\alpha_{1}}}
... \frac{\partial x^{\alpha'_{m}}}{\partial x^{\alpha_{m}}}
T^{\alpha_{1}...\alpha_{m}}_{\rho_{1}...\rho_{n}}(x)\right)_{x=x(x')}$\\ \\
\hline
\end{tabular}
\end{eqnarray}
they are said to be the components of a tensorial field; in both cases, we will speak about components of tensorial fields (specifying only where needed).

Since the transformations are a group, then we get an equivalence relation between the different sets of components of a tensorial field; the equivalence relation induces a partition in a quotient, whose elements are called Tensorial Fields.

Upper indices can also be called Controvariant indices and lower indices can also be called Covariant indices; the total number of indices $i+j$ is invariant under coordinates transformations, and so it is well defined, and it is called the Rank of the tensorial field: in particular, tensorial fields of rank $1$ are said Vectorial Fields, and tensorial fields of rank $0$ are said Scalar fields, or Invariant Fields.
\end{dfn}

In the following, since the aim of this geometry is that it has to be local, so point-dependent, we will not consider the special case in which tensorial fields are constants, and so we can unambiguously refer to them, simply, as Tensors.

Now, we give the following, fundamental result
\begin{thm} 
A tensor has all the components vanishing in all the points of a given system of reference if and only if it has all the components vanishing in all the points of any system of reference.
\end{thm}

Now we see why tensors represent perfectly the principle of covariance: writing down properties of physical quantities in the form of vanishing of tensors means that if the property of a given physical quantity is true in a system of reference then it is true in all the systems of reference, and so it does not depend on the system of reference itself.

In this way the principle of covariance can be mathematically translated in the Principle of Tensoriality, namely
\begin{pre}[Tensoriality]
The properties of physical quantities have to be mathematically expressed in the form of vanishing tensors.
\end{pre}

Given the fundamental definition of tensor, and showed why this is precisely the mathematical tool we need in absolute relativity, we can now proceed, giving some of its properties.

First of all, let's see some application of this definition:
\begin{itemize}
\item[A)] Example of Tensors
\begin{itemize}
\item[0)] Tensors of rank $0$ - Scalars - Invariants. The transformation law is given as follow
\begin{eqnarray}\nonumber
T'(x')=(T(x))_{x=x(x')}
\end{eqnarray} 
i.e. they do not transform. An example of this kind of tensor are the constants.
\item[1)] Tensors of rank $1$. They are separated into two different groups, according to the position of the index, which can be upper or lower:  
\begin{eqnarray}\nonumber
T'_{\rho'}(x')=\left(\frac{\partial x^{\rho}}{\partial x^{\rho'}}T_{\rho}(x)\right)_{x=x(x')}\\
\nonumber
T'^{\alpha'}(x')=\frac{\partial x^{\alpha'}}{\partial x^{\alpha}}(T^{\alpha}(x))_{x=x(x')}.
\end{eqnarray}
An examples of these tensors can be easily given, considering the gradient of scalars and the differential of the position: from the chain rule we get 
\begin{eqnarray}
\frac{\partial f}{\partial x^{\rho'}}=\frac{\partial x^{\rho}}{\partial x^{\rho'}}\frac{\partial f}{\partial x^{\rho}}
\end{eqnarray}
and
\begin{eqnarray}
dx'^{\mu}\equiv \frac{\partial x^{\mu}}{\partial x^{\nu}} dx^{\nu}.
\label{dx}
\end{eqnarray}
\item[2)] Tensors of rank $2$. In this case we can also have tensors with mixed indices, i.e. one upper and one lower index: the transformation law is then
\begin{eqnarray}\nonumber
T'^{\alpha' \beta'}(x')=\frac{\partial x^{\alpha'}}{\partial x^{\alpha}}
\frac{\partial x^{\beta'}}{\partial x^{\beta}}\left(T^{\alpha \beta}(x)\right)_{x=x(x')}\\
\nonumber
T'_{\rho' \sigma'}(x')=\left(\frac{\partial x^{\rho}}{\partial x^{\rho'}}
\frac{\partial x^{\sigma}}{\partial x^{\sigma'}}T_{\rho \sigma}(x)\right)_{x=x(x')}\\
\nonumber
T'^{\alpha'}_{\sigma'}(x')=\frac{\partial x^{\alpha'}}{\partial x^{\alpha}}
\left(\frac{\partial x^{\sigma}}{\partial x^{\sigma'}}T^{\alpha}_{\sigma}(x)\right)_{x=x(x')}.
\end{eqnarray}
A very particular example of these tensors is, in fact, the mixed tensor defined as follow
\begin{dt}
The quantity
\begin{eqnarray}
\begin{tabular}{|c|}
\hline\\
$\delta^{\alpha}_{\phantom{\alpha}\beta}=\delta_{\beta}^{\phantom{\beta}\alpha}
=\delta_{\beta}^{\alpha}
=\left\{\begin{array}{ll}
0, & \textrm{if $\alpha \neq \beta$}\\
1, & \textrm{if $\alpha=\beta$}
\end{array} \right.$\\ \\
\hline
\end{tabular}
\end{eqnarray}
is a tensor, which has the same components in any reference system, called Delta of Kronecker.
\end{dt}
\end{itemize}
\item[B)] Examples of non-Tensors
\item[1)] The only example of non-tensors we will bring is given by the position; the set of $4$ quantities $x^{\mu}$ has a transformation law which is not the one required to be a tensor. This fact is particularly important, because the criterion for which physical laws should not contain explicitly the position comes out to be automatically satisfied, once we decide to deal only with tensors! We remark that, even if the position is not a tensor, the differential of the position is a tensor, as we said above, and showed by the relation (\ref{dx}).
\end{itemize}

Moving forward, it is possible to endow this geometry with some additional structures, defining operations on tensors and their properties. 

First of all, given two tensors of the same rank and disposition of indices, their linear combination is defined as the linear combination of all their single components, and it is a tensor of the same rank and disposition of indices; in this way, the set of all the tensors of the same rank and disposition of indices is endowed with a structure of linear space. 

Given a tensor of rank $r$ and a tensor of rank $s$, their product is defined as the product of all their single components, and it is a tensor of rank $r+s$; in this way, the linear space of tensors is endowed with a structure of (non abelian) group, i.e. it is a (non abelian) algebra.

An important operation we can define is the Contraction, for which, given a tensor of rank $r$ with $r>2$ and at least one upper and one lower index, it is possible to consider one of the upper and one of the lower indices, and to force them to have the same value, performing the sum over all the possible values the indices can get; doing so, we get a tensor of rank $r-2$ called the contraction of the given tensor. Obviously, we can repeat the contraction till when we reach tensors without upper or lower indices to contract, or till we reach a tensor whose contraction is identically zero: in this case the tensor is said to be Irreducible.

Now, in the definition of tensors, any index is characterized by two features: its upper or lower position and its ``horizontal'' position, that is its position in the row of columns. 

For this last position, we have the possibility to switch two indices; given a tensor with at least $2$ upper or lower indices, we can switch two indices, getting a tensor called the Transposition of the original tensor in those two indices; if the transposition of a certain tensor is equal to the same tensor up to the sign, the tensor is said to be Symmetric or Antisymmetric in those indices according to the fact that the sign is plus or minus, respectively; a tensor symmetric or antisymmetric in any couple of indices is said to be Completely Symmetric or Completely Antisymmetric, respectively.

For the upper-lower case, instead, we cannot move an index, since this is not, in general, a tensorial operation, i.e. the final result is not, in general, a tensor (for example, the Kronecker delta with two upper or lower indices is not a tensor). 

This operation can be made possible by the introduction of an additional rank $2$ tensor $g$. In fact, if we multiply a generic tensor, let's say, a covector $A_{\mu}$ with $g^{\alpha\beta}$ and we contract the index of $A$ with one of the two indices of $g$, then we get $A_{\mu}g^{\mu\beta}$, which is a vector; analogously, we could have used the same idea to lower down the index of a vector.

Nevertheless, such a procedure is not well defined yet: there is still a problem concerning which one of the two indices of the rank $2$ tensor $g$ has to be kept free.

Let's consider, for example, the vector $A^{\mu}$, and a tensor $g_{\alpha\beta}$: it is clear that the tensor $A^{\mu}g_{\mu\nu}$ is a covector, which can be defined to be the vector $A_{\nu}$ after the lowering procedure; but we can also define $A_{\nu}=A^{\mu}g_{\nu\mu}$. Even worse, we can decide to raise the previously lowered index to the initial position, by using another tensor $g^{\alpha\beta}$, getting a vector $A^{\mu}$ different from the original one!

In order to avoid such a situation, we can think to postulate the following Axiom
\begin{axm}
\label{a1}
Any tensor decomposed into a given configuration of indices must be unique; in particular, an eventual raising-lowering procedure must keep the unicity of the tensor considered, so that raising up and lowering down or lowering down and raising up the same index will leave the tensor unchanged.
\end{axm}

With this axiom we can prove the following
\begin{thm}
An eventual raising tensor $g^{\nu \mu}$ and lowering tensor $g_{\nu \theta}$ must be non-degenerate (i.e. its representative matrix must have a determinant different from zero) and symmetric; seen as matrices, they will be one the inverse of the other, so that
\begin{eqnarray}\nonumber
g^{\nu \mu}g_{\nu \theta}=\delta^{\theta}_{\nu}
\end{eqnarray}
where $\delta^{\theta}_{\nu}$ is the Delta of Kronecker.\\
$\square$
\footnotesize
\textit{Proof.} In general, if we suppose that two different contractions are allowed, we can label the two tensors in two different ways, according to which one of the two indices is contracted. For example, we will define
\begin{eqnarray}\nonumber
(A^{\mu})_{S}=A_{\nu}g^{\nu \mu}\\
\nonumber
(A^{\mu})_{D}=A_{\nu}g^{\mu \nu}.
\end{eqnarray}
According to the axiom, if we raise and then lower the index, the final result is nothing else but the vector we started with; two different procedures for raising and lowering indices give in total four different cases of the raise-lower, or lower-raise, procedure, listed as follow
\begin{eqnarray}\nonumber
A_{\mu}g^{\mu \sigma}g_{\sigma \kappa}=A_{\kappa}\\
\nonumber
A_{\mu}g^{\mu \sigma}g_{\kappa \sigma}=A_{\kappa}\\
\nonumber
A_{\mu}g^{\sigma \mu}g_{\sigma \kappa}=A_{\kappa}\\
\nonumber
A_{\mu}g^{\sigma \mu}g_{\kappa \sigma}=A_{\kappa}.
\end{eqnarray}
Taking the differences of the previous ones, we get relationships, which has to be valid for any vector, or in general, any tensor we consider, giving
\begin{eqnarray}\nonumber
g^{\mu \sigma}(g_{\sigma \kappa}-g_{\kappa \sigma})=0\\
\nonumber
(g^{\sigma \mu}-g^{\mu \sigma})g_{\sigma \kappa}=0\\
\nonumber
g^{\sigma \mu}(g_{\sigma \kappa}-g_{\kappa \sigma})=0\\
\nonumber
(g^{\sigma \mu}-g^{\mu \sigma})g_{\kappa \sigma}=0
\end{eqnarray}
and we also have
\begin{eqnarray}\nonumber
g^{\mu \sigma}g_{\sigma \kappa}=\delta^{\mu}_{\kappa}\\
\nonumber
g^{\mu \sigma}g_{\kappa \sigma}=\delta^{\mu}_{\kappa}\\
\nonumber
g^{\sigma \mu}g_{\sigma \kappa}=\delta^{\mu}_{\kappa}\\
\nonumber
g^{\sigma \mu}g_{\kappa \sigma}=\delta^{\mu}_{\kappa}.
\end{eqnarray}
So
\begin{eqnarray}\nonumber
0=g_{\mu \alpha}(g^{\mu \sigma}(g_{\sigma \kappa}-g_{\kappa \sigma}))
=(g_{\mu \alpha}g^{\mu \sigma})(g_{\sigma \kappa}-g_{\kappa \sigma})\\
\nonumber
=\delta_{\alpha}^{\sigma}(g_{\sigma \kappa}-g_{\kappa \sigma})
=g_{\alpha \kappa}-g_{\kappa \alpha}
\end{eqnarray}
and then
\begin{eqnarray}\nonumber
g_{\alpha \kappa}\equiv g_{\kappa \alpha}
\end{eqnarray}
and in the same way we can prove that
\begin{eqnarray}\nonumber
g^{\alpha \kappa}\equiv g^{\kappa \alpha}.
\end{eqnarray}
Finally, we can see that these two tensors, seen as matrices, are one the inverse of the other; in particular, their determinant must be different from zero.
\normalsize
$\blacksquare$
\end{thm}

Now that we have discussed all the features that raising and lowering procedures must have in order to be able to raise and lower tensorial indices without pathologies, we can give the following
\begin{dfn}
A non-degenerate symmetric tensor $g_{\alpha \beta}$ and a non-degenerate symmetric tensor $g^{\alpha \beta}$ are called respectively Fundamental Lowering Tensor and Fundamental Raising Tensor; they verify the relation
\begin{eqnarray}
\begin{tabular}{|c|}
\hline\\
$g_{\mu \nu}g^{\nu \alpha}=\delta_{\mu}^{\alpha}$\\ \\
\hline
\end{tabular}
\label{inv}
\end{eqnarray}
and their determinants will be written as $g$ and $g^{-1}$.
\end{dfn}

In order to formalize what we have said before, we have the fundamental 
\begin{dt}
Given a tensor with at least one upper index, it is always possible to use the fundamental lowering tensor in order to lower that index down; given a tensor with at least one lower index, it is always possible to use the fundamental raising tensor in order to raise that index up: in any case, the rank of the tensor will not be modified. This procedure is called, respectively, lowering and raising procedure of tensorial indices. 
\end{dt}    

An important thing we have to think about, is to find a possible tensor with all those properties that would make it suitable to represent the fundamental tensor $g$.

It is quite easy to see that for a space endowed with a metric, we have the possibility to build up a fundamental tensor with the following prescription: consider the First Fundamental Form of a metric space $ds^{2}$, and write it in the form $ds^{2}=a_{\mu\nu}dx^{\mu}dx^{\nu}$; since $ds^{2}$ is an invariant of the space, and since from the chain rule $dx^{\mu}$ transforms as a vector, as in equation (\ref{dx}), then $a_{\mu\nu}$ has to be a rank $2$ tensor; also, it is non degenerate and symmetric by construction: so we can identify $a_{\mu\nu}=g_{\mu\nu}$, and then we can build $g^{\mu\nu}$ as the unique tensor which satisfies the relation (\ref{inv}).

In this way, we can consider the metric tensor as the fundamental lowering tensor and its inverse as the fundamental raising tensor.

As we just saw, the symmetry of the metric tensor provides many important properties, but it is not enough to provide one of the most important ones.

The diagonalizability, which is ensured by symmetry in a space in which in at least one reference system the metric tensor is constant, or in spaces with at most three dimensions, is not ensured any longer, if the space we consider has more than three dimensions and the metric tensor depends on the point. Then, in general, it is not possible to diagonalize the metric tensor, even if it is symmetric.

However, if we drop the need to have a global diagonalizability, locally we have the result
\begin{thm}
It is always possible to find a reference system in which in a given point the metric tensor is diagonalizable, and in which the elements in the diagonal are unitary positive or negative elements, up to their order.
\end{thm}

It could be possible to prove that
\begin{thm}
The number of negative - and so positive - elements in the diagonal does not change, changing the point in a given reference system or changing the reference system: that number will be said Signature of the metric tensor.
\end{thm}

In the following, since we shall talk about physical theories, time and space have to be represented by coordinates with opposite sign in the metric, and we will always use the convention to choose the temporal coordinate with positive sign, i.e. the temporal component of the metric to be positive: we will then chose $s=3$ for the signature of the metric.

Symmetry allows the metric tensor to verify many algebraic relationships.

\begin{thm}
We have the following identities
\begin{eqnarray}
\frac{\partial g}{\partial g_{\mu \nu}}=gg^{\mu \nu}\\
\frac{\partial g}{\partial g^{\mu \nu}}=-gg_{\mu \nu}
\label{detg}
\end{eqnarray}
and so
\begin{eqnarray}
\partial_{\alpha}g=-gg_{\mu\nu}\partial_{\alpha}g^{\mu\nu}=gg^{\mu\nu}\partial_{\alpha}g_{\mu\nu}.
\end{eqnarray}
\end{thm}
As we can see, the determinant of the metric and all the formulae in which it is involved are \emph{not} tensorial.

In fact, we have the transformation rules given as 
\begin{thm}
The determinant of the metric tensor transforms as
\begin{eqnarray}\nonumber
g'(x')=(J^{-2}g(x))_{x=x(x')}
\end{eqnarray}
where $J$ is the Jacobian of the transformation.
\end{thm}

We have that
\begin{thm}
In the spacetime, we have that
\begin{eqnarray}
\nonumber
\mathrm{sign}(g)=-1
\end{eqnarray}
and it is a scalar.
\end{thm}

The fact that the determinant of the metric is not a tensor does not prevent us to consider another important non-tensorial quantity, in order to build up a tensor.
\begin{dfn}
The set of coefficients
\begin{eqnarray}
\nonumber
\epsilon_{i_{1}i_{2}i_{3}i_{4}}=\epsilon_{\sigma(1)\sigma(2)\sigma(3)\sigma(4)}\equiv \epsilon_{\sigma(1234)}=\mathrm{sign}(\sigma)
\end{eqnarray}
is such that it is zero in all the cases in which $\sigma(1234)$ is not a permutation of $(1234)$, otherwise it has a unitary value with positive or negative sign, according to the fact that the permutation is even or odd, respectively; it is called Set of Coefficients Epsilon of Ricci-Levi-Civita.
\end{dfn}
We have directly from the definition that this set of coefficients does not transform as a tensor; but they transform in such a particular way, that they can be coupled together with the determinant of the metric, to get tensors; we have that
\begin{dt}
In the spacetime, the quantity
\begin{eqnarray}
\nonumber
\varepsilon_{\alpha\beta\gamma\delta}(x)=\epsilon_{\alpha\beta\gamma\delta}\sqrt{|g(x)|}
\end{eqnarray} 
is a pseudo-tensor of rank $4$ completely antisymmetric; with it, we can define the pseudo-tensor of rank $4$ completely antisymmetric 
\begin{eqnarray}
\nonumber
\varepsilon^{abcd}=
-g^{\alpha a}g^{\beta b}g^{\gamma c}g^{\delta d}\varepsilon_{\alpha\beta\gamma\delta}
\end{eqnarray}
and they are called Pseudo-Tensor Epsilon of Ricci-Levi-Civita. 

They verify the following properties
\begin{eqnarray}
\nonumber
\varepsilon^{abcd}\varepsilon_{\alpha\beta\gamma\delta}=
-\mathrm{det}\left( \begin{array}{cccc}
\delta^{a}_{\alpha}&\delta^{b}_{\alpha}&\delta^{c}_{\alpha}&\delta^{d}_{\alpha}\\
\delta^{a}_{\beta}&\delta^{b}_{\beta}&\delta^{c}_{\beta}&\delta^{d}_{\beta}\\
\delta^{a}_{\gamma}&\delta^{b}_{\gamma}&\delta^{c}_{\gamma}&\delta^{d}_{\gamma}\\
\delta^{a}_{\delta}&\delta^{b}_{\delta}&\delta^{c}_{\delta}&\delta^{d}_{\delta}
\end{array}\right),
\end{eqnarray}
in particular
\begin{eqnarray}
\nonumber
\varepsilon^{\rho bcd}\varepsilon_{\rho \beta\gamma\delta}=
-\mathrm{det}\left( \begin{array}{ccc}
\delta^{b}_{\beta}&\delta^{c}_{\beta}&\delta^{d}_{\beta}\\
\delta^{b}_{\gamma}&\delta^{c}_{\gamma}&\delta^{d}_{\gamma}\\
\delta^{b}_{\delta}&\delta^{c}_{\delta}&\delta^{d}_{\delta}
\end{array}\right),
\end{eqnarray}
in particular
\begin{eqnarray}
\nonumber
\varepsilon^{\rho \sigma cd}\varepsilon_{\rho \sigma \gamma\delta}=
-2 \mathrm{det} \left( \begin{array}{cc}
\delta^{c}_{\gamma}&\delta^{d}_{\gamma}\\
\delta^{c}_{\delta}&\delta^{d}_{\delta}
\end{array}\right),
\end{eqnarray}
in particular
\begin{eqnarray}
\nonumber
\varepsilon^{\rho \sigma \tau d}\varepsilon_{\rho \sigma \tau \delta}=-6\delta^{d}_{\delta},
\end{eqnarray}
in particular
\begin{eqnarray}
\nonumber
\varepsilon^{\rho \sigma \tau \eta}\varepsilon_{\rho \sigma \tau \eta}=-24.
\end{eqnarray}
\end{dt}

Finally, we give the following
\begin{dt}
Let be $T$ a tensor of rank $k$ completely antisymmetric; the quantity
\begin{eqnarray}\nonumber
(\ast T)^{\alpha_{k+1}...\alpha_{4}}=
\frac{1}{k!}\varepsilon^{\alpha_{1}...\alpha_{4}}T_{\alpha_{1}...\alpha_{k}}
\end{eqnarray}
is a tensor of rank $(4-k)$ completely antisymmetric called Dual of the previous tensor, and the operator $\ast$ is an automorphism in the subspace of the antisymmetric tensors called Hodge Operator. We also have that if $T$ is a tensor (pseudo-tensor) then $\ast T$ is a pseudo-tensor (tensor).
\end{dt}

We can also apply more times the Hodge operator, with the following result
\begin{thm}
In the spacetime, let be $T$ an antisymmetric tensor of rank $k$: then we have the identity
\begin{eqnarray}\nonumber
(\ast(\ast T))=(T)[(-1)^{k+1}].
\end{eqnarray}
\end{thm}

\subsection{Covariant Differentiation on Tensors}
Everything we developed here above consists in the introduction of tensors.

But a physical theory has to be written in terms of differential equations, and it has to be built up on the fundamental concept of derivative; hence, we need to develop also a differential theory for tensors.

The problem we have to face is defining a differentiation acting on tensors, which gives as a final result a tensor.

It is obvious that the derivative we are looking for, has to be a generalization of the usual partial derivative, since the partial derivative transform according to the law
\begin{eqnarray}\nonumber
\frac{\partial T'^{\alpha'}}{\partial x^{\mu'}}
=\frac{\partial x^{\alpha'}}{\partial x^{\alpha}}\frac{\partial x^{\mu}}{\partial x^{\mu'}}
\frac{\partial T^{\alpha}}{\partial x^{\mu}}
+T^{\kappa}\frac{\partial^{2}x^{\alpha'}}{\partial x^{\kappa}\partial x^{\mu'}}
\end{eqnarray}
which is not the transformation law for tensors.

By adding to the usual partial derivative another field in the following way
\begin{eqnarray}\nonumber
D_{\iota}V_{\alpha}=\partial_{\iota}V_{\alpha}+V_{\kappa}C^{\kappa}_{\iota \alpha}      
\end{eqnarray}
we ensure to the generalized derivation the linearity.

It could be possible to prove that, by requiring this generalized derivative to verify the tensorial transformation law, we can see how the compensating field has to transform; the transformation law is given as 
\begin{eqnarray}\nonumber
C'^{\kappa}_{\alpha \beta}
=\frac{\partial x'^{\kappa}}{\partial x^{\rho}}\frac{\partial x^{\pi}}{\partial x'^{\alpha}}
\frac{\partial x^{\sigma}}{\partial x'^{\beta}}C^{\rho}_{\pi \sigma}
-\frac{\partial x'^{\kappa}}{\partial x^{\rho}}
\frac{\partial^{2} x^{\rho}}{\partial x'^{\alpha} \partial x'^{\beta}}
\end{eqnarray}
and so, the transformation law for the generalized derivative is
\begin{eqnarray}\nonumber
D'_{\rho'}V_{\sigma'}
=\frac{\partial x^{\rho}}{\partial x^{\rho'}}
\frac{\partial x^{\sigma}}{\partial x^{\sigma'}}
D_{\rho}V_{\sigma}
\end{eqnarray}
that is, it is actually the tensorial transformation law we were looking for.

Furthermore, it could be possible to prove that the same transformation law for the coefficients $C^{\alpha}_{\iota \kappa}$ is enough to ensure the fact that also the generalization of the previous definition to tensors of any rank
\begin{eqnarray}\nonumber
D_{\iota}T^{\alpha...\mu}_{\rho...\omega}
=\partial_{\iota} T^{\alpha...\mu}_{\rho...\omega}
+(T^{\kappa...\mu}_{\rho...\omega}\Gamma^{\alpha}_{\kappa \iota}...      
+T^{\alpha...\kappa}_{\rho...\omega}\Gamma^{\mu}_{\kappa \iota})\\
\nonumber
-(T^{\alpha...\mu}_{\kappa...\omega}\Gamma^{\kappa}_{\rho \iota}...      
+T^{\alpha...\mu}_{\rho...\kappa}\Gamma^{\kappa}_{\omega \iota}),
\end{eqnarray}
in which also the Leibniz rule is ensured, is a tensor.

Leibniz rule beside linearity makes this definition the definition of a derivative; the transformation law for the compensating field gives the tensoriality: the previous is, then, the most general expression we can have for a tensorial differentiation.
%%%%%%%%%%%%%%%%%%%%%%%%%%%%%%%%%%%%%%%%%%%%%%%%%%%%%%%%%%%%%%%%%%%%%%%%%%%%%%%%%%%%%%%%%%%%%%%%%%%
\subsubsection{The Coefficients of Connection}
\begin{dfn}
A set of coefficients that transforms according to the law
\begin{eqnarray}
\begin{tabular}{|c|}
\hline\\
$\Gamma'^{\alpha'}_{\mu' \nu'}(x')=
\frac{\partial x^{\mu}}{\partial x'^{\mu'}}
\frac{\partial x^{\nu}}{\partial x'^{\nu'}}
\left(\frac{\partial x'^{\alpha'}}{\partial x^{\alpha}}\Gamma^{\alpha}_{\mu\nu}(x)\right)_{x=x(x')}
+\frac{\partial ^2 x^{\kappa}}{\partial x'^{\mu'} \partial x'^{\nu'}}             
\left(\frac{\partial x'^{\alpha'}}{\partial x^{\kappa}}\right)_{x=x(x')}$\\ \\
\hline
\end{tabular}
\end{eqnarray}
is called set of the Coefficients of Connection, or simply Connection.
\end{dfn}

An obvious remark is that, from its own structure, the connection is not, in general, a tensor (it turns out to be a tensor if and only if the transformation between the two systems of reference is linear); thus, we can ask whether it exists a system of reference in which it is equal to zero, even if it is not vanishing in other reference systems.

Another thing we can notice is that, in general, the connection has no symmetry in the two lower indices. So, we can separate the connection as a sum of two parts
\begin{eqnarray}
\nonumber
\Gamma^{\alpha}_{\mu \nu}=\frac{1}{2}\left(
\Gamma^{\alpha}_{\mu \nu}+\Gamma^{\alpha}_{\nu \mu}\right)+
\frac{1}{2}\left(
\Gamma^{\alpha}_{\mu \nu}-\Gamma^{\alpha}_{\nu \mu}\right),
\end{eqnarray}
which are symmetric and antisymmetric in the lower indices: the antisymmetric part \emph{is} actually a tensor, and the characteristic to transform as the connection is completely given in inheritance to the symmetric part; moreover, the two different parts transform without mixing between each other, hence, they are two formally independent parts. Then, we can define 
\begin{dt}
Given the connection, the following quantity
\begin{eqnarray}
\begin{tabular}{|c|}
\hline\\
$\Gamma^{\alpha}_{\mu \nu}-\Gamma^{\alpha}_{\nu \mu}=F^{\alpha}_{\phantom{\alpha}\mu \nu}$\\ \\
\hline
\end{tabular}
\end{eqnarray}
is a tensor called Cartan Tensor, and it is antisymmetric in the second and third index.
\end{dt}
And the connection is now writable as
\begin{eqnarray}\nonumber
\Gamma^{\alpha}_{\mu \nu}=\frac{\Gamma^{\alpha}_{\mu \nu}+
\Gamma^{\alpha}_{\nu \mu}}{2}+\frac{1}{2}F^{\alpha}_{\phantom{\alpha}\mu \nu}.
\end{eqnarray}

The issue of the existence of a frame in which the connection vanishes and the symmetry properties in the two lower indices are related between each other. To see this, we can start noticing that the connection does contain an antisymmetric, tensorial part, which, being tensorial, prevents the whole connection to be zero in a given frame. On the other hand, when Cartan tensor vanishes, then we could ask ourselves whether it is possible to get a frame in which the connection vanishes too; the answer, however, is only partial, since in this case we can actually find such a frame, but in it the connection will only be locally equal to zero, as expressed by the following
\begin{thm}[Weyl]
If the Cartan tensor is zero then it always exists a reference system in which at least in one point the whole connection is zero.
\end{thm}

A following step consists then in finding some criterion for which we can extend this result to a global statement.

To get this extension, it is clear that we will have to consider also derivatives of the connection; the reason for this is that a differential structure on the connection would allow us to extend the vanishing of the connection in a given point to a whole neighborhood of that point, more or less in the same way in which a function that is zero in a point and its derivatives are zero in that point is equal to zero locally around that point.

The idea is then to look for a quantity built up by using the derivatives of the connection.

The way in which such a quantity can be suggested, and in which a general theorem about the global vanishing of the connection can be proven, passes through the Frobenius theorem of Analysis, which reads as follow
\begin{thm}[Frobenius]
Let us consider the set of functions
\begin{eqnarray}\nonumber
F_{i j}: \mathbb{R}^{n}\times \mathbb{R}^{p} \longrightarrow \mathbb{R}
\end{eqnarray}
in which $i=1, \ldots, n$ e $j=1, \ldots, p$, and the differential equations for the function
\begin{eqnarray}\nonumber
y_{j}=y_{j}(x^{i})
\end{eqnarray}
given by
\begin{eqnarray}\nonumber
\frac{\partial y_{j}}{\partial x^{i}}=F_{j i}(x; y).
\end{eqnarray}

If we have that
\begin{eqnarray}\nonumber
\frac{\partial F_{j i}(x; y)}{\partial x^{k}}+
\sum_{r} \frac{\partial F_{j i}(x; y)}{\partial y_{r}}F_{r k}(x; y)=\\
\nonumber
=\frac{\partial F_{j k}(x; y)}{\partial x^{i}}+
\sum_{r} \frac{\partial F_{j k}(x; y)}{\partial y_{r}}F_{r i}(x; y)
\end{eqnarray}
then, given a set of initial conditions, there exists a unique solution of the differential equations
\begin{eqnarray}\nonumber
y_{j}=y_{j}(x^{i}).
\end{eqnarray}
\end{thm}

The main idea of the theorem is that given the functions in the form
\begin{eqnarray}\nonumber
F_{j i}(x; y)=\frac{\partial y_{j}}{\partial x^{i}}
\end{eqnarray}
then
\begin{eqnarray}\nonumber
\frac{\partial F_{j i}(x; y)}{\partial x^{k}}+
\sum_{r} \frac{\partial F_{j i}(x; y)}{\partial y_{r}}F_{r k}(x; y)=\\
\nonumber
=\frac{\partial F_{j i}(x; y(x))}{\partial x^{k}}+
\sum_{r} \frac{\partial F_{j i}(x; y(x))}{\partial y_{r}}\frac{\partial y_{r}}{\partial x^{k}}=\\
\nonumber
=\frac{\partial F_{j i}(x)}{\partial x^{k}}
\end{eqnarray}
so that the condition to satisfy is
\begin{eqnarray}\nonumber
\frac{\partial F_{j i}(x)}{\partial x^{k}}=\frac{\partial F_{j k}(x)}{\partial x^{i}}
\end{eqnarray}
or, in an equivalent way
\begin{eqnarray}\nonumber
\frac{\partial }{\partial x^{k}}\frac{\partial y_{j}}{\partial x^{i}}
=\frac{\partial }{\partial x^{i}}\frac{\partial y_{j}}{\partial x^{k}}
\end{eqnarray}
which is the same to require that the partial derivatives commutes: it is an integrability condition, the one we need to extend the local vanishing of the connection to a global result. 

We have then a corollary, given in the linear case, as follow
\begin{thm}[Frobenius]
Let us consider the set of functions
\begin{eqnarray}\nonumber
F_{i j}: \mathbb{R}^{n}\times \mathbb{R}^{p} \longrightarrow \mathbb{R}
\end{eqnarray}
in which $i=1, \ldots, n$ e $j=1, \ldots, p$, and the differential equations for the function
\begin{eqnarray}\nonumber
y_{j}=y_{j}(x^{i})
\end{eqnarray}
given by
\begin{eqnarray}\nonumber
\frac{\partial y_{j}}{\partial x^{i}}=\sum_{k} F^{k}_{ji}(x)y_{k}.
\end{eqnarray}

If we have that
\begin{eqnarray}
\nonumber
\frac{\partial F^{a}_{j i}(x)}{\partial x^{k}}-\frac{\partial F^{a}_{j k}(x)}{\partial x^{i}}
+F^{a}_{r k}(x)F^{r}_{j i}(x)-F^{a}_{r i}(x)F^{r}_{j k}(x)=0
\end{eqnarray}
then, given a set of initial conditions, there exists a unique solution of the differential equations
\begin{eqnarray}\nonumber
y_{j}=y_{j}(x^{i}).
\end{eqnarray}
\end{thm}

Now, we can give the following, fundamental
\begin{dt}
Given the connection, the following quantity
\begin{eqnarray}
\begin{tabular}{|c|}
\hline\\
$\partial_{k} \Gamma^{a}_{j i}-\partial_{i} \Gamma^{a}_{j k}+
\Gamma^{a}_{r k}\Gamma^{r}_{j i}-\Gamma^{a}_{r i}\Gamma^{r}_{j k}=F^{a}_{\phantom{a} j k i}$\\ \\
\hline
\end{tabular}
\end{eqnarray}
is a tensor called Riemann Tensor, and it is antisymmetric in the third and fourth index.
\end{dt}

And now, it is possible to prove that
\begin{thm}
It is always possible to find a system of reference in which the connection is (globally) zero if and only if both Cartan and Riemann tensors vanish.\\
$\square$
\footnotesize
\textit{Proof.} First, let us take into account Frobenius theorem, for which if
\begin{eqnarray}\nonumber
\frac{\partial C^{a}_{k \rho}}{\partial x^{i}}+C^{a}_{k r}C^{r}_{i \rho}
-\frac{\partial C^{a}_{i \rho}}{\partial x^{k}}-C^{a}_{i r}C^{r}_{k \rho}=0
\end{eqnarray}
then
\begin{eqnarray}\nonumber
\frac{\partial Y^{a}}{\partial x^{k}}=C^{a}_{k \rho}Y^{\rho};
\end{eqnarray}
this is true also for $C^{a}_{i r}=-\Gamma^{a}_{i r}$, and in the case in which $Y^{j}=\frac{\partial x^{j}}{\partial x'^{\mu}}|_{x'=x}$, so that, if we have the condition 
\begin{eqnarray}\nonumber
\frac{\partial \Gamma^{a}_{i \rho}}{\partial x^{k}}+\Gamma^{a}_{k r}\Gamma^{r}_{i \rho}
-\frac{\partial \Gamma^{a}_{k \rho}}{\partial x^{i}}-\Gamma^{a}_{i r}\Gamma^{r}_{k \rho}=0
\end{eqnarray}
then we also have that
\begin{eqnarray}\nonumber
\frac{\partial }{\partial x^{k}}\frac{\partial x^{a}}{\partial x'^{\mu}}+
\Gamma^{a}_{k \rho}\frac{\partial x^{\rho}}{\partial x'^{\mu}}=0.
\end{eqnarray}

Now, we have the following (non-tensorial) identity
\begin{eqnarray}\nonumber
F^{a}_{\phantom{a} jik}
+\frac{\partial F^{a}_{ji}}{\partial x^{k}}-\frac{\partial F^{a}_{jk}}{\partial x^{i}}
+F^{a}_{r k}\Gamma^{r}_{ij}+\Gamma^{a}_{kr}F^{r}_{j i}
-F^{a}_{ri}\Gamma^{r}_{kj}-\Gamma^{a}_{ir}F^{r}_{j k}-\\
\nonumber
-F^{a}_{ri}F^{r}_{j k}+F^{a}_{r k}F^{r}_{j i}=
\frac{\partial \Gamma^{a}_{kj}}{\partial x^{i}}-\frac{\partial \Gamma^{a}_{ij}}{\partial x^{k}}
+\Gamma^{a}_{ir}\Gamma^{r}_{kj}-\Gamma^{a}_{kr}\Gamma^{r}_{ij}.
\end{eqnarray}

So, if the Riemann and Cartan tensors are zero, then the left side of the identity above is zero, giving
\begin{eqnarray}\nonumber
\frac{\partial \Gamma^{a}_{kj}}{\partial x^{i}}-\frac{\partial \Gamma^{a}_{ij}}{\partial x^{k}}
+\Gamma^{a}_{ir}\Gamma^{r}_{kj}-\Gamma^{a}_{kr}\Gamma^{r}_{ij}\equiv0
\end{eqnarray}
which is the hypothesis for the Frobenius's theorem that gives the thesis
\begin{eqnarray}\nonumber
\frac{\partial }{\partial x^{k}}\frac{\partial x^{a}}{\partial x'^{\mu}}+
\Gamma^{a}_{k \rho}\frac{\partial x^{\rho}}{\partial x'^{\mu}}=0,
\end{eqnarray}
and this relation implies that the connection transforms as 
\begin{eqnarray}\nonumber
\left(\Gamma^{\alpha}_{\beta \mu}\right)'=
\frac{\partial x'^{\alpha}}{\partial x^{\nu}}\frac{\partial x^{\theta}}{\partial x'^{\beta}}
\left(\frac{\partial x^{\rho}}{\partial x'^{\mu}}\Gamma^{\nu}_{\theta \rho}
+\frac{\partial}{\partial x^{\theta}}\frac{\partial x^{\nu}}{\partial x'^{\mu}}\right)=
\frac{\partial x'^{\alpha}}{\partial x^{\nu}}\frac{\partial x^{\theta}}{\partial x'^{\beta}}0=0,
\end{eqnarray}
so, it exists a system of reference in which the connection is zero, identically.

On the other hand, if it exists a system of reference in which the connection is zero, then Riemann and Cartan tensors are zero in that frame, and so they are zero in any frame, which means that they are zero tensors.
\normalsize
$\blacksquare$
\end{thm}

So, we do have a criterion for the global vanishing of the connection.

We see that Cartan tensor plays the role of a tensorial connection and Riemann tensor plays the role of a tensorial derivative of the connection, and if they are both zero, then the connection itself is zero, at least in one system of reference; this, roughly speaking, allows us to say that the connection is somehow equivalent to Cartan and Riemann tensors considered together.  

We want to underline that Cartan and Riemann tensors are tensors related to the connection, so that, since many inequivalent connections can be \emph{a priori} defined for a given space, then many different Cartan and Riemann tensors can be defined in it.
%%%%%%%%%%%%%%%%%%%%%%%%%%%%%%%%%%%%%%%%%%%%%%%%%%%%%%%%%%%%%%%%%%%%%%%%%%%%%%%%%%%%%%%%%%%%%%%%%%%
\subsubsection{The Covariant Derivation}
Given the connection, we can determine the covariant derivation by giving the following
\begin{dt}
Let be given a connection; if $T$ is a tensor then the quantity
\begin{eqnarray}
\begin{tabular}{|c|}
\hline\\
$D_{\mu}T^{\alpha_{1}...\alpha_{p}}_{\rho_{1}...\rho_{q}}
=\partial_{\mu} T^{\alpha_{1}...\alpha_{p}}_{\rho_{1}...\rho_{q}}
+\sum_{j=1}^{j=p}T^{\alpha_{1}...\kappa...\alpha_{p}}_{\rho_{1}...\rho_{q}}
\Gamma^{\alpha_{j}}_{\kappa \mu}
-\sum_{j=1}^{j=q}T^{\alpha_{1}...\alpha_{p}}_{\rho_{1}...\kappa...\rho_{q}}
\Gamma^{\kappa}_{\rho_{j} \mu}$\\ \\
\hline
\end{tabular}
\end{eqnarray}
is a tensor called Covariant Derivative of the tensor $T$ with respect to the $\mu$ coordinate.
\end{dt}

Then, it is possible to define a covariant directional derivative, by using a vectorial field
\begin{dfn}
Let be $A$ a vectorial field, that is a direction; if $T$ is a tensor then the Covariant Directional Derivative along the direction of the vector field $A$ is defined by
\begin{eqnarray}
\nonumber
A^{\iota}[\frac{\partial T^{\alpha...\mu}_{\rho...\omega}}{\partial x^{\iota}}
+(T^{\kappa...\mu}_{\rho...\omega}\Gamma^{\alpha}_{\kappa \iota}...      
+T^{\alpha...\kappa}_{\rho...\omega}\Gamma^{\mu}_{\kappa \iota})
-(T^{\alpha...\mu}_{\kappa...\omega}\Gamma^{\kappa}_{\rho \iota}...      
+T^{\alpha...\mu}_{\rho...\kappa}\Gamma^{\kappa}_{\omega \iota})]\equiv\\
\nonumber
\equiv A^{\iota}D_{\iota}T^{\alpha...\mu}_{\rho...\omega}=D_{A}T^{\alpha...\mu}_{\rho...\omega}.
\end{eqnarray}
\end{dfn}

Again, we can then define the directional derivative of the vectorial field itself, having the expression
\begin{eqnarray}\nonumber
D_{A}A^{\mu}=A^{\iota}D_{\iota}A^{\mu}
\end{eqnarray}
so that we can define
\begin{dfn}
If a vectorial field $A$ is such that 
\begin{eqnarray}\nonumber
D_{A}A^{\mu}=0
\end{eqnarray}
then the curve to which the vector field is always tangent is said to be Autoparallel.
\end{dfn}

Given a covariant derivative with respect to an index, it is possible to calculate the commutator of covariant derivatives with respect to two different indices; the final result is something we can write in term of quantities we already know, and it is given as 
\begin{thm}
We have that
\begin{eqnarray}
\nonumber
[D_{\alpha},D_{\beta}]T^{\alpha_{1}...\alpha_{p}}_{\rho_{1}...\rho_{q}}
=F^{\eta}_{\alpha \beta}D_{\eta}T^{\alpha_{1}...\alpha_{p}}_{\rho_{1}...\rho_{q}}
+\sum_{j=1}^{j=p}T^{\alpha_{1}...\kappa...\alpha_{p}}_{\rho_{1}...\rho_{q}}
F^{\alpha_{j}}_{\kappa \alpha \beta}\\
-\sum_{j=1}^{j=q}T^{\alpha_{1}...\alpha_{p}}_{\rho_{1}...\kappa...\rho_{q}}
F^{\kappa}_{\rho_{j}\alpha \beta}
\end{eqnarray}
for any tensor $T$.
\end{thm}
From the Jacobi identities we get the following identities for the Cartan and Riemann tensors
\begin{thm}[Jacobi-Bianchi identities]
Cartan and Riemann tensors are such that they verify the structural differential identities
\begin{eqnarray}\nonumber
F^{\rho}_{\phantom{\rho}\kappa \nu \mu}
+F^{\rho}_{\phantom{\rho}\mu \kappa \nu}
+F^{\rho}_{\phantom{\rho}\nu \mu \kappa}\equiv\\
\nonumber
\equiv D_{\kappa}F^{\rho}_{\phantom{\rho}\mu \nu}
+D_{\nu}F^{\rho}_{\phantom{\rho} \kappa \mu}
+D_{\mu}F^{\rho}_{\phantom{\rho} \nu \kappa}+\\
+F^{\pi}_{\phantom{\pi} \nu \kappa}F^{\rho}_{\phantom{\rho}\mu \pi}
+F^{\pi}_{\phantom{\pi}\mu \nu}F^{\rho}_{\phantom{\rho}\kappa \pi}
+F^{\pi}_{\phantom{\pi}\kappa \mu}F^{\rho}_{\phantom{\rho}\nu \pi}
\end{eqnarray}
and
\begin{eqnarray}\nonumber
D_{\mu}F^{\nu}_{\phantom{\nu}\iota \kappa \rho}
+D_{\kappa}F^{\nu}_{\phantom{\nu}\iota \rho \mu}
+D_{\rho}F^{\nu}_{\phantom{\nu}\iota \mu \kappa}+\\
+F^{\nu}_{\phantom{\nu}\iota \beta \mu}F^{\beta}_{\phantom{\beta}\rho \kappa}
+F^{\nu}_{\phantom{\nu}\iota \beta \kappa}F^{\beta}_{\phantom{\beta}\mu \rho}
+F^{\nu}_{\phantom{\nu}\iota \beta \rho}F^{\beta}_{\phantom{\beta}\kappa \mu}\equiv0.
\end{eqnarray}
\end{thm}
%%%%%%%%%%%%%%%%%%%%%%%%%%%%%%%%%%%%%%%%%%%%%%%%%%%%%%%%%%%%%%%%%%%%%%%%%%%%%%%%%%%%%%%%%%%%%%%%%%%
\subsubsection{The metric structure of the Coefficients of Connection}
So far, we have underlined the existence of two very particular quantities, whose relationship has to be investigated at a deeper level: the metric and the connection.

We have remarked that the metric tensor, since it must be non-degenerate, it has to be different from zero in any reference system; on the other hand, the connection is not a tensor: for both, we have that they will always depend on the reference system they refer to.

This common feature of theirs makes them more related between each other than how one could have thought \emph{a priori}, and it is worth to go further in the study of their interconnections.

Let us consider the procedure of raising indices; we have \emph{by definition} that $A^{\mu}=A_{\nu}g^{\nu\mu}$. Now, this should be true for any vector or covector, no matter what their explicit expression is: so, in particular, if the vector is the derivative of a scalar $A_{\nu}=D_{\nu}\phi$, then we should have $D^{\mu}\phi=g^{\nu\mu}D_{\nu}\phi$ for any scalar $\phi$, and so formally $D^{\mu}=g^{\nu\mu}D_{\nu}$, which defines the Controvariant Derivative; and in particular, if the vector is the directional derivative of a vector $A_{\nu}=D_{B}V_{\nu}$, then we should have $D_{B}V^{\mu}=g^{\nu\mu}D_{B}V_{\nu}$, thus $B^{\alpha}D_{\alpha}V^{\mu}=g^{\nu\mu}B^{\alpha}D_{\alpha}V_{\nu}$ for any vector $B$, which is $D_{\alpha}V^{\mu}=g^{\nu\mu}D_{\alpha}V_{\nu}$. But on the other side, we also have, as usual, $V^{\mu}=g^{\nu\mu}V_{\nu}$. The two relations put together should then give
\begin{eqnarray}
\nonumber
g^{\nu\mu}D_{\alpha}V_{\nu} \equiv D_{\alpha}V^{\mu} \equiv D_{\alpha}(g^{\nu\mu}V_{\nu})=D_{\alpha}g^{\nu\mu}V_{\nu}+g^{\nu\mu}D_{\alpha}V_{\nu}
\end{eqnarray}
from which we get
\begin{eqnarray}
\nonumber
D_{\alpha}g^{\nu\mu}V_{\nu}=0
\end{eqnarray}
for any vector $V$, and so
\begin{eqnarray}
D_{\alpha}g^{\nu\mu}=0,
\label{dg=0}
\end{eqnarray}
which seems to suggest that the covariant derivative of the metric tensor should be zero.

This condition states that the metric can pass through the covariant differential operator, i.e. the procedure of raising/lowering indices and the covariant differentiation commute: this means that the algebra of the covariant differentiations has the tensor $g$, in all its indices dispositions ($g^{\alpha}_{\beta}=\delta^{\alpha}_{\beta}$, $g^{\alpha\beta}$ and $g_{\alpha\beta}$), as its constant.

Under the point of view of its geometrical interpretation, instead, the condition expressed in equation (\ref{dg=0}) states that the procedure of parallel displacement preserves the metric: if we want that constant vectors have constant lengths and constant angles between them, then the covariant derivative of $g$ must be equal to zero.

Being concerned about the issue of preserving the metric properties of the geometry, or thinking that the raising/lowering procedure should commute with the covariant derivative, we shall assume hereafter the condition given in equation (\ref{dg=0}), or Metricity condition, to hold, and in this case we say that the connection is a Metric connection.

A discussion is, at this point, due; as we said before, for a given space we can define many inequivalent connections, each of them with its own Cartan and Riemann tensors, and also, obviously, with its own form for the covariant derivative of the metric $Dg$. But the previous discussion lead us to postulate the validity of the condition $Dg=0$.

Form a logical viewpoint, we have to distinguish between two cases: among all the connections we can define for a given space, and which can be metric, the first case occurs when some of them are metric while the others are not, the second case occurs when all of them are metric. Suppose now that the case that actually occurs is the first, i.e. that some connection are not metric: it is obvious that this case does not represent the metricity condition, since, even without requiring metricity, it is possible that some connection are metric, by chance; conversely, if we require metricity because we want to work only with metric connections, but we require it in such a way that some of them are allowed to be non-metric, then it could be possible that we pick up a connection that is not metric, which is not what we want to work with: hence, requiring metricity can not mean that we require it only for some connection of the space. The only case left is that we have to postulate metricity for all the connections we can define for a given space (after all, if we postulate metricity because we want that the raising/lowering procedure can act through the covariant differential operator, since the raising/lowering procedure is related to the metric and not to the specific differential operator, then it is natural that the metricity condition should hold, given the metric, for any differential operator, thus for any connection -- the same considerations are true also if we postulate metricity because we want to preserve the metric properties of the space).

In this way, a more precise form, free from ambiguities, of the metricity condition can be stated, and we will postulate hereafter that any connection we can define in a given space is a metric connection; we give this axiom 
\begin{axm}
\label{a2}
Given the metric of the space, any connection has to define a covariant derivative such that
\begin{eqnarray}
\begin{tabular}{|c|}
\hline\\
$D_{\alpha}g_{\rho \omega}\equiv0$\\ \\
\hline
\end{tabular}.
\end{eqnarray}
\end{axm}

Having this axiom, an important result can be stated for the Cartan tensor; but before we have to state the following
\begin{dt}
The set of quantities
\begin{eqnarray}
\begin{tabular}{|c|}
\hline\\
$\Lambda^{\eta}_{\alpha \beta}\equiv\frac{1}{2}g^{\eta \mu}
\left(\partial_{\beta}g_{\alpha\mu}+\partial_{\alpha}g_{\mu\beta}-\partial_{\mu}g_{\alpha\beta}\right)$\\ \\
\hline
\end{tabular}
\end{eqnarray}
verifying $\Lambda^{\eta}_{\alpha \beta}\equiv\Lambda^{\eta}_{\beta \alpha}$ is a symmetric connection, called Symmetric Connection, or Levi-Civita Connection, and it is the only connection verifying the symmetry in the two lower indices.\\
$\square$
\footnotesize
\textit{Proof.} First of all, it is easy to see that this is a connection, which is symmetric, and indeed it verifies the condition of metricity; vice versa, if one suppose the connection to be symmetric, the symmetry property (together with the axiom \ref{a2}) gives that the connection is precisely given by $\Lambda$, which is the only connection uniquely defined in terms of the metric tensor $g$.
\normalsize
$\blacksquare$
\end{dt}

Of course, this connection defines a covariant derivative $\nabla_{\mu}$ such that we have indeed $\nabla_{\mu}g_{\alpha\beta}=0$.

Given this result, we have that 
\begin{thm}
\label{completelyantisymmetricCartantensor}
Cartan tensor is completely antisymmetric; thus Cartan tensor is irreducible.\\
$\square$
\footnotesize
\textit{Proof.} Let us consider the form of the connection decomposed in symmetric and antisymmetric parts
\begin{eqnarray}\nonumber
\Gamma^{\alpha}_{\mu \nu}=\frac{\Gamma^{\alpha}_{\mu \nu}+
\Gamma^{\alpha}_{\nu \mu}}{2}+\frac{1}{2}F^{\alpha}_{\phantom{\alpha}\mu \nu}.
\end{eqnarray}
Of course, also the quantity
\begin{eqnarray}\nonumber
\Gamma^{\alpha}_{\mu \nu}=\frac{\Gamma^{\alpha}_{\mu \nu}+\Gamma^{\alpha}_{\nu \mu}}{2}
\end{eqnarray}
is a connection, which is symmetric, and for the axiom \ref{a2}, and the previous theorem, it is necessary the Levi-Civita connection $\Lambda$; then, we get
\begin{eqnarray}\nonumber
\Gamma^{\alpha}_{\mu\nu}=\Lambda^{\alpha}_{\mu\nu}+\frac{1}{2}F^{\alpha}_{\phantom{\alpha}\mu\nu}.
\end{eqnarray}
Finally, since also this connection has to verify the axiom \ref{a2}, then we have 
\begin{eqnarray}\nonumber
0=D_{\alpha}g_{\rho \omega}=
\partial_{\alpha}g_{\rho \omega}
-g_{\kappa \omega}\Gamma^{\kappa}_{\rho\alpha}
-g_{\rho \kappa}\Gamma^{\kappa}_{\omega\alpha}=\\
\nonumber
=\partial_{\alpha}g_{\rho \omega}
-g_{\kappa \omega}\Lambda^{\kappa}_{\rho\alpha}
-\frac{1}{2}g_{\kappa \omega}F^{\kappa}_{\phantom{\kappa}\rho\alpha}
-g_{\rho \kappa}\Lambda^{\kappa}_{\omega\alpha}
-\frac{1}{2}g_{\rho \kappa}F^{\kappa}_{\phantom{\kappa}\omega\alpha}=\\
\nonumber
=\partial_{\alpha}g_{\rho \omega}
-g_{\kappa \omega}\Lambda^{\kappa}_{\rho\alpha}
-g_{\rho \kappa}\Lambda^{\kappa}_{\omega\alpha}
-\frac{1}{2}g_{\kappa \omega}F^{\kappa}_{\phantom{\kappa}\rho\alpha}
-\frac{1}{2}g_{\rho \kappa}F^{\kappa}_{\phantom{\kappa}\omega\alpha}=\\
\nonumber
=\left(\partial_{\alpha}g_{\rho \omega}
-g_{\kappa \omega}\Lambda^{\kappa}_{\rho\alpha}
-g_{\rho \kappa}\Lambda^{\kappa}_{\omega\alpha}\right)
-\frac{1}{2}\left(F_{\omega\rho\alpha}+F_{\rho\omega\alpha}\right)=
-\frac{1}{2}\left(F_{\omega\rho\alpha}+F_{\rho\omega\alpha}\right)
\end{eqnarray}
that is 
\begin{eqnarray}\nonumber
F_{\omega\rho\alpha}+F_{\rho\omega\alpha}=0
\end{eqnarray}
and so Cartan tensor is antisymmetric in the first pair of indices; but it is antisymmetric also in the second pair of indices: and thus it is completely antisymmetric.
\normalsize
$\blacksquare$
\end{thm}

It is possible to see that 
\begin{dt}
The quantity
\begin{eqnarray}
\nonumber
V^{\alpha}\equiv(\ast F)^{\alpha}
=\frac{1}{6}F_{\mu\rho\sigma}\varepsilon^{\mu\rho\sigma\alpha}
\end{eqnarray}
is a pseudo-vector called Cartan pseudo-Vector.
\end{dt}
Finally, we have 
\begin{thm}
In the spacetime, let be $F$ the Cartan tensor: then we have the identity
\begin{eqnarray}\nonumber
(\ast(\ast F))=F.
\end{eqnarray}
\end{thm}

Finally, we give the definition
\begin{dfn}
A space in which Cartan tensor is identically zero is called Riemann space.
\end{dfn}
We want to stress that Cartan tensor is the only source from which we can get many inequivalent connections for a given space; when Cartan tensor is zero, no multiple, inequivalent connections arise, and the only connection we can define is the (symmetric) Levi-Civita connection, written in terms of the only metric we have in a given space.

As we have already seen in the proof for the theorem \ref{completelyantisymmetricCartantensor}, we have that
\begin{thm}
Given a completely antisymmetric Cartan tensor and the metric, or equivalently the Levi-Civita symmetric connection, the most general connection we can define is of the form
\begin{eqnarray}
\begin{tabular}{|c|}
\hline\\
$\Gamma^{\alpha}_{\mu \nu}=
\Lambda^{\alpha}_{\mu \nu}+\frac{1}{2}F^{\alpha}_{\phantom{\alpha}\mu \nu}$\\ \\
\hline
\end{tabular}.
\end{eqnarray}
\end{thm}

We give another definition
\begin{dt}
Given the metric, or the Levi-Civita symmetric connection, the following tensor
\begin{eqnarray}
\begin{tabular}{|c|}
\hline\\
$R^{a}_{\phantom{a} j k i}=
\partial_{k} \Lambda^{a}_{j i}-\partial_{i} \Lambda^{a}_{j k}+
\Lambda^{a}_{r k}\Lambda^{r}_{j i}-\Lambda^{a}_{r i}\Lambda^{r}_{j k}$\\ \\
\hline
\end{tabular},
\end{eqnarray}
or equivalently
\begin{eqnarray}
\begin{tabular}{|c|}
\hline\\
$R_{\iota \kappa \rho \mu}\!=\!\frac{1}{2}\!\left(
\partial^2_{\kappa \rho}g_{\iota \mu}\!+\!\partial^2_{\iota \mu}g_{\kappa \rho}\!-\!
\partial^2_{\kappa \mu}g_{\iota \rho}\!-\!\partial^2_{\iota \rho}g_{\kappa \mu}\right)
\!+\!g_{\nu \pi}\!\left[\Lambda^{\nu}_{\kappa \rho}\Lambda^{\pi}_{\iota \mu}\!-\!
\Lambda^{\nu}_{\kappa \mu}\Lambda^{\pi}_{\iota \rho}\right]$\\ \\
\hline
\end{tabular}
\end{eqnarray}
is called Riemann Curvature Tensor, and it is antisymmetric in the first couple of indices and in the second couple of indices, and it is symmetric for a switch of the first and the second couple of indices; also, it verifies the three indices identity $R^{\rho}_{\phantom{\rho}\kappa\nu\mu}+R^{\rho}_{\phantom{\rho}\mu\kappa\nu}+R^{\rho}_{\phantom{\rho}\nu\mu\kappa}\equiv0$.
\end{dt}

And now, it is possible to prove that
\begin{thm}
It is always possible to find a reference system in which the metric is constant if and only if it is always possible to find a reference system in which the symmetric connection is (globally) zero if and only if the Riemann curvature tensor vanishes.
\end{thm}

\begin{dt}
Riemann tensor has only one independent contraction $R^{a}_{\phantom{a} j a i}=R_{j i}$, which is symmetric in the two indices $R_{j i}=R_{i j}$, and it is called Ricci curvature tensor, which admits a contraction $R=g^{j i}R_{i j}$, called Ricci curvature scalar.
\end{dt}

Finally, we have that
\begin{dt}
Riemann tensor can be decomposed in terms of Riemann curvature tensor and Cartan tensor as follow
\begin{eqnarray}
\begin{tabular}{|c|}
\hline\\
$F^{a}_{\phantom{a} j k i}=R^{a}_{\phantom{a} j k i}+
\frac{1}{4}(F^{a}_{\phantom{a}rk}F^{r}_{\phantom{r}ji}-F^{a}_{\phantom{a}ri}F^{r}_{\phantom{r}jk})+
\frac{1}{2}(\nabla_{k}F^{a}_{\phantom{a}j i}-\nabla_{i}F^{a}_{\phantom{a}j k})$\\ \\
\hline
\end{tabular},
\end{eqnarray}
which admits only one independent contraction $F^{a}_{\phantom{a} j a i}=F_{j i}$, called Ricci tensor, that can be decomposed in terms of Ricci curvature tensor and Cartan tensor as follow 
\begin{eqnarray}
\begin{tabular}{|c|}
\hline\\
$F_{ji}=R_{ji}+\frac{1}{4}F^{a}_{\phantom{a}rj}F^{r}_{\phantom{r}ai}+
\frac{1}{2}\nabla_{a}F^{a}_{\phantom{a}ji}$\\ \\
\hline
\end{tabular},
\end{eqnarray}
which admits a contraction $F=g^{j i}F_{i j}$, called Ricci scalar, that can be decomposed in terms of Ricci curvature scalar and Cartan tensor as follow
\begin{eqnarray}
\begin{tabular}{|c|}
\hline\\
$F=R-\frac{1}{4}F_{abc}F^{abc}$\\ \\
\hline
\end{tabular}.
\label{ricciscalar}
\end{eqnarray}
\end{dt}

So, Riemann tensor (and its contractions) can be decomposed in terms of the Cartan tensor and the metric, which are supposed to be independent, and thus, they are the two fundamental quantities of this geometry.
%%%%%%%%%%%%%%%%%%%%%%%%%%%%%%%%%%%%%%%%%%%%%%%%%%%%%%%%%%%%%%%%%%%%%%%%%%%%%%%%%%%%%%%%%%%%%%%%%%%
\subsubsection{The metric structure of the Covariant Derivation}
What we have just done is to determine the metric structure of the connection; doing so, we have implicitly determined the metric structure of the covariant derivative as 
\begin{eqnarray}\nonumber
D_{\iota}T^{\alpha...\mu}_{\rho...\omega}
=\nabla_{\iota} T^{\alpha...\mu}_{\rho...\omega}
+\frac{1}{2}(T^{\kappa...\mu}_{\rho...\omega}F^{\alpha}_{\kappa \iota}...      
+T^{\alpha...\kappa}_{\rho...\omega}F^{\mu}_{\kappa \iota})\\
\nonumber
-\frac{1}{2}(T^{\alpha...\mu}_{\kappa...\omega}F^{\kappa}_{\rho \iota}...      
+T^{\alpha...\mu}_{\rho...\kappa}F^{\kappa}_{\omega \iota})
\end{eqnarray}
then, for the covariant directional derivative
\begin{eqnarray}\nonumber
D_{a}T^{\alpha...\mu}_{\rho...\omega}=a^{\iota}D_{\iota}T^{\alpha...\mu}_{\rho...\omega}=\\
\nonumber
=a^{\iota}[\nabla_{\iota}T^{\alpha...\mu}_{\rho...\omega}
+\frac{1}{2}(T^{\kappa...\mu}_{\rho...\omega}F^{\alpha}_{\kappa \iota}...      
+T^{\alpha...\kappa}_{\rho...\omega}F^{\mu}_{\kappa \iota})\\
\nonumber
-\frac{1}{2}(T^{\alpha...\mu}_{\kappa...\omega}F^{\kappa}_{\rho \iota}...      
+T^{\alpha...\mu}_{\rho...\kappa}F^{\kappa}_{\omega \iota})].
\end{eqnarray}
In particular, for the autoparallel curve we have that
\begin{eqnarray}\nonumber
0=\nabla_{A}A^{\mu}+\frac{1}{2}A^{\beta}A^{\alpha}F^{\mu}_{\phantom{\mu}\alpha \beta}\equiv
\nabla_{A}A^{\mu}
\end{eqnarray}
because of the complete antisymmetry of Cartan tensor. 

Then
\begin{thm}
The autoparallel curve verifies the equation 
\begin{eqnarray}\nonumber
\nabla_{A}A^{\mu}=0
\end{eqnarray}
and in this form it is equivalent to the curve which minimizes the action given as the length of the curve between two fixed point, called geodesic; we can therefore say that the autoparallel curve coincides with the geodesic curve between two fixed point, in a space of a given metric.
\end{thm}

The commutator of two covariant derivatives is given as 
\begin{thm}
We have that
\begin{eqnarray}
\nonumber
[\nabla_{\alpha},\nabla_{\beta}]T^{\gamma...\mu}_{\rho...\omega}=
+(R^{\gamma}_{\kappa \alpha \beta}T^{\kappa...\mu}_{\rho...\omega}...
+R^{\mu}_{\kappa \alpha \beta}T^{\gamma...\kappa}_{\rho...\omega})\\
-(R^{\kappa}_{\rho \alpha \beta}T^{\gamma...\mu}_{\kappa...\omega}...
+R^{\kappa}_{\omega \alpha \beta}T^{\gamma...\mu}_{\rho...\kappa})
\end{eqnarray} 
for any tensor $T$.
\end{thm}
and from it, we can see that the Jacobi-Bianchi identities are
\begin{thm}[Jacobi-Bianchi identities]
We have that Riemann curvature tensor verifies the structural differential identities
\begin{eqnarray}
\nabla_{\mu}R^{\nu}_{\phantom{\nu}\iota \kappa \rho}
+\nabla_{\kappa}R^{\nu}_{\phantom{\nu}\iota \rho \mu}
+\nabla_{\rho}R^{\nu}_{\phantom{\nu}\iota \mu \kappa}\equiv0.
\end{eqnarray}
\end{thm}

Finally, the set of Jacobi-Bianchi identities can be contracted till they reach an irreducible form; the fully contracted identities are
\begin{thm}[Fully Contracted Jacobi-Bianchi identities]
\begin{eqnarray}\nonumber
F_{\mu\nu}-F_{\nu \mu}\equiv \nabla_{\rho}F^{\rho}_{\phantom{\rho}\mu \nu}
\end{eqnarray}
and
\begin{eqnarray}\nonumber
D_{\mu}(F^{\mu\kappa}-\frac{1}{2}g^{\mu\kappa}F)=
-\frac{1}{2}F^{\mu\rho \beta \kappa}F_{\beta\mu \rho}-F_{\mu\beta}F^{\beta\kappa \mu}.
\end{eqnarray}
\end{thm}
For the metric identities we have 
\begin{thm}[Fully contracted Jacobi-Bianchi identities]
\begin{eqnarray}\nonumber
\nabla_{\mu}(R^{\mu\kappa}-\frac{1}{2}g^{\mu\kappa}R)\equiv0.
\end{eqnarray}
\end{thm}

The fully contracted Jacobi-Bianchi identities are used to suggest the form of modified tensors, which can simplify the form of the identities themselves; we can in fact give the
\begin{dfn}
We define the tensor
\begin{eqnarray}
R^{\mu\kappa}-\frac{1}{2}g^{\mu\kappa}R=E^{\mu\kappa},
\end{eqnarray}
which has the same symmetry properties of Ricci curvature tensor, called Einstein curvature tensor.
\end{dfn}
and also
\begin{dfn}
We define the tensor 
\begin{eqnarray}
F^{\mu\kappa}-\frac{1}{2}g^{\mu\kappa}F=G^{\mu\kappa}
\end{eqnarray}
called Einstein tensor.
\end{dfn}

With these modified tensors, the fully contracted Jacobi-Bianchi identities are also modified as follow 
\begin{thm}
We have the following identities
\begin{eqnarray}
\begin{tabular}{|c|}
\hline\\
$\nabla_{\rho}F^{\rho\mu \nu}\equiv G^{\mu\nu}-G^{\nu \mu}$\\ \\
\hline
\end{tabular}
\label{JBC}
\end{eqnarray}
and
\begin{eqnarray}
\begin{tabular}{|c|}
\hline\\
$D_{\mu}G^{\mu \rho}=
-G_{\mu\sigma}F^{\mu\sigma\rho}-\frac{1}{2}F_{\alpha\beta\mu}F^{\alpha\beta\mu\rho}$\\ \\
\hline
\end{tabular}.
\label{JB}
\end{eqnarray}
\end{thm}

And also
\begin{thm}
We have the identity
\begin{eqnarray}
\begin{tabular}{|c|}
\hline\\
$\nabla_{\mu}E^{\mu\kappa}\equiv0$\\ \\
\hline
\end{tabular}.
\label{JBm}
\end{eqnarray}
\end{thm}

We need to discuss about a feature of these equations; equation (\ref{JBC}) states a particular characteristic of the Ricci tensor: Ricci tensor is \emph{not} symmetric in general, but its antisymmetric part is determined \emph{only} by a metric derivative of Cartan tensor. This looks like the behaviour of the energy-momentum and spin tensors we have in Quantum Field Theory, in which the divergence of the spin is proportional to the antisymmetric part of the energy-momentum tensor. 

These similarities induced Einstein first (\cite{e}), and then Sciama and Kibble (\cite{k} and \cite{sc}), to think that (the modified) Ricci tensor and Cartan tensor could be related to the energy-momentum tensor $T$ and the spin tensor $S$ of Quantum Field Theory as
\begin{eqnarray}
\nonumber
E^{\mu\nu}=k \ T^{\mu\nu}\\
\nonumber
F^{\alpha\mu\nu}=k \ S^{\alpha\mu\nu}
\end{eqnarray}
in a more unitary theory called Einstein-Cartan-Sciama-Kibble (for a general discussion about Einstein-Cartan-Sciama-Kibble theory see, for example, the discussion of Trautman in \cite{t}, and of Hehl, von der Heyde, Kerlick and Nester in \cite{h-vdh-k-n}), and hence, they were lead to see in these relationships the fundamental equations that rule over the dynamics of those fields, called Einstein-Cartan-Sciama-Kibble's equations.
%%%%%%%%%%%%%%%%%%%%%%%%%%%%%%%%%%%%%%%%%%%%%%%%%%%%%%%%%%%%%%%%%%%%%%%%%%%%%%%%%%%%%%%%%%%%%%%%%%%
\paragraph{The principle of equivalence and the equation of motion.} The covariant geometry developed right above is what should be the necessary environment for \emph{any} physical theory.

In order to achieve the passage from geometry to a physical theory, we need to implement in it principles that rule over the dynamical behaviour of fields. 

The dynamics is split into two complementary descriptions: one tells us, given a configuration of bodies, which field they produce; the other tells us, given an underlying field, which motion a body will follow interacting with it. 

The rules defining the dynamics of fields are the Field Equations; the rules defining the dynamics of bodies are the Equations of Motion. In general, these equations have to be solved simultaneously, and since the effects of the motion of a body will change the field that interacts with the body itself, general effects of non-linearity will appear, making this resolution a difficult task to accomplish.

That's why, in the following, studying the equation of motion of a body, we will consider the body unable to change the field it is interacting with; such a body will be called Test Body, and it will be considered in general as a point-like particle of matter.

The equation that defines the dynamical behaviour of test bodies is Newton's Law; considering the first fundamental form of a metric space $ds^{2}$, and building up the quantity $u^{\mu}=\frac{dx^{\mu}}{ds}$, which turns out to be a tensor of rank $1$, and so a vector, called Velocity, we can give the covariant expression of Newton's Law in the form
\begin{eqnarray}\nonumber
F^{\mu}=m\nabla_{u}u^{\mu}
\end{eqnarray}
where $m$ is the mass of the particle; such an equation admits a variational formulation as 
\begin{eqnarray}\nonumber
F_{\alpha}=\frac{d}{ds}\frac{\partial T}{\partial u^{\alpha}}-\frac{\partial T}{\partial x^{\alpha}}
\end{eqnarray}
where $T=\frac{1}{2}mu^{\nu}u_{\nu}$, and, in the case in which we can write $F$ in terms of a potential in the form $F_{\alpha}=\frac{d}{ds}\frac{\partial V}{\partial u^{\alpha}}-\frac{\partial V}{\partial x^{\alpha}}$ \footnote{Note that this is precisely the case for the Electromagnetism. In fact we can see that $V=eA_{\mu}u^{\mu}$ gives $F_{\alpha}=eu^{\nu}(\partial_{\nu}A_{\alpha}-\partial_{\alpha}A_{\nu})$, which is precisely the expression for the Lorentz force.}, then we can define the Lagrangian $L=T-V$, and this equation gets the form
\begin{eqnarray}
\nonumber
\frac{d}{ds}\frac{\partial L}{\partial u^{\alpha}}-\frac{\partial L}{\partial x^{\alpha}}=0.
\end{eqnarray}

A few remarks are due. First of all, the covariant expression of Newton's Law makes useless to specify whether we are in an inertial reference system or not; this means that all the inertial forces must already be considered somewhere in the expression of the equation of motion: and this is in fact the case, since a direct calculation can show that they are completely defined in terms of the Levi-Civita connection. 

Second, in both the formulations of Newton's Law, the term $F^{\mu}$, or equivalently its potential $V$, is supposed to contain all the informations about all the non-inertial, or better, physical forces acting on the particle.

In this way, the equation of motion can be decomposed as
\begin{eqnarray}\nonumber
m\frac{du^{\mu}}{ds}=-m\Lambda^{\mu}_{\alpha\beta}u^{\alpha}u^{\beta}+F^{\mu}
\end{eqnarray} 
in which in the right-hand-side, the first term contains the inertial forces, and the second term the physical ones.

It is in this situation that Einstein's Principle of Equivalence finds place.

This principle can be expressed in many different ways; one of the best is to say that 
\begin{pre}[of Equivalence]
Let be given a test body moving under the action of the gravitational force alone; the reference system at rest with respect to the test body is a frame in which the local effects of the gravitational force vanish. 
\end{pre}

This principle of equivalence between inertial and gravitational forces, removing, at a local level, any distinction between inertia and gravity, can create a sort of fuzziness in the classification of the forces in Newton's dynamics, because now the gravitational force, which is of course a physical force, can be made to vanish with a proper choice of frame, which gives to it the status of an inertial force, at least locally. Actually, this fuzziness can be cleared up if we pay attention to the adjective \emph{local} in the statement of the Equivalence Principle; in fact gravitational forces can be made to vanish only locally around the test body, while an inertial force can be made to vanish everywhere in the spacetime around the body itself: this keeps gravity different from inertial forces.

Now that we have made clear the difference between inertial forces, gravity and other physical forces, we have to find a way to place gravity into the equation of motion. 

Obviously, since gravitational force is not a physical, tensorial force, it cannot be put into the term $F^{\mu}$; thus, the only choice we have is to put gravitational interaction into the $\Lambda$ term. But in this way, we need a tool to distinguish between gravitational interaction and inertial forces, for they are represented by the very same term in the equation of motion.

This tool is provided by the Riemann curvature tensor. Riemann curvature tensor is written in terms of the Levi-Civita connection, which is supposed to contain both inertial and gravitational effects, and so Riemann tensor should contain both inertial and gravitational effects as well; but inertial effects are due to the frame only, and so, although they can be present into the connection (because it is not a tensor), they can not appear in the expression of Riemann tensor, which will contain then only the gravitational informations, acting like a sort of filter.

Thinking Riemann curvature tensor able to contain only gravitational effects, its non-vanishing can be seen as the presence of gravitational field, and we can think that gravity is the expression of the curvature of the spacetime.

If Riemann curvature tensor is different from zero then it is different from zero in any frame, and so in all of them there is a gravitational field, and gravity has a physical appearance; but we can find anyway a frame in which the Levi-Civita symmetric connection is locally zero, and in this frame the effects of the acceleration on a body are locally negligible: in this frame the gravitational field is present, but the inertial effects will compensate it, and this compensation will be better and better as we consider them more and more local, and, in the limit of an infinitesimal neighborhood, we get a null total effect on the motion of the test body.

So, although the presence of gravity is given through the curvature of the spacetime, the gravitational forces are to be described side by side to the inertial effects into the Levi-Civita connection, while all the other, physical interactions will find place into the $F^{\mu}$ term. 

For our purposes, we won't deal with other physical field (leaving this issue to the next chapter); thus, for what it will be our interest in the present chapter, we will restrict our attention to the case in which in Newton's Law of Motion of test bodies there is no force, and the dynamics is determined from a purely geometrical point of view. In this sense, we can say that the theory of gravitation is the theory of a free dynamics in a curved spacetime.

A remark is due. The reason for which the gravitational force can be put into the Levi-Civita connection is that it acts proportionally to the gravitational mass, whose value is experimentally known to be equal to that of the inertial mass: this principle is what is commonly called Weak Equivalence Principle, and it is the reason for which we can actually simplify the mass from the equation of motion, getting the geometrization of gravity.

So, we can write the law of motion in presence of a gravitational field as 
\begin{eqnarray}
\nonumber
\frac{du^{\mu}}{ds}=-\Lambda^{\mu}_{\alpha\beta}u^{\alpha}u^{\beta}, \ \ R^{\mu}_{\phantom{\mu}\alpha\beta\gamma}\neq0
\end{eqnarray}
in which in the right-hand-side, the $\Lambda$ term contains the inertial effects and the gravitational force, and we can know about the existence of the gravitational field by watching the Riemann curvature tensor; if it does not vanish, then we can not find a frame in which the symmetric connection is zero everywhere, and we can only find a frame in which the connection is locally zero: we can then say that we can remove the effects of the gravitational field only locally, but in a domain of space large enough, the gravitational interaction can not be completely removed, showing its physical presence. Otherwise, if it is equal to zero, we have the free motion simply given as 
\begin{eqnarray}\nonumber
\frac{du^{\mu}}{ds}=-\Lambda^{\mu}_{\alpha\beta}u^{\alpha}u^{\beta}, \ \ R^{\mu}_{\phantom{\mu}\alpha\beta\gamma}=0
\end{eqnarray}
for which we can actually have a frame in which the connection is globally zero and, in that frame, we can write 
\begin{eqnarray}\nonumber
\frac{du^{\mu}}{ds}=0.
\end{eqnarray}

The expression for Newton's Law is then 
\begin{eqnarray}\nonumber
\nabla_{u}u^{\mu}=0, \ \ R^{\mu}_{\phantom{\mu}\alpha\beta\gamma}\neq0,
\end{eqnarray}
and the variational formulation of this equation is given as 
\begin{eqnarray}\nonumber
\frac{d}{ds}\frac{\partial T}{\partial u^{\alpha}}-\frac{\partial T}{\partial x^{\alpha}}=0, \ \ R^{\mu}_{\phantom{\mu}\alpha\beta\gamma}\neq0
\end{eqnarray}
where $T=\frac{1}{2}mu^{\nu}u_{\nu}$. In this formulation we recognize the autoparallel equation and the geodesic equation, respectively; the fact that they are equivalent allows us to see how the AG paradox discussed by Fiziev in \cite{f} (see also Garecki in \cite{ga}) can be solved, since no distinction between autoparallel and geodesic equation can be experienced.

We wish to finish this paragraph with the following remark. Under the point of view of the equation of motion of test bodies, Cartan tensor never appears: this situation have an explanation if we consider that in the ECSK picture, Cartan tensor is supposed to represent the spin of matter field, which is a pure quantum mechanical effect; thus, the absence of Cartan tensor is directly related to the absence of quantum mechanical effects in the equation of motion for test particles. 

In fact, the equation of motion is supposed to describe the motion of a test particle, which is thought to be a point-like particles only because we needed to have a body moving in a space of given geometry without being able to influence that geometry; but even if this recalls the idea of microscopic bodies, it does not mean at all that a test body takes into account the microscopic structure of matter field, and no quantum mechanical effects are present in the equation of motion for test bodies.

In a more realistic theory for matter fields, quantum mechanical effects shall be considered, and in this case we will see that Cartan tensor will take part in the description of the dynamics of matter fields.
%%%%%%%%%%%%%%%%%%%%%%%%%%%%%%%%%%%%%%%%%%%%%%%%%%%%%%%%%%%%%%%%%%%%%%%%%%%%%%%%%%%%%%%%%%%%%%%%%%%
\paragraph{Field equations in the ECSK theory.}
Talking about the ECSK theory, we saw how the geometrical background provided a set of identities (the fully contracted Jacobi-Bianchi identities), whose form is analogous to the form of some relations holding in the context of the general dynamics for physical fields; and we saw how this similarities suggested the identification of the geometrical entities, on one hand, and the physical quantities, on the other hand. This identification lead to the equations 
\begin{eqnarray}
E^{\mu\nu}=k \ T^{\mu\nu}\\
F^{\alpha\mu\nu}=k \ S^{\alpha\mu\nu}
\label{ECSK}
\end{eqnarray}
which can be considered to be the fundamental equations relating physics and geometry, in the ECSK theory.

In this picture, the energy-momentum content of matter fields curves the spacetime, by changing its metric, while the spin content of matter fields twists the spacetime.

However, few points need to be discussed. Let us consider the first of these two field equations
\begin{eqnarray}
E^{\mu\nu}=k \ T^{\mu\nu},
\label{EC}
\end{eqnarray}
and suppose there is no mass, or energy, in a given point of the space; in that point $T^{\mu\nu}=0$ and since field equations are differential, so point dependent, in the same point there will be $0=E^{\mu\nu}=R^{\mu\nu}-\frac{1}{2}g^{\mu\nu}R$, which gives the contraction $R=0$, and so $R^{\mu\nu}=0$, but no further deduction can be made on the curvature of the spacetime, in particular Riemann curvature is not necessary zero. 

This means that in the point in which there is no mass there might be gravitational field anyway, which is precisely what we observe in nature. But if the Universe were empty, field equations (\ref{EC}) would tell us that that Universe would have a vanishing Ricci curvature everywhere, but it might have had a non-vanishing Riemann tensor. 

This issue can be understood by considering that in an empty region there can be a gravitational field produced outside the region, while this can not occur if we consider the region as the whole universe, for which it does not exist an ``outside''; this discrepancy lies in the fact that Einstein's field equations are differential equations, and so they need boundary conditions not contained into the field equations themselves.

On the other side, the other field equations are 
\begin{eqnarray}
F^{\alpha\mu\nu}=k \ S^{\alpha\mu\nu},
\label{CSK}
\end{eqnarray}
and if we suppose there is no spin in a given point of the space, in that point we will have $S^{\alpha\mu\nu}=0$, and so $F^{\alpha\mu\nu}=0$; this occurs because Cartan tensor is linked to the spin via an algebraic equation: physicists are used to talk about this issue as the fact that Cartan tensor does not propagate in the vacuum (see, for example, Garecki \cite{ga}).

Actually, this is not a problem in the ECSK theory, because the fact that the Cartan tensor does not ``go out'' of the matter is essentially the reason for which, up to now, no observation has detected it in a region of free space.

A theory in which Cartan tensor can propagate outside matter should explain, in order to be compatible with experimental measures, why Cartan tensor is allowed to propagate out of matter, but only with a value too small to be measured.

In the picture of ECSK theories, then, metric and Cartan tensor are the fundamental tensors the theory possesses; these two geometrical fields are intimately related to the physical fields of energy and spin: so, if we know the metric of a space and its Cartan tensor, we can have informations about the energy and the spin, or the mass and the spin of the matter fields that produce them.

Mass and spin, together with charge, are the three labels that identify elementary particles; but charge is related to the property of the particle to interact with fields of other nature (fields that we are not going to consider now), and so mass and spin are the only two quantum numbers that characterize the free particle, and they do it in terms of unitary irreducible representations of the Poincar\'e group, as discussed by Wigner in \cite{wig}. 

The fact that both mass and spin characterize particles tells us that we have to take into account the spin beside the mass in the description of the dynamics of matter fields, as well as the geometry can not avoid to consider Cartan tensor beside the metric for a complete description of the geometrical background.

We wish to conclude with a curiosity. The fact that Wigner classified particles according to their values of mass and spin allows the possibility to think about spin and mass as related to the spacetime rototranslations; on the other hand, a gauge theory of the Poincar\'e group (see for example Obukhov in \cite{ob}) produces the result for which Cartan tensor comes from the translational part, while curvature comes from the rotational part of the group: we have then that mass is related to translations that are related to Cartan tensor, and spin is related to rotations that are related to curvature. From this viewpoint, we could conclude that we have the relationships ``mass-Cartan tensor'' and ``spin-curvature'', while in ECSK theory we have the opposite relationship ``spin-Cartan tensor'' and ``mass-curvature''. 
%%%%%%%%%%%%%%%%%%%%%%%%%%%%%%%%%%%%%%%%%%%%%%%%%%%%%%%%%%%%%%%%%%%%%%%%%%%%%%%%%%%%%%%%%%%%%%%%%%%

Finally, all the general results here obtained can also be found in the general review of Dubrovin, Fomenko and Novikov in \cite{d-f-n}, and by de Sabbata and Sivaram in \cite{ds-s}; in more details, they are considered in \cite{s}, \cite{a-p}, \cite{w-h}, and \cite{c-l-s} by Shapiro, Arcos and Pereira, Watanabe and Hayashi, and Capozziello, Lambiase and Stornaiolo.
%%%%%%%%%%%%%%%%%%%%%%%%%%%%%%%%%%%%%%%%%%%%%%%%%%%%%%%%%%%%%%%%%%%%%%%%%%%%%%%%%%%%%%%%%%%%%%%%%%%
\section{The Principle of Minimal Action:\\ Gravitation}
The dynamics of a physical theory of gravitation has been taken into account through considerations about the general structure of the underlying geometry in the ECSK theory.

As well as the equation of motion has been written as Euler-Lagrange equation for some Lagrangian, we can wonder whether it is possible to write also field equations for ECSK's theory in terms of a principle of minimal action.

In order to try to give an answer to this problem, we have to be able to calculate the variation of a given action.

In the geometry we have developed, we saw that there are two fundamental quantities: the metric and the connection; axiom \ref{a2}, then, enables a decomposition of the connection in terms of the metric and the Cartan tensor, and the metric tensor can be defined in terms of the connection; further, the connection is somehow equivalent to Cartan and Riemann tensors: thus, an action that wants to describe the dynamics of the structure of the spacetime has to be written in terms of Cartan and Riemann tensors, and being able to calculate the variation of this action means being able to calculate the variation of Cartan and Riemann tensors themselves.

However, we already saw that Riemann tensor can be decomposed into two parts, one written in terms of Cartan tensor and the other written in terms of the Riemann curvature tensor; hence, to calculate the variation of Cartan and Riemann tensors we need to calculate the variation of Cartan tensor and the variation of Riemann curvature tensor.

Riemann curvature tensor, however, is not a fundamental quantity, being written in terms of the metric tensor; finally, we need to calculate the variation of Riemann curvature tensor in terms of the variation of the metric.

The Palatini Method is the method developed in order to calculate the variations of curvature tensors, when we perform a variation of the metric tensor.

Now, the variation of the metric is a tensor, and we will write it as $\delta g_{\mu \rho}$; from the equation (\ref{inv}) we have that 
\begin{eqnarray}
\delta g^{\mu \rho}=-g^{\alpha \rho}g^{\mu \beta}\delta g_{\alpha \beta};
\end{eqnarray}
moreover, as we will need later, we have from equation (\ref{detg}) that
\begin{eqnarray}
\delta g=-gg_{\mu \beta}\delta g^{\mu \beta}
\end{eqnarray}
so
\begin{eqnarray}
\delta \sqrt{|g|}=-\frac{1}{2}\sqrt{|g|}g_{\mu \beta}\delta g^{\mu \beta}.
\end{eqnarray}

We can now calculate the variation of the connection; the computation is really mechanical, and gives the final result
\begin{eqnarray}
\begin{tabular}{|c|}
\hline\\
$\delta \Gamma^{\lambda}_{\mu \nu}=\frac{1}{2}g^{\lambda \rho}
(\nabla_{\nu}\delta g_{\mu \rho}+
\nabla_{\mu}\delta g_{\nu \rho}-
\nabla_{\rho}\delta g_{\mu \nu})$\\ \\
\hline
\end{tabular}.
\end{eqnarray}

We have just to pay attention to the fact that all the quantities involved are tensors, so we cannot switch the covariant derivative $\nabla$ with the variation $\delta$; also, we can see directly from its expression that, although the connection is not a tensor, the variation of the connection is a tensor (we already know that the variation of an object which is not a tensor can be a tensor: for example, although the position is not a tensor, the variation of position is a tensor, as showed in equation (\ref{dx})).

Finally, we can calculate the variation of all the curvatures, starting from the Riemann curvature tensor
\begin{eqnarray}
\begin{tabular}{|c|}
\hline\\
$\delta R^{\alpha}_{\phantom{\alpha} \beta \mu \nu}=
\nabla_{\mu}\delta \Gamma^{\alpha}_{\beta \nu}-
\nabla_{\nu}\delta \Gamma^{\alpha}_{\beta \mu}$\\ \\
\hline
\end{tabular}
\end{eqnarray}
or equivalently
\begin{eqnarray}\nonumber
\delta R_{\rho \beta \mu \nu}=
\frac{1}{2}(R^{\theta}_{\phantom{\theta} \beta \mu \nu}\delta g_{\rho \theta}
-R^{\theta}_{\phantom{\theta} \rho \mu \nu}\delta g_{\beta \theta})+\\
\nonumber
+\frac{1}{2}(\nabla_{\mu}\nabla_{\beta}\delta g_{\rho \nu}-\nabla_{\nu}\nabla_{\beta}\delta g_{\rho \mu}+
\nabla_{\nu}\nabla_{\rho}\delta g_{\beta \mu}-\nabla_{\mu}\nabla_{\rho}\delta g_{\beta \nu}).
\end{eqnarray}

Then Ricci curvature tensor is given by
\begin{eqnarray}
\begin{tabular}{|c|}
\hline\\
$\delta R_{\alpha \beta}=\frac{1}{2}g^{\mu \rho}
(\nabla_{\mu}\nabla_{\alpha}\delta g_{\rho \beta}+\nabla_{\mu}\nabla_{\beta}\delta g_{\alpha \rho}
-\nabla_{\alpha}\nabla_{\beta}\delta g_{\rho \mu}-\nabla_{\mu}\nabla_{\rho}\delta g_{\beta \alpha})$\\ \\
\hline
\end{tabular}.
\end{eqnarray}

And the scalar curvature is finally obtained by the contraction with the metric tensor as well
\begin{eqnarray}\nonumber
\delta R=R_{\alpha \beta}\delta g^{\beta \alpha}+\\
\nonumber
+\frac{1}{2}g^{\mu \rho}g^{\alpha \beta}
(\nabla_{\mu}\nabla_{\alpha}\delta g_{\rho \beta}+\nabla_{\mu}\nabla_{\beta}\delta g_{\alpha \rho}
-\nabla_{\alpha}\nabla_{\beta}\delta g_{\rho \mu}-\nabla_{\mu}\nabla_{\rho}\delta g_{\beta \alpha}).
\end{eqnarray}

Now, given the variation of the metric tensor, we can calculate the variation of the Riemann curvature tensor, and of all its contractions; finally, we can calculate the variation of any action we can write for the gravitational field.

We know that, in the limit of ECSK theory in which Cartan tensor goes to zero, the field equations are simply
\begin{eqnarray}
\nonumber
E^{\mu\kappa}=kT^{\mu\kappa}.
\end{eqnarray}

These equations can be obtained also from a variational principle using the methods explained right above.

If we consider the Lagrangian given as the simplest scalar we can have, namely, Ricci scalar
\begin{eqnarray}
\nonumber
\mathscr{L}=R
\end{eqnarray}
we have that the action is   
\begin{eqnarray}
\nonumber
S=\int R\sqrt{|g|}dV
\end{eqnarray}
and its variation is given as 
\begin{eqnarray}
\nonumber
\delta S=\int \delta(R\sqrt{|g|})dV=
\int (\delta(R_{\alpha\beta}g^{\alpha\beta})\sqrt{|g|}
-\frac{1}{2}Rg_{\alpha\beta}\delta g^{\alpha\beta}\sqrt{|g|})dV=\\
\nonumber
=\int (\delta R_{\alpha\beta}g^{\alpha\beta}\sqrt{|g|}+
R_{\alpha\beta}\delta g^{\alpha\beta}\sqrt{|g|}
-\frac{1}{2}Rg_{\alpha\beta}\delta g^{\alpha\beta}\sqrt{|g|})dV
\end{eqnarray}
and so
\begin{eqnarray}
\nonumber
\delta S=\int (\nabla_{\mu}[g^{\alpha\beta}\delta \Gamma^{\mu}_{\beta \alpha}-
g^{\mu\beta}\delta \Gamma^{\rho}_{\beta \rho}]+
[R_{\alpha\beta}-\frac{1}{2}Rg_{\alpha\beta}]\delta g^{\alpha\beta})\sqrt{|g|}dV;
\end{eqnarray}
since the first term is a divergence of some vector, it can be neglected in the expression of the integral over a volume, and the equivalent action is
\begin{eqnarray}
\nonumber
\delta S=\int (R_{\alpha\beta}-\frac{1}{2}Rg_{\alpha\beta})\delta g^{\alpha\beta}\sqrt{|g|}dV.
\end{eqnarray}

The principle of minimal action $\delta S=0$ implies
\begin{eqnarray}
\nonumber
\int (R_{\alpha\beta}-\frac{1}{2}Rg_{\alpha\beta})\delta g^{\alpha\beta}\sqrt{|g|}dV=0
\end{eqnarray}
for any volume, which is 
\begin{eqnarray}
\nonumber
(R_{\alpha\beta}-\frac{1}{2}Rg_{\alpha\beta})\delta g^{\alpha\beta}=0
\end{eqnarray}
and finally
\begin{eqnarray}
\nonumber
R_{\alpha\beta}-\frac{1}{2}Rg_{\alpha\beta}=0
\end{eqnarray}
which means 
\begin{eqnarray}
\nonumber
E_{\alpha\beta}=0
\end{eqnarray}
and they can be recognized to be Einstein's field equations for gravity in the vacuum.

An easy generalization can indeed show that an action of the form
\begin{eqnarray}
\nonumber
S=\int (R+k\mathscr{L}_{p})\sqrt{|g|}dV
\end{eqnarray}
gives, defining
\begin{eqnarray}
T_{\mu\nu}=\frac{2}{\sqrt{|g|}}\frac{\delta (\mathscr{L}_{p} \sqrt{|g|})}{\delta g^{\mu\nu}},
\end{eqnarray}
the field equations as
\begin{eqnarray}
E_{\alpha\beta}=-\frac{k}{2}T_{\alpha\beta}
\end{eqnarray}
and they are recognized to be Einstein's field equations for gravity.

The most general action we can have can contain invariant products of many Riemann curvature tensors; since Riemann curvature contains $2$ derivatives of the metric tensor, an action containing $N$ Riemann curvatures will contain $2N$ derivatives of the metric tensor, and it will be called a $2N^{\mathrm{th}}$-Order action: in this case we will talk about Higher-Order theories of gravitation. 

Of course, the final generalization is given by considering also invariant products of Cartan tensor in the action as well.

At this stage, we do not have any other information that tells us how to build up an action.

However, we can state some general properties the field equations will have; it is possible to prove that
\begin{thm}
Let $S$ be an action of the form
\begin{eqnarray}
S=\int (\mathscr{L}+k\mathscr{L}_{p})\sqrt{|g|}dV;
\end{eqnarray}
defining
\begin{eqnarray}
A_{\mu\nu}=\frac{2}{\sqrt{|g|}}\frac{\delta (\mathscr{L} \sqrt{|g|})}{\delta g^{\mu\nu}}; \ \ 
B_{\alpha}^{\phantom{\alpha}\mu \nu}=
2\frac{\delta \mathscr{L}}{\delta F^{\alpha}_{\phantom{\alpha}\mu \nu}}
\end{eqnarray}
and also
\begin{eqnarray}
T_{\mu\nu}=\frac{2}{\sqrt{|g|}}\frac{\delta (\mathscr{L}_{p} \sqrt{|g|})}{\delta g^{\mu\nu}}; \ \ 
S_{\alpha}^{\phantom{\alpha}\mu\nu}=
2\frac{\delta\mathscr{L}_{p}}{\delta F^{\alpha}_{\phantom{\alpha}\mu\nu}},
\end{eqnarray}
we have that the principle of minimal action leads to the field equations
\begin{eqnarray}
A_{\alpha\beta}=-kT_{\alpha\beta}\\
B_{\rho\sigma\mu}=-kS_{\rho\sigma\mu}.
\end{eqnarray}
\end{thm}
%%%%%%%%%%%%%%%%%%%%%%%%%%%%%%%%%%%%%%%%%%%%%%%%%%%%%%%%%%%%%%%%%%%%%%%%%%%%%%%%%%%%%%%%%%%%%%%%%%%
\subsection{Second-Order Gravity: the ECSK theory}
The very first case is given (as we just said) with only one curvature in the action: this case is the very-well-known Einstein-Cartan-Sciama-Kibble gravity.

Since only one Riemann tensor is allowed in the action, the Lagrangian cannot be other than its contraction, Ricci scalar, plus other terms describing physical fields.

\begin{dfn}
Let be given the action of the second-order as 
\begin{eqnarray}
S=\int([R-\frac{1}{4}F_{\mu\rho\sigma}F^{\mu\rho\sigma}]+k\mathscr{L}_{p})\sqrt{|g|}dV
\label{action}
\end{eqnarray}
in which the constant $k$ has to be determined.
\end{dfn}

\begin{thm}[Field Equations]
The action is minimized i.e. it is stationary with respect any variation of the metric and of Cartan tensor if and only if we have that
\begin{eqnarray}
G_{\alpha\beta}-\frac{1}{2}\nabla_{\rho}F^{\rho}_{\phantom{\rho}\alpha\beta}
=-\frac{k}{2}T_{\alpha\beta}
\label{se1}
\end{eqnarray}
and
\begin{eqnarray}
F^{\alpha}_{\phantom{\alpha}\mu \nu}=k S^{\alpha}_{\phantom{\alpha}\mu \nu};
\label{se2}
\end{eqnarray}
equations (\ref{se1}) and (\ref{se2}) are the fundamental field equations of the second-order gravity, also known as Einstein-Cartan-Sciama-Kibble theory.
\end{thm}

We stress a couple of points; in the vacuum of physical fields, we have that  
\begin{eqnarray}
\nonumber
G_{\alpha\beta}-\frac{1}{2}\nabla_{\rho}F^{\rho}_{\phantom{\rho}\alpha\beta}=0
\end{eqnarray}
and
\begin{eqnarray}
\nonumber
F^{\alpha}_{\phantom{\alpha}\mu \nu}=0
\end{eqnarray}
so that 
\begin{eqnarray}
\nonumber
R_{\alpha\beta}=0\\
\nonumber
F^{\alpha}_{\phantom{\alpha}\mu \nu}=0
\end{eqnarray}
and field equations would reduce to field equations in a spacetime with no Cartan tensor and in which Ricci curvature tensor is zero; and yet, the Riemann curvature of the space, until we fix boundary conditions, would remain, in general, different from zero.

We also want to stress that the extra piece that seems to make the first equation pretty ugly is, indeed, what is necessary to make the left-hand-side symmetric, as well as Belinfante procedure would do for the right-hand-side; the second equation manifests its presence also in the first equation through this extra term.

In a Riemannian spacetime we simply have 
\begin{dfn}
In a Riemannian spacetime, the action of the second-order is 
\begin{eqnarray}
S=\int [R+k\mathscr{L}_{p}]\sqrt{|g|}dV
\label{action1}
\end{eqnarray}
in which the constant $k$ has to be determined.
\end{dfn}

So, field equations are
\begin{thm}[Field Equations]
In a Riemannian spacetime, the action is minimized i.e. it is stationary with respect any variation of the metric tensor if and only if we have that
\begin{eqnarray}
E_{\alpha\beta}=-\frac{k}{2}T_{\alpha\beta};
\label{se}
\end{eqnarray}
equations (\ref{se}) are the fundamental field equations of the second-order gravity in a Riemannian spacetime, also known as Einstein theory.
\end{thm}
%%%%%%%%%%%%%%%%%%%%%%%%%%%%%%%%%%%%%%%%%%%%%%%%%%%%%%%%%%%%%%%%%%%%%%%%%%%%%%%%%%%%%%%%%%%%%%%%%%%
\paragraph{Second-Order Gravity and Dust.} The case of Dust is the case in which we consider the test body discussed above in a huge amount of unities; each and everyone of them is a point-like particle, and they are assumed to be non-interacting between each other, so to remove any consideration about pressure: thus, it behaves as a fluid of dust.

The equation of motion for such a body has already been discussed, and it has been given in the form $\nabla_{u}u^{\mu}=0$, and the extension to the field of dust is achieved by considering that the distribution of all the particles looks like a fluid that we can describe through a function assigning the value of the density of mass to any point of the spacetime, as $\mu=\mu(x)$.

Such a mass distribution has to be constrained by the conservation law of the mass given as $\nabla_{\alpha}(\mu u^{\alpha})=0$; using this, we see that $\nabla_{\alpha}(\mu u^{\alpha}u^{\beta})=\nabla_{\alpha}(\mu u^{\alpha})u^{\beta}+\mu u^{\alpha}\nabla_{\alpha}u^{\beta}=0+0=0$, and so, we can define the tensor $T^{\alpha \beta}=\mu u^{\alpha}u^{\beta}$ such that $T^{\alpha \beta}=T^{\beta \alpha}$ and $\nabla_{\alpha}T^{\alpha \beta}=0$, which can be considered as the energy-momentum tensor of the field of dust. 

So, we have that
\begin{dt}
Given a distribution of Dust described by $\mu=\mu(x)$, and being $u^{\beta}$ the field of velocity, we can define the tensor 
\begin{eqnarray}
\nonumber
T^{\alpha \beta}=\mu u^{\alpha}u^{\beta}
\end{eqnarray}
which is symmetric, its trace is equal to $\mu$ and it is divergenceless, and we will call it the Energy-Momentum Tensor of the field of dust.
\end{dt}

Field equations for a field of dust read
\begin{eqnarray}
\nonumber
R^{\alpha \beta}-\frac{1}{2}Rg^{\alpha \beta}=-\frac{k}{2}\mu u^{\alpha}u^{\beta}
\end{eqnarray}
from which it follows that
\begin{eqnarray}
\nonumber
R=\frac{k}{2}\mu.
\end{eqnarray}

This equation is very important because it is through it that we can have informations about the value of the constant $k$; considering the fact that a mass distribution is always positive, and that an attractive gravity as a negative $R$, then it follows that $k$ has to be negative in the case of a dust distribution.

We will finish with a remark about a particular physical situation we can have, the one concerning maximally symmetric spaces; in these spacetimes we have that the Riemann curvature tensor can be written (see \cite{we}) as $R_{\alpha\beta\mu\nu}=Ag_{\alpha[\mu}g_{\nu]\beta}$ with $A$ a constant. We have that the solution is given as  
\begin{eqnarray}
\nonumber
\mu=\frac{24A}{k}
\end{eqnarray}
and 
\begin{eqnarray}
\nonumber
u^{\alpha}u^{\beta}=\frac{1}{4}g^{\alpha \beta}
\end{eqnarray}
from which we can calculate the expression for the gravitational field in terms of Riemann curvature to be
\begin{eqnarray}
\nonumber
R_{\alpha\beta\mu\nu}=
\frac{k}{24}\mu\left(g_{\alpha\mu}g_{\nu\beta}-g_{\alpha\nu}g_{\mu\beta}\right)=
\frac{2k}{3}\mu\left(u_{\alpha}u_{\mu}u_{\nu}u_{\beta}-u_{\alpha}u_{\nu}u_{\mu}u_{\beta}\right)\equiv 0
\end{eqnarray}
and so, no solution, unless trivial, is actually given in maximally symmetric spaces for dust distribution.
%%%%%%%%%%%%%%%%%%%%%%%%%%%%%%%%%%%%%%%%%%%%%%%%%%%%%%%%%%%%%%%%%%%%%%%%%%%%%%%%%%%%%%%%%%%%%%%%%%%
\subsection{Fourth-Order Gravity}
The second case is given with two curvatures in the action.

Since two Riemann tensors are allowed in the action, the Lagrangian can contain more pieces; for the sake of simplicity, we will consider only Riemannian spacetimes.

Thus, in a Riemannian spacetime we would have that all the different terms actually reduce, via symmetry properties of Riemann tensor, to $3$ terms only.

\begin{dfn}
In a Riemannian spacetime, the action of the fourth-order is 
\begin{eqnarray}
S_{(a,b,c)}=\int
[(aR^{2}+bR^{\alpha\beta}R_{\alpha\beta}+cR^{\alpha\theta\mu\rho}R_{\alpha\theta\mu\rho})
+k\mathscr{L}_{p}]\sqrt{|g|}dV
\label{action2}
\end{eqnarray}
in which the constant $k$ has to be determined, and the parameters $a,b,c$ have to be chosen.
\end{dfn}

So, field equations are
\begin{thm}[Field Equations]
In a Riemannian spacetime, the action is minimized i.e. it is stationary with respect any variation of the metric tensor if and only if we have that
\begin{eqnarray}
\nonumber
(b+4c)\nabla^{2}R_{\mu \nu}+(2a+\frac{b}{2})g_{\mu \nu}\nabla^{2}R
-(2a+b+2c)\nabla_{\mu}\nabla_{\nu}R-\\
\nonumber
-2c(\frac{1}{4}g_{\mu \nu}R_{\alpha \beta \rho \sigma}R^{\alpha \beta \rho \sigma}-
R_{\nu \beta \rho \sigma}R_{\mu}^{\phantom{\mu} \beta \rho \sigma})
+(4c+2b)R_{\nu \beta \mu \sigma}R^{\beta \sigma}-\\
-(\frac{b}{2}g_{\mu \nu}R_{\alpha \beta}R^{\alpha \beta}
+4cR_{\nu \beta}R_{\mu}^{\phantom{\mu} \beta})
+2aRR_{\nu \mu}-\frac{a}{2}g_{\nu \mu}R^{2}=-\frac{k}{2} T_{\mu \nu};
\label{f}
\end{eqnarray}
equations (\ref{f}) are the fundamental field equations of the fourth-order gravity in a Riemannian spacetime.
\end{thm}
%%%%%%%%%%%%%%%%%%%%%%%%%%%%%%%%%%%%%%%%%%%%%%%%%%%%%%%%%%%%%%%%%%%%%%%%%%%%%%%%%%%%%%%%%%%%%%%%%%%
\paragraph{Fourth-Order Gravity and Dust.}
We saw that a field of dust can be described by an energy-momentum tensor of the form $T^{\alpha \beta}=\mu u^{\alpha}u^{\beta}$; in the framework of fourth-order gravity, we change the structure of the geometrical side, but not the structure of the energetic side of the field equations, obtaining a change of the reciprocal relations between curvature and energy.

If we consider the contracted field equations we get
\begin{eqnarray}
\nonumber
(3a+b+c)\nabla^{2}R=\frac{k}{2}\mu
\end{eqnarray}
and we can see how the behaviour of $R$ can not give information about the sign of the value of the constant $k$, since in general, we have to fix at least the sign of the coefficient $3a+b+c$. 
%%%%%%%%%%%%%%%%%%%%%%%%%%%%%%%%%%%%%%%%%%%%%%%%%%%%%%%%%%%%%%%%%%%%%%%%%%%%%%%%%%%%%%%%%%%%%%%%%%%
\subsection{Sixth-Order Gravity}
The last case we are going to consider is given with three Riemann tensors in the action.

When three Riemann tensors are allowed in the action we have that, in a Riemannian spacetime, all the different terms reduce to $11$ terms, $8$ coming from the triplet of curvatures and $3$ coming from the doublet of covariant derivatives of curvatures. 

\begin{dfn}
In a Riemannian spacetime, the action of the sixth-order is 
\begin{eqnarray}
\nonumber
S_{(\Phi,\Psi,\Theta,A,B,C,D,E,F,G,H)}=\int [(\Phi\nabla_{\rho}R\nabla^{\rho}R+
\Psi\nabla_{\rho}R_{\alpha \beta}\nabla^{\rho}R^{\alpha \beta}+\\
\nonumber
\Theta\nabla_{\rho}R_{\alpha \beta \mu \nu}\nabla^{\rho}R^{\alpha \beta \mu \nu}+AR^{3}+\\
\nonumber
BRR_{\alpha \beta}R^{\alpha \beta}+
CRR_{\alpha \beta \mu \nu}R^{\alpha \beta \mu \nu}+\\
\nonumber
DR_{\alpha \beta}R^{\beta \gamma}R^{\alpha}_{\gamma}+
ER^{\alpha \beta}R^{\gamma \delta}R_{\alpha \gamma \beta \delta}+\\
\nonumber
FR^{\alpha \beta}R_{\alpha \gamma \delta \mu}R_{\beta}^{\phantom{\beta}\gamma\delta\mu}+
GR^{\alpha\beta\gamma\delta}R_{\alpha\beta\mu\nu}R_{\gamma\delta}^{\phantom{\gamma\delta}\mu\nu}+\\
HR_{\alpha\beta\gamma\delta}R^{\beta\mu\delta\nu}
R^{\alpha}_{\phantom{\alpha}\mu}{}^{\gamma}_{\phantom{\gamma}\nu})+k\mathscr{L}_{p}]\sqrt{|g|}dV
\label{action3}
\end{eqnarray}
in which the constant $k$ has to be determined, and the parameters $\Phi,\Psi,\Theta$ and $A,B,C,D,E,F,G,H$ have to be chosen.
\end{dfn}

And consequently we can determine for this action the corresponding field equations, which are presented in appendix.
%%%%%%%%%%%%%%%%%%%%%%%%%%%%%%%%%%%%%%%%%%%%%%%%%%%%%%%%%%%%%%%%%%%%%%%%%%%%%%%%%%%%%%%%%%%%%%%%%%%
%%%%%%%%%%%%%%%%%%%%%%%%%%%%%%%%%%%%%%%%%%%%%%%%%%%%%%%%%%%%%%%%%%%%%%%%%%%%%%%%%%%%%%%%%%%%%%%%%%%
\chapter{Spin Theories}
%%%%%%%%%%%%%%%%%%%%%%%%%%%%%%%%%%%%%%%%%%%%%%%%%%%%%%%%%%%%%%%%%%%%%%%%%%%%%%%%%%%%%%%%%%%%%%%%%%%
\section{Orthonormal Bases, Complex\\ Representations and Spinors}
Regarding the definition of tensor we gave at the beginning, it is possible to be a little skeptical, in the sense that such a definition seems to need a system of coordinates, leading us to think that, even if relationships between tensors are coordinates independent, tensors might be not.

If we want tensors to be independent on coordinates, as well as relationships between tensors are, we need to introduce a notation in which this independence is manifest; in this formalism, it will be clear for tensors, and consequently even more for relationships between tensors, that they pre-exist the choice of a system of coordinates.

How to achieve this notation can be better understood from this example.

Let us consider the easiest example of space, namely, a $2$-dimensional space with a vanishing connection; in it, the $2$-dimensional metric can get the polar form 
\begin{eqnarray}
\nonumber
g=\left(\begin{array}{cc}
1 & 0\\
0 & r^{2}
\end{array}\right)
\end{eqnarray}
for which the only non-vanishing components of the connection are 
\begin{eqnarray}
\nonumber
\Lambda^{r}_{\theta\theta}=-r \ ; \ \ \  
\Lambda^{\theta}_{\theta r}=\frac{1}{r}.
\end{eqnarray}

The acceleration in these coordinates is given as $A^{\mu}=\nabla_{u}u^{\mu}$, that is 
\begin{eqnarray}
\nonumber
A^{r}=\ddot{r}-r \dot{\theta}^{2}\\
\nonumber 
A^{\theta}=\ddot{\theta}+\frac{2}{r}\dot{r}\dot{\theta}
\end{eqnarray}
where $\dot{a}=\frac{da}{ds}$, being $ds^{2}=dr^{2}+r^{2}d\theta^{2}$ the line element. 

We see that this form does \emph{not} correspond to what we know to be the real components of the acceleration in polar coordinates, and, moreover, its components are not even homogeneous in the dimensions, and $A^{\theta}$ does not even have the dimension of an acceleration. 

In order to have all those components with the correct expression for the acceleration, we will introduce a basis of orthonormal vectors $\{e^{(j)}_{k}\}$, so that we can define the vector $A^{(\alpha)}=A^{\mu}e_{\mu}^{(\alpha)}$: it is easy to see that if the basis is given as 
\begin{eqnarray}
\nonumber
e^{(r)}_{r}=1 \ , \ \ 
e^{(\theta)}_{r}=0 \ , \ \ 
e^{(r)}_{\theta}=0 \ , \ \ 
e^{(\theta)}_{\theta}=r
\end{eqnarray}
then the acceleration is 
\begin{eqnarray}
\nonumber
A^{(r)}=A^{\alpha}e_{\alpha}^{(r)}=A^{r}e_{r}^{(r)}+A^{\theta}e_{\theta}^{(r)}=A^{r}=
\ddot{r}-r\dot{\theta}^{2}
\end{eqnarray}
and also
\begin{eqnarray}
\nonumber 
A^{(\theta)}=A^{\alpha}e_{\alpha}^{(\theta)}=A^{r}e_{r}^{(\theta)}+A^{\theta}e_{\theta}^{(\theta)}
=rA^{\theta}=r\ddot{\theta}+2\dot{r}\dot{\theta}
\end{eqnarray}
that is 
\begin{eqnarray}
\nonumber
A^{(r)}=\ddot{r}-r \dot{\theta}^{2}\\
\nonumber 
A^{(\theta)}=r\ddot{\theta}+2\dot{r}\dot{\theta}
\end{eqnarray}
which is the right expression for the acceleration in polar coordinates.

Of course, this procedure is valid also for the velocity, or any other physical field.

So, a suitable choice for the basis of vectors can actually give us all the expressions in the form we want them to have; the problem is then how can we get such a basis without any \emph{ad hoc} choice.

We notice that the basis of the previous example can be written as
\begin{eqnarray}
\nonumber
\vec{e}_{r}=
\left(\begin{array}{c}
1\\
0
\end{array}\right), \ \ \ \ 
\vec{e}_{\theta}=
\left(\begin{array}{c}
0\\
r
\end{array}\right)
\end{eqnarray}
and it verifies the relationships
\begin{eqnarray}
\nonumber
\vec{e}_{r}\cdot \vec{e}_{r}=1=g_{rr} \ , \ \ 
\vec{e}_{\theta}\cdot \vec{e}_{\theta}=r^{2}=g_{\theta\theta} \ , \ \ 
\vec{e}_{r}\cdot \vec{e}_{\theta}=0=g_{r\theta}
\end{eqnarray}
so that we can write
\begin{eqnarray}
\nonumber
\vec{e}_{\alpha}\cdot \vec{e}_{\beta}=g_{\alpha\beta}
\end{eqnarray}
from which we see that the basis is not introduced \emph{ad hoc}, but we can define it in terms of the metric of the space.

Obviously, this procedure does not apply only to polar coordinates in $2$ dimensions, but for any general $n$-dimensional metrics defined in a space.

So, since the metric is given, it is possible to calculate from it a basis of vectors, through which we can define components for the fields that have a real, physical meaning. 

Let us introduce then the general formalism of the ``Vierbein'', or Bases of Tetrads; from here on, we will keep Greek letters to label tensorial indices, and Latin letters to label indices of basis, or Lorentz indices.

Let be given a set of $4$ vectors labelled by an index of basis (in brackets) $\{\xi^{\mu}_{(a)}\}$; with them, we can re-label lower tensorial indices to be Lorentz indices, defining them to be $A_{\mu}\xi^{\mu}_{(a)}=A_{(a)}$. Now $A_{(a)}$ are a column of scalars, of invariants, which means that they have an intrinsic, physical meaning!

Since it is natural to extend this definition also to upper tensorial indices, we can introduce a set of $4$ covectors labelled by $\{\xi_{\mu}^{(a)}\}$ such that
\begin{eqnarray}
\xi^{(a)}_{\mu}\xi_{(b)}^{\mu}=\delta^{a}_{b}
\end{eqnarray}
and
\begin{eqnarray}
\xi^{(m)}_{\mu}\xi_{(m)}^{\nu}=\delta^{\mu}_{\nu}
\end{eqnarray}
in which $\delta$ is the Kronecker $4$-dimensional tensor; using them, we can define Lorentz components also for tensors with upper indices.

Now, we said before that we will use an orthonormal basis; in fact, we can see that it is always possible, via the process of orthonormalization of Graham-Schmidt, to orthonormalize a basis of vector, to get it such that
\begin{eqnarray}
\xi^{(a)}_{\mu}\xi^{(b)}_{\nu}g^{\mu\nu}=\eta^{ab}
\end{eqnarray}
where the matrix
\begin{eqnarray}
(\eta^{ab})=\left(\begin{array}{cccc}
1 & 0 & 0 & 0\\
0 & -1 & 0 & 0\\ 
0 & 0 & -1 & 0\\
0 & 0 & 0 & -1
\end{array}\right)
\end{eqnarray}
is the Minkowskian Matrix; since an orthonormal basis is a very useful tool to calculate geometrical quantities, and important properties come form it, and since the restriction to orthonormal basis is of no physical meaning, we can postulate that we shall deal only with orthonormal bases, meaning that once we get the orthonormal basis we are looking for, we will allow only those transformations of basis which will preserve its form.

We have that 
\begin{axm}
\label{a3}
The basis used to describe physical quantities will always be consider orthonormal, and the only transformations between different bases will always be such that they preserve the orthonormality.
\end{axm}

If we are in an orthonormal basis, the transformations that preserve the orthonormality of the bases are the very-well-known Lorentz transformations $\Lambda$.

Lorentz transformations can be written in these terms
\begin{dt}
The transformation laws allowed by the axiom \ref{a3} are all the transformations represented by the matrix $\Lambda$ such that
\begin{eqnarray}
\eta=\Lambda\eta \Lambda^{T}
\end{eqnarray}
and they are called Lorentz Transformations; they form a group called Lorentz Group $O(1,3;\mathbb{R})$, whose subgroup of all the matrices of unitary determinant is called Special Lorentz Group $SO(1,3;\mathbb{R})$.
\end{dt}

We wish to point out that for our physical purposes, we will consider only the subgroup which preserves the orientations of the basis, namely, the subgroup which does not change a left basis into a right one, which is the special Lorentz group.

Further, we would like to stress that, since no restriction has been made on them, we can also consider complex physical fields; thus, we have to take into account also complex representations of the special Lorentz group. 

Finally, we will admit the possibility that Lorentz group could act depending on the point we are considering, namely, that the parameters of the transformation are point dependent, and the transformation locally defined, so to have a local, complex representation of the special Lorentz group.

A complex representation of the special Lorentz group will be given through the matrix $S$, and belongs to the so called Group of the Spinorial representations of the special Lorentz group.

Both for the $\Lambda$ and for the $S$ matrices, we have that their expansions in terms of the generators give the same commutators.

To see this fact, let us consider the Lorentz algebra $so(1,3;\mathbb{R})$; we have that the commutation relationships between infinitesimal generators are
\begin{eqnarray}
\nonumber
\left[R_{j},R_{b}\right]&=&\epsilon_{jba}R_{a}\\
\nonumber
\left[B_{j},B_{b}\right]&=&-\epsilon_{jba}R_{a}\\
\nonumber
\left[R_{j},B_{b}\right]&=&\epsilon_{jba}B_{a};
\end{eqnarray}
the complex representation of the Lorentz algebra can be given considering the set of Pauli matrices (see \cite{g} for details) given as $\{\sigma_{j}, j=1,2,3\}$ where
\begin{eqnarray}
\nonumber
\sigma_{1}=\left(\begin{array}{cc}
0 & 1\\        
1 & 0
\end{array}\right); \ \ 
\sigma_{2}=\left(\begin{array}{cc}
0 & -i\\
i & 0
\end{array}\right); \ \ 
\sigma_{3}=\left(\begin{array}{cc}
1 & 0\\       
0 & -1
\end{array}\right),
\end{eqnarray}
and verifying the relationships
\begin{eqnarray}
\nonumber
\left[\sigma_{j},\sigma_{b}\right]=2i\epsilon_{jba}\sigma_{a},
\end{eqnarray}
and considering the set formed with the matrices $\{J_{j}=-\frac{i}{2}\sigma_{j}, j=1,2,3\}$ and $\{K_{j}=\frac{1}{2}\sigma_{j}, j=1,2,3\}$ that gives
\begin{eqnarray}
\nonumber
\left[J_{j},J_{b}\right]&=&\epsilon_{jba}J_{a}\\
\nonumber
\left[K_{j},K_{b}\right]&=&-\epsilon_{jba}J_{a}\\
\nonumber
\left[J_{j},K_{b}\right]&=&\epsilon_{jba}K_{a},
\end{eqnarray}
which is the expression for the Lorentz algebra with complex $2 \times 2$ matrices $sl(2;\mathbb{C})$ for a semispinorial $2$-dimensional representation; the correspondent $4$-dimensional spinorial representation is given by the set of matrices
\begin{eqnarray}
\nonumber
R^{\rm{ch}}_{a}=\left(\begin{array}{cc}
J_{a} & 0\\        
0 & J_{a}
\end{array}\right); \ \ 
B^{\rm{ch}}_{a}=\left(\begin{array}{cc}
K_{a} & 0\\  
0 & -K_{a}
\end{array}\right)
\end{eqnarray}
or else, 
\begin{eqnarray}
\nonumber
R^{\rm{st}}_{a}=\left(\begin{array}{cc}
J_{a} & 0\\        
0 & J_{a}
\end{array}\right); \ \ 
B^{\rm{st}}_{a}=\left(\begin{array}{cc}
0 & K_{a}\\  
K_{a} & 0
\end{array}\right)
\end{eqnarray}
respectively called Chiral and Standard spinorial representations, and in both cases we get 
\begin{eqnarray}
\nonumber
\left[R_{j},R_{b}\right]&=&\epsilon_{jba}R_{a}\\
\nonumber
\left[B_{j},B_{b}\right]&=&-\epsilon_{jba}R_{a}\\
\nonumber
\left[R_{j},B_{b}\right]&=&\epsilon_{jba}B_{a},
\end{eqnarray}
which is a $4 \times 4$ spinorial representation of the Lorentz algebra, and that corresponds to the set of generators for the matrix $S$ of the spinorial representation of the special Lorentz group; these commutation relationships can also be written in the $4$-dimensional notation as
\begin{eqnarray}
\nonumber
\epsilon_{ija}R_{a}=\sigma_{ij}\\
\nonumber
B_{k}=\sigma_{0k}
\end{eqnarray}
so that
\begin{eqnarray}
\left[\sigma_{tr},\sigma_{pq}\right]=
\left(\eta_{tp}\sigma_{qr}-\eta_{tq}\sigma_{pr}
+\eta_{rq}\sigma_{pt}-\eta_{rp}\sigma_{qt}\right).
\end{eqnarray}

Given these two representations, taking the (local) parameter of the transformation as the set of $\theta^{ab}$, antisymmetric in the two indices, we have that an infinitesimal Lorentz transformation is given as 
\begin{eqnarray}
\Lambda^{a}_{b}=\delta^{a}_{b}+\delta\theta^{a}_{\phantom{a}b}
\end{eqnarray}
while an infinitesimal spinorial transformation is given as 
\begin{eqnarray}
S=\mathbb{I}+\frac{1}{2}\delta\theta^{ab}\sigma_{ab}.
\end{eqnarray}

The transformations represented by $\Lambda$ and by $S$ are the transformation laws we were looking for.

Before proceedings, we will give some result. 

First of all, we will give this definition
\begin{dt}
The quantity
\begin{eqnarray}
[\xi_{(a)},\xi_{(b)}]^{\alpha}=
\xi^{\kappa}_{(a)}\partial_{\kappa}\xi^{\alpha}_{(b)}
-\xi^{\kappa}_{(b)}\partial_{\kappa}\xi^{\alpha}_{(a)}
\end{eqnarray}
is a vector called Commutator, and it verifies:
\begin{itemize}
\item[1)] 
\begin{eqnarray}
\nonumber
[\xi_{(a)},\xi_{(a)}]=0
\end{eqnarray}
or
\begin{eqnarray}
\nonumber
[\xi_{(a)},\xi_{(b)}]+[\xi_{(b)},\xi_{(a)}]=0
\end{eqnarray}
antisymmetry
\item[2)] 
\begin{eqnarray}
\nonumber
[\xi_{(a)},[\xi_{(b)},\xi_{(c)}]]+
[\xi_{(c)},[\xi_{(a)},\xi_{(b)}]]+
[\xi_{(b)},[\xi_{(c)},\xi_{(a)}]]=0
\end{eqnarray}
Jacobi identity.
\end{itemize}
It is always possible to find a set of coefficients $C_{ab}^{\phantom{ab}c}$ verifying:
\begin{itemize}
\item[1)] 
\begin{eqnarray}
\nonumber
C_{ab}^{\phantom{ab}c}+C_{ba}^{\phantom{ba}c}=0
\end{eqnarray}
\item[2)] defining $\partial_{a} \equiv \xi_{(a)}^{\mu}\partial_{\mu}$ we have
\begin{eqnarray}
\nonumber
(C_{ij}^{\phantom{ij}m}C_{mk}^{\phantom{mk}l}+
C_{jk}^{\phantom{jk}m}C_{mi}^{\phantom{mi}l}+
C_{ki}^{\phantom{ki}m}C_{mj}^{\phantom{mj}l})+\\
\nonumber
+(\partial_{k}C_{ij}^{\phantom{ij}l}+
\partial_{i}C_{jk}^{\phantom{jk}l}+
\partial_{j}C_{ki}^{\phantom{ki}l})=0
\end{eqnarray}
\end{itemize}
called Structure Tensors, such that the commutator can be written in a unique way as 
\begin{eqnarray}
[\xi_{(a)},\xi_{(b)}]^{\alpha}=C_{ab}^{\phantom{ab}c}\xi_{(c)}^{\alpha}.
\end{eqnarray}
If it is possible to have
\begin{eqnarray}
\nonumber
C_{ab}^{\phantom{ab}c}=0
\end{eqnarray}
then the basis is called Holonomic, Anholonomic otherwise.
\end{dt}
This definition is important since an orthonormal basis is, in general, anholonomic.

Defined an anholonomic orthonormal basis as above, we can introduce for them the procedure of covariant derivation; we have that we can calculate the quantity
\begin{dfn}
We have that the quantity
\begin{eqnarray}
\begin{tabular}{|c|}
\hline\\
$\omega^{k}_{\phantom{k}a\beta}=\xi_{\alpha}^{k}D_{\beta}\xi_{a}^{\alpha}$\\ \\
\hline
\end{tabular}
\end{eqnarray}
is a vector called Spin Connection.
\end{dfn}
Through it, we can re-write Riemann tensor as
\begin{thm}
We have the identity
\begin{eqnarray}
\partial_{a]} \omega^{k}_{\phantom{k} c[b}+\omega^{k}_{\phantom{k}j[a}\omega^{j}_{\phantom{j}c[b}
-\omega^{k}_{\phantom{k}ci}C_{ab}^{\phantom{ab}i}=F^{k}_{\phantom{k} c ab}.
\end{eqnarray}
\end{thm}

Although Riemann tensor is defined with all tensorial indices, here it has been written as $F^{k}_{\phantom{k} c ab}=F^{\mu}_{\phantom{\mu} \nu \alpha\beta}\xi^{(k)}_{\mu}\xi^{\nu}_{(c)}\xi^{\alpha}_{(a)}\xi^{\beta}_{(b)}$, i.e. in the form in which all its components have a real, physical meaning; anyway, it is possible to give another form, in which indices are mixed, as follow
\begin{thm}
We have that
\begin{eqnarray}
\begin{tabular}{|c|}
\hline\\
$\partial_{[\alpha}\omega^{k}_{\phantom{k} j[\beta}+
\omega^{k}_{\phantom{k}a[\alpha} \omega^{a}_{\phantom{a}j[\beta}
=F^{k}_{\phantom{k}j\alpha \beta}$\\ \\
\hline
\end{tabular}.
\end{eqnarray}
\end{thm}

The reason for such a definition can be understood by considering the following transformation law
\begin{dfn}
A basis of vectors $\{\xi_{(m)}^{\mu}\}$ can be transformed into another basis of vectors $\{\xi'^{\mu}_{(m)}\}$ via the transformations
\begin{eqnarray}
\nonumber
\xi'^{\mu}_{(a)}=(\Lambda^{-1}){}^{b}_{a}\xi_{(b)}^{\mu}
\end{eqnarray}
and a basis of covectors $\{\xi^{(m)}_{\mu}\}$ can be transformed into another basis of covectors $\{\xi'^{(m)}_{\mu}\}$ via the transformations
\begin{eqnarray}
\nonumber
\xi'^{(a)}_{\mu}=(\Lambda)^{a}_{b}\xi^{(b)}_{\mu}.
\end{eqnarray}
\end{dfn}

With this transformation, we have that
\begin{thm}
For a change of basis, the spin connection defined as 
\begin{eqnarray}
\nonumber
\omega^{k}_{\phantom{k}j \beta}=(\omega_{\beta})^{k}_{\phantom{k}j} 
\end{eqnarray}
transforms as 
\begin{eqnarray}
\nonumber
\omega_{\beta}'=\Lambda(\omega_{\beta}-\Lambda^{-1}\partial_{\beta}\Lambda)\Lambda^{-1},
\end{eqnarray}
and Riemann tensor defined as
\begin{eqnarray}
\nonumber
F^{k}_{\phantom{k}j\alpha \beta}=(F_{\alpha \beta})^{k}_{\phantom{k}j}
\end{eqnarray}
transforms as
\begin{eqnarray}
\nonumber
F_{\alpha \beta}'=\Lambda F_{\alpha \beta}\Lambda^{-1}.
\end{eqnarray}
\end{thm}

In the forms given before (the ones with all the tensorial indices, or the one with all Lorentzian indices) these tensors would have not transformed in this way; somehow, we can say that it is due to this very disposition of indices that we can have both a tensorial and a Lorentz covariant form for Riemann tensor.

The reason for which we do not consider Cartan tensor in the discussion about which index should be put in the basis form and which should be kept tensorial is the following. 

The condition of metricity expressed by the axiom \ref{a2} gives to Riemann and Cartan tensor the antisymmetry with respect to the first couple of indices, while their definition gives them antisymmetry with respect to the last couple of indices. Riemann tensor has four indices, and for this the properties of antisymmetry do not mix an index belonging to the first with an index belonging to the second couple, nor we have symmetry properties for which we can switch the two couples between each other: in Riemann tensor the first couple of indices and the second couple of indices have two different characters. On the other side, Cartan tensor has three indices, and so the second index belongs both to the first and to the second couple of indices, and the properties of antisymmetry allow all the indices to mix between each other: for Cartan tensor all the three indices have the same character. So, while it should not surprise that Riemann tensor has a form with the first couple of Lorentz indices and the second couple with tensorial ones, Cartan tensor should have either all the indices in the tensorial form or all the indices in the Lorentzian form.

For Cartan tensor with all Lorentzian indices we have
\begin{thm}
Cartan tensor is such that 
\begin{eqnarray}
F^{k}_{\phantom{k} ab}=\omega^{k}_{\phantom{k}[a,b]}+C_{ab}^{\phantom{ab}k}.
\end{eqnarray}
\end{thm}

For the spin connection and Riemann tensor, we have that there is the fundamental property, coming directly from axioms \ref{a2} and \ref{a3}, that has to be mentioned; we have that 
\begin{thm}
We have that the spin connection operator $\omega_{\beta}$ and the Riemann tensor operator $F_{\alpha \beta}$ are antisymmetric operators.\\
$\square$
\footnotesize
\textit{Proof.} In fact, it is possible to calculate the expression of $D_{\mu}g^{\alpha\beta}$ in terms of the basis and the spin connection; in an orthonormal basis, and since $D_{\mu}g^{\alpha\beta}=0$, the thesis for the spin connection is immediately given. For Riemann tensor it is analogous.
\normalsize
$\blacksquare$
\end{thm}

Now, we have everything we need to introduce and develop the theory of spinors. 
%%%%%%%%%%%%%%%%%%%%%%%%%%%%%%%%%%%%%%%%%%%%%%%%%%%%%%%%%%%%%%%%%%%%%%%%%%%%%%%%%%%%%%%%%%%%%%%%%%%
\subsection{Spinors}
We start by giving the following, fundamental
\begin{dfn}
Let be given a Spinorial representation of the special Lorentz group $S$; a given column of functions $\psi$ whose transformation law is given as  
\begin{eqnarray}
\begin{tabular}{|c|}
\hline\\
$\psi'=S\psi$\\ \\
\hline
\end{tabular}
\end{eqnarray}
is called Spinorial Field.
\end{dfn}

As well as for tensors, we will simply refer to spinorial fields as Spinors. 

And as well as what we have done for tensors, we can write down the principle of Spinoriality, namely
\begin{pre}[Spinoriality]
The properties of physical quantities have to be mathematically expressed in the form of vanishing spinors.
\end{pre}
%%%%%%%%%%%%%%%%%%%%%%%%%%%%%%%%%%%%%%%%%%%%%%%%%%%%%%%%%%%%%%%%%%%%%%%%%%%%%%%%%%%%%%%%%%%%%%%%%%%
\subsection{Covariant Differentiation on Spinors}
As we did for tensors, also for spinors we can give the definition of coefficients of spinorial connection, to get a spinorial covariant derivative.
\subsubsection{The Coefficients of Connection}
We have that
\begin{dfn}
A set of coefficients that transforms according to the law
\begin{eqnarray}
\begin{tabular}{|c|}
\hline\\
$\mathbf{\Omega}_{\beta}'=S(\mathbf{\Omega}_{\beta}-S^{-1}\partial_{\beta}S)S^{-1}$\\ \\
\hline
\end{tabular}
\end{eqnarray}
is called set of the Spinorial Connection Coefficients.
\end{dfn}

We would like to stress this important fact: this set of coefficients of spinorial connection has indices of different type (namely, one is tensorial (Greek) and the other two -- the hidden ones -- are spinorial); for this reason, this connection in very different from the tensorial connection we defined for tensors, and no permutation of a spinorial index with a tensorial one is allowed: so, no Cartan tensor could possibly be defined for it!

On the other hand, instead, we can define the analogous of Riemann tensor, as 
\begin{dt}
Given the spinorial connection, the quantity
\begin{eqnarray}
\begin{tabular}{|c|}
\hline\\
$\partial_{[\alpha} \mathbf{\Omega}_{\beta]}+[\mathbf{\Omega}_{\alpha},\mathbf{\Omega}_{\beta}]
=\mathbf{F}_{\alpha \beta}$\\ \\
\hline
\end{tabular}
\end{eqnarray}
transforms as 
\begin{eqnarray}
\begin{tabular}{|c|}
\hline\\
$\mathbf{F}'_{\alpha\beta}=S\mathbf{F}_{\alpha\beta}S^{-1}$\\ \\
\hline
\end{tabular}
\end{eqnarray}
and it is antisymmetric in its indices, and it is called Riemann Spinor.
\end{dt}

We wish to stress that the transformation laws for these quantities are similar to those of the spin connection and Riemann tensor in basis indices; so, naturally we have that
\begin{thm}
It is possible to decompose the spinorial connection in terms of the spin connection as 
\begin{eqnarray}
\mathbf{\Omega}_{\beta}=\frac{1}{2}\omega^{ab}_{\phantom{ab}\beta}\sigma_{ab}
\end{eqnarray}
and Riemann spinor in terms of Riemann tensor as 
\begin{eqnarray}
\mathbf{F}_{\alpha\beta}=\frac{1}{2}F^{ab}_{\phantom{ab}\alpha\beta}\sigma_{ab},
\end{eqnarray}
where $\sigma_{ab}$ are antisymmetric operators belonging to the spinorial representation of the Lorentz algebra.
\end{thm}
%%%%%%%%%%%%%%%%%%%%%%%%%%%%%%%%%%%%%%%%%%%%%%%%%%%%%%%%%%%%%%%%%%%%%%%%%%%%%%%%%%%%%%%%%%%%%%%%%%%
\subsubsection{The Covariant Derivation}
Given the spinorial connection, we can determine the spinorial covariant derivation by giving the following
\begin{dt}
Given a spinorial connection, if $\psi$ is a spinor then the quantity
\begin{eqnarray}
\begin{tabular}{|c|}
\hline\\
$\mathbf{D}_{\beta}\psi=D_{\beta}\psi+\mathbf{\Omega}_{\beta}\psi$\\ \\
\hline
\end{tabular}
\end{eqnarray}
transforms as 
\begin{eqnarray}
\begin{tabular}{|c|}
\hline\\
$(\mathbf{D}_{\beta}\psi)'=S\mathbf{D}_{\beta}\psi$\\ \\
\hline
\end{tabular}
\end{eqnarray}
and is called Spinorial Covariant Derivative of the spinor $\psi$ with respect to the coordinate $\beta$.
\end{dt}

Given a covariant derivative with respect to a given index, it is possible to calculate the commutator of two covariant derivatives with respect to two different indices; the final result is something we can write in term of quantities we already know, and it is given as 
\begin{thm}
We have that
\begin{eqnarray}
[\mathbf{D}_{\alpha},\mathbf{D}_{\beta}]\psi=
F^{\lambda}_{\phantom{\lambda}\alpha\beta}\mathbf{D}_{\lambda}\psi+\mathbf{F}_{\alpha \beta}\psi
\end{eqnarray} 
for any spinor $\psi$.
\end{thm}

For the introduction to Dirac spinor field in curved-twisted spacetimes see the original work of Fock and Ivanenko \cite{f-i}.
%%%%%%%%%%%%%%%%%%%%%%%%%%%%%%%%%%%%%%%%%%%%%%%%%%%%%%%%%%%%%%%%%%%%%%%%%%%%%%%%%%%%%%%%%%%%%%%%%%%
\section{The Principle of Minimal Action:\\ Matter fields}
The whole geometrical environment that is supposed to contain physical systems has been discussed. In it, when the gravitational action is given, we are able to have field equations for the gravitational field, which tell us how matter acts on the spacetime; the further step is to know how gravitational fields can rule over the evolution of matter fields.

The construction of a Quantum Matter Field Theory has been a long process, whose roots lie in the background of Mechanics. 

Historically, Mechanics, in its Newtonian, Lagrangian or Hamiltonian formulation, is the theory that provides us the description of the motion of point-like particles of matter, given a potential containing the informations about the actions of external fields of force on the particle itself. However, the description mechanics provided had some weak point.

First of all, it had to be written in a covariant formalism, which, introducing naturally the concept of locality, made the description of matter through point-like particles inadequate.

The relativistic description of nature requires particles to be ``spread up'' to extended fields; but even after the generalization of Mechanics up to Field Theory, we didn't have the right description of nature yet, since at small scales Heisenberg Principle of Uncertainty rules.

This principle states that there are in nature couples of variables conjugate in such a way that the higher precision we get for one the less precision we get for the other in the process involving their measurements: in formulae
\begin{eqnarray}
\nonumber
\Delta A \Delta B \geqslant \frac{\hbar}{2}.
\end{eqnarray}

Now, there's a theorem stating that, for linear Hermitian operators, a similar expression can actually be obtained (See the textbook of Gasiorowicz \cite{g}) in the form 
\begin{eqnarray}
\nonumber
\Delta A \Delta B \geqslant \frac{1}{2}|\left\langle [A,B]\right\rangle|.
\end{eqnarray}

According to both the principle and the result, we can say that, if the couple of dynamical variables are conjugate as to verify $|\left\langle [A,B]\right\rangle|=\hbar$, then the quantum condition is verified. 

It turns out that observations tell us that the position of a particle in a given direction and its momentum along that direction do have this property of conjugation, and they should then verify the condition
\begin{eqnarray}
\nonumber
|\left\langle [x,p]\right\rangle|=\hbar.
\end{eqnarray}

This condition can be solved in many equivalent ways, called representations, but for our purposes we prefer to give the solution 
\begin{eqnarray}
\nonumber
x=x\\
\nonumber
p=-i\hbar \partial_{x}.
\end{eqnarray}

This expression of the momentum operator holds, of course, also for the other coordinates, and moreover for energy, allowing the further generalization to
\begin{eqnarray}
\nonumber
x^{\mu}=x^{\mu}\\
\nonumber
p_{\alpha}=-i\hbar \partial_{\alpha}.
\end{eqnarray}

This solution then gives
\begin{eqnarray}
\nonumber
|\left\langle [x^{\mu},p_{\alpha}]\right\rangle|=\delta^{\mu}_{\alpha}\hbar
\end{eqnarray}
and so
\begin{eqnarray}
\nonumber
\Delta x^{\mu} \Delta p_{\alpha} \geqslant \delta^{\mu}_{\alpha}\frac{\hbar}{2} 
\end{eqnarray}
which is the Lorentz invariant form of the Heisenberg Principle of Uncertainty.

The expressions
\begin{eqnarray}
\nonumber
x^{\mu}=x^{\mu}\\
\nonumber
p_{\alpha}=-i\hbar \partial_{\alpha}
\end{eqnarray}
or simply, the expressions
\begin{eqnarray}
\nonumber
p_{\alpha}=-i\hbar \partial_{\alpha}
\end{eqnarray}
are what allows us the passage to quantum physics, and matter field equations written with this constraint are quantum matter field equations.

The only condition relativity gives in order to describe some constraints on the motion of matter fields comes from the definition of velocity $u^{\mu}=\frac{dx^{\mu}}{ds}$, for which we have
\begin{eqnarray}
\nonumber
u^{\mu}u_{\mu}=u^{\mu}u^{\nu}g_{\nu\mu}=g_{\nu\mu}\frac{dx^{\mu}}{ds}\frac{dx^{\nu}}{ds}=
\frac{g_{\nu\mu}dx^{\mu}dx^{\nu}}{dsds}=\frac{ds^{2}}{ds^{2}}=1;
\end{eqnarray}
multiplying both sides by the constant $m^{2}$ we get
\begin{eqnarray}
\nonumber
mu^{\mu}mu_{\mu}=m^{2}
\end{eqnarray}
and then, the definition of momentum $mu_{\mu}=p_{\mu}$ gives
\begin{eqnarray}
\nonumber
p^{\mu}p_{\mu}=m^{2}.
\end{eqnarray}

Applying to this constraint on the motion of matter fields the condition of quantization discussed above we get 
\begin{eqnarray}
\nonumber
\left(g^{\nu\mu} \partial_{\nu}\partial_{\mu}+\frac{m^{2}}{\hbar^{2}}\right)\phi=0
\end{eqnarray}
as the equations for the quantum matter field $\phi$, and it is the so-called Klein-Gordon equation.

This equation, however, had something Dirac was disappointed about; it is, in fact, second order derivative of the field $\phi$, while Dirac thought that equations of motion should be first order derivative of the quantum matter field (see the discussion of Dirac about the order of derivative in \cite{d}).

At this point, Dirac saw that Klein-Gordon equation could be written as
\begin{eqnarray}
\nonumber
\left(-\hat{p}^{2}+m^{2}\right)\phi=0
\end{eqnarray}
where $\hat{p}^{2}=\hat{p}^{\mu}\hat{p}_{\mu}=g^{\mu\nu}\hat{p}_{\mu}\hat{p}_{\nu}$, and $\hat{p}_{\mu}=-i\hbar\partial_{\mu}$ are operators, and that this form can actually be split up into two parts, as well as we can write $a^{2}-b^{2}=(a-b)(a+b)$, with the only difference that here we deal with sums of products of operators.

Dirac's idea was to introduce an extra quantity that could have made possible to split up the squared of momenta into a product of operators, and not a sum of products of operators; this condition can in fact be realized by the introduction of the so-called Dirac matrices $\gamma^{\mu}$, such that $\{\gamma^{\mu},\gamma^{\nu}\}=2\mathbb{I}g^{\mu\nu}$, and by defining the momentum operator as $\hat{P}=\gamma^{\mu}\hat{p}_{\mu}$: we can see that
\begin{eqnarray}
\nonumber
\hat{P}^{2}=\hat{P}\hat{P}=
\gamma^{\mu}\gamma^{\nu}\hat{p}_{\mu}\hat{p}_{\nu}\equiv
\frac{1}{2}\{\gamma^{\mu},\gamma^{\nu}\}\hat{p}_{\mu}\hat{p}_{\nu}=
\mathbb{I}g^{\mu\nu}\hat{p}_{\mu}\hat{p}_{\nu}=\mathbb{I}\hat{p}^{2}
\end{eqnarray} 
which is, up to the $4 \times 4$ identity matrix, what we needed. 

In this way, supposing that the matter field $\phi$ is a $4$ component vector in the linear space in which Dirac matrices are defined, Dirac has been able to write down Klein-Gordon equation as 
\begin{eqnarray}
\nonumber
0=(\hat{p}^{2}-m^{2})\psi=(\mathbb{I}\hat{p}^{2}-\mathbb{I}m^{2})\psi=
(\hat{P}^{2}-\mathbb{I}m^{2})\psi=(\hat{P}-\mathbb{I}m)(\hat{P}+\mathbb{I}m)\psi;
\end{eqnarray}
through this expression, we can see that a solution for the equation 
\begin{eqnarray}
\nonumber
(\hat{P}+\mathbb{I}m)\psi=0
\end{eqnarray}
is also a solution for the Klein-Gordon equation, but the last equation is first order derivative of the field $\psi$.

The equation
\begin{eqnarray}
\nonumber
(\hat{P}+\mathbb{I}m)\psi=0
\end{eqnarray}
can be written explicitly as
\begin{eqnarray}
\nonumber
i\hbar\gamma^{\mu}\partial_{\mu}\psi-m\psi=0
\end{eqnarray}
where the field $\psi$ is, as we said, a $4$ dimensional vector in the space of the Dirac matrices; since for a spinorial transformation the generic Dirac matrix $\gamma^{\mu}$ transforms as $\gamma'^{\mu}=S\gamma^{\mu}S^{-1}$, then $\psi$ has to transform as $\psi'=S\psi$ (where $S$ is a complex representation of the special Lorentz group), in order to give the Lorentz invariance of the whole equation.

Making the transformation $S$ local, and replacing the partial derivative with the spinorial derivative defined above, we get, in the spirit of a local gauge theory for the special Lorentz group, the generally covariant expression
\begin{eqnarray}
\nonumber
i\hbar\gamma^{\mu}\mathbf{D}_{\mu}\psi-m\psi=0
\end{eqnarray}
called Dirac Spinorial Equation.

Dirac Spinor is what we need to describe quantum matter fields, and Dirac Spinorial Equation is the fundamental equation of quantum matter field theory.

Now that we know which kind of equation we have to work with, it is possible to derive it from a variational principle.

In order to write the action, we have to say which fundamental fields we are going to use in the Lagrangian; beside Dirac Spinor field $\psi$, we have to take into account its Dirac Complex Conjugate Spinor field $\overline{\psi}$, which can be obtained as $\overline{\psi}=\psi^{\dagger}D$, where $D$ is an Hermitian matrix such that $\gamma^{\mu\dagger}=D\gamma^{\mu}D^{-1}$.

We would like to stress that it is possible to have explicit choices for the expression of the $\gamma$ matrices; in particular, we can have the Chiral representation
\begin{eqnarray}
\nonumber
\gamma_{\rm{ch}}^{0}=
\left(\begin{array}{c|c}
0 & 1\\
\hline
1 & 0
\end{array}\right); \ \ 
\gamma_{\rm{ch}}^{a}=
\left(\begin{array}{c|c}
0 & \sigma_{a}\\
\hline
-\sigma_{a} & 0
\end{array}\right), \ a=1,2,3
\end{eqnarray}
and Standard representation
\begin{eqnarray}
\nonumber
\gamma_{\rm{st}}^{0}=
\left(\begin{array}{c|c}
1 & 0\\
\hline
0 & -1
\end{array}\right); \ \ 
\gamma_{\rm{st}}^{a}=
\left(\begin{array}{c|c}
0 & \sigma_{a}\\
\hline
-\sigma_{a} & 0
\end{array}\right), \ a=1,2,3
\end{eqnarray}
and which, both in the Chiral and in the Standard representation, verify the fundamental identity
\begin{eqnarray}
\gamma_{a}\gamma_{b}\gamma_{c}=\gamma_{a}\eta_{bc}-\gamma_{b}\eta_{ca}+\gamma_{c}\eta_{ab}
+i\gamma^{P}\epsilon_{abcd}\gamma^{d}
\label{propgamma}
\end{eqnarray}
where $\gamma^{P}=i\gamma^{0}\gamma^{1}\gamma^{2}\gamma^{3}$ is called Dirac pseudo-matrix\footnote{We stress that the explicit form for the Dirac pseudo-matrix is given as 
\begin{eqnarray}
\nonumber
\gamma_{\rm{st}}^{P}=
\left(\begin{array}{c|c}
0 & 1\\
\hline
1 & 0
\end{array}\right); \ \ 
\gamma_{\rm{ch}}^{P}=
\left(\begin{array}{c|c}
1 & 0\\
\hline
0 & -1
\end{array}\right)
\end{eqnarray}
in the Standard and Chiral representations.}, since it changes sign for a parity odd transformation; we want to stress that the Minkowskian matrix is used to move the Lorentz index $\gamma^{i}\eta_{ij}=\gamma_{j}$, while the basis is used to pass to tensorial indices $\gamma^{i}\xi_{(i)}^{\mu}=\gamma^{\mu}$. Finally, since the $\gamma$ matrices transform as $\gamma'^{i}=S\Lambda^{i}_{j}\gamma^{j}S^{-1}\equiv \gamma^{i}$, then their spin covariant derivative is zero $\mathbf{D}_{\mu}\gamma^{\alpha}\equiv0$; and this condition is compatible with the metricity condition expressed in axiom \ref{a2}.

So, we can give the expression of the Lagrangian as
\begin{eqnarray}
\nonumber
\mathscr{L}=\frac{i\hbar}{2}
(\overline{\psi}\gamma^{\mu}\mathbf{D}_{\mu}\psi-\overline{\mathbf{D}_{\mu}\psi}\gamma^{\mu}\psi)
-m\overline{\psi}\psi
\end{eqnarray}
being $m$ the mass of the matter field.

The volume element is given as 
\begin{eqnarray}
\nonumber
g \equiv \mathrm{det}\mathnormal{(g_{\mu\nu})}=
\mathrm{det}\mathnormal{(\eta_{ab}\xi^{(a)}_{\mu}\xi^{(b)}_{\nu})}=
-[\mathrm{det}(\mathnormal{\xi^{(a)}_{\mu}})]^{2}\equiv -\xi^{2}
\end{eqnarray}
and so $\sqrt{|g|}=|\xi|$, for which we get that 
\begin{eqnarray}
\delta \xi=\xi \xi^{\mu}_{(a)}\delta \xi_{\mu}^{(a)},
\end{eqnarray}
which gives the factor of the infinitesimal volume element of an integral in terms of the vierbein.

Finally, the action is given as 
\begin{dfn}
The action for the Dirac spinor field and the conjugate Dirac spinor field is
\begin{eqnarray}
S=\int\left[\frac{i\hbar}{2}
(\overline{\psi}\gamma^{\mu}\mathbf{D}_{\mu}\psi-\overline{\mathbf{D}_{\mu}\psi}\gamma^{\mu}\psi)
-m\overline{\psi}\psi\right]\xi \ dV.
\end{eqnarray}
\end{dfn}

This action does give the right field equation 
\begin{thm}
Dirac spinorial field equation for the Dirac spinor field $\psi$ is
\begin{eqnarray}
i\hbar\gamma^{\mu}\mathbf{D}_{\mu}\psi-m\psi=0,
\end{eqnarray}
and for the Dirac conjugate spinor field $\overline{\psi}$ is
\begin{eqnarray}
i\hbar\overline{\mathbf{D}_{\mu}\psi}\gamma^{\mu}+m\overline{\psi}=0.
\end{eqnarray}
$\square$
\footnotesize
\textit{Proof.} Let us consider a variation of the action with respect to the conjugate field $\delta \overline{\psi}$; we have that 
\begin{eqnarray}
\nonumber
\delta S=\int\left[\frac{i\hbar}{2}
(\delta \overline{\psi}\gamma^{\mu}\mathbf{D}_{\mu}\psi
-\mathbf{D}_{\mu}\delta \overline{\psi}\gamma^{\mu}\psi)
-m\delta \overline{\psi}\psi\right]\xi \ dV \equiv \\
\nonumber
\equiv \int\left[\frac{i\hbar}{2}
(\delta \overline{\psi}\gamma^{\mu}\mathbf{D}_{\mu}\psi
-\mathbf{D}_{\mu}(\delta \overline{\psi}\gamma^{\mu}\psi)
+\delta \overline{\psi}\gamma^{\mu}\mathbf{D}_{\mu}\psi)
-m\delta \overline{\psi}\psi\right]\xi \ dV = \\
\nonumber
= \int\left[\frac{i\hbar}{2}
(2\delta \overline{\psi}\gamma^{\mu}\mathbf{D}_{\mu}\psi
-\mathbf{D}_{\mu}(\delta \overline{\psi}\gamma^{\mu}\psi))
-m\delta \overline{\psi}\psi\right]\xi \ dV= \\
\nonumber
= \int\left[ \delta \overline{\psi}(i\hbar\gamma^{\mu}\mathbf{D}_{\mu}\psi-m\psi)
-\mathbf{D}_{\mu}(\frac{i\hbar}{2}\delta \overline{\psi}\gamma^{\mu}\psi)\right]\xi \ dV
\end{eqnarray}
where the spinorial derivative is decomposed as follow
\begin{eqnarray}
\nonumber
\mathbf{D}_{\mu}(\frac{i\hbar}{2}\delta \overline{\psi}\gamma^{\mu}\psi)\equiv
D_{\mu}(\frac{i\hbar}{2}\delta \overline{\psi}\gamma^{\mu}\psi)\equiv\\
\nonumber
\equiv \nabla_{\mu}(\frac{i\hbar}{2}\delta \overline{\psi}\gamma^{\mu}\psi)+\frac{1}{2}
(\frac{i\hbar}{2}\delta \overline{\psi}\gamma^{\alpha}\psi)F^{\mu}_{\phantom{\mu}\alpha\mu}
\end{eqnarray}
and, because of the irreducibility of Cartan tensor, it is 
\begin{eqnarray}
\nonumber
\mathbf{D}_{\mu}(\frac{i\hbar}{2}\delta \overline{\psi}\gamma^{\mu}\psi)\equiv
D_{\mu}(\frac{i\hbar}{2}\delta \overline{\psi}\gamma^{\mu}\psi)\equiv
\nabla_{\mu}(\frac{i\hbar}{2}\delta \overline{\psi}\gamma^{\mu}\psi).
\end{eqnarray}
This term give rise to a quadridivergence, for which the volume integral can be transformed into an integral over the surface on which the fields vanish; the final variation of the action is then
\begin{eqnarray}
\nonumber
\delta S= \int \delta \overline{\psi}(i\hbar\gamma^{\mu}\mathbf{D}_{\mu}\psi-m\psi) \xi \ dV, 
\end{eqnarray}
and the principle of minimal action gives 
\begin{eqnarray}
\nonumber
\int \delta \overline{\psi}(i\hbar\gamma^{\mu}\mathbf{D}_{\mu}\psi-m\psi) \xi \ dV=0
\end{eqnarray}
for any volume, which means 
\begin{eqnarray}
\nonumber
\delta \overline{\psi}(i\hbar\gamma^{\mu}\mathbf{D}_{\mu}\psi-m\psi)=0
\end{eqnarray}
for any variation, which finally means
\begin{eqnarray}
\nonumber
i\hbar\gamma^{\mu}\mathbf{D}_{\mu}\psi-m\psi=0
\end{eqnarray}
for the field equation of the spinor $\psi$. The field equation for the conjugate spinor $\overline{\psi}$ is obtained by varying the action with respect to $\delta\psi$ and following the same calculations.
\normalsize
$\blacksquare$
\end{thm}

Using the action, and having these field equations, we can derive the expression for the conserved quantities; since we have to link field equations, written in terms of $\gamma$, with the conserved quantities related to the spinorial representation of the symmetry group, the matrices $\sigma$, we need to express the matrices $\sigma$ in terms of $\gamma$, and we see that we can do this using the relationships 
\begin{eqnarray}
\sigma_{ab}=\frac{1}{4}[\gamma_{a},\gamma_{b}].
\end{eqnarray}

Now, we can write the expressions for the most fundamental quantities related to Dirac spinor fields, and for which the fact that Dirac spinor verifies Dirac spinorial field equation leads to the fact that these quantities will consequently verify some sort of conservation laws.

The expressions for the (non-symmetric) energy-momentum tensor and for the spin are given as 
\begin{dt}
Dirac spinorial field's energy-momentum tensor is given as
\begin{eqnarray}
T_{\mu\nu}=\frac{i\hbar}{2}(\overline{\psi}\gamma_{\mu}\mathbf{D}_{\nu}\psi-
\overline{\mathbf{D}_{\nu}\psi}\gamma_{\mu}\psi)
\end{eqnarray}
and Dirac spinorial field's spin tensor is given as 
\begin{eqnarray}
S^{\alpha\mu\rho}=
\frac{i\hbar}{4}\overline{\psi}\{\gamma^{\alpha},\sigma^{\mu\rho}\}\psi\equiv
-\frac{\hbar}{4}\overline{\psi}\gamma_{\sigma}\varepsilon^{\sigma\alpha\mu\rho}\gamma^{P}\psi
\end{eqnarray}
and it is completely antisymmetric.\\
$\square$
\footnotesize
\textit{Proof.} Taking into account the property of the gamma matrices expressed in equation (\ref{propgamma}), it is easy to prove the fact that in the last equation the former expression is equal to the latter one, which is manifestly antisymmetric.
\normalsize
$\blacksquare$
\end{dt}

And with these quantities we have that
\begin{thm}
Dirac spinorial field's energy-momentum tensor and spin verify the relationships
\begin{eqnarray}
\nabla_{\alpha}S^{\alpha\mu\nu}=-\frac{1}{2}(T^{\mu\nu}-T^{\nu\mu})
\label{conservationlawSpin}
\end{eqnarray}
and 
\begin{eqnarray}
D_{\mu}T^{\mu\nu}=-T_{\alpha\rho}F^{\alpha\rho\nu}+S_{\alpha\rho\sigma}F^{\alpha\rho\sigma\nu}
\label{conservationlawEnergy}
\end{eqnarray}
where $F^{\alpha\rho\sigma}$ and $F^{\alpha\rho\sigma\nu}$ are Cartan and Riemann tensors.
\end{thm}
From the last of these relationships we can see that energy-momentum tensor couples to Cartan tensor and spin couples to Riemann tensor, and these two couplings are what produces the failure in the conservation law for the energy-momentum tensor itself.  

This is not to be considered a weak point of Dirac field since it is precisely this failure that allows the relationships (\ref{conservationlawEnergy}) to be analogous to the Jacobi-Bianchi identities for Einstein tensor; in fact, we can see that the conservation laws (\ref{conservationlawEnergy}) and (\ref{conservationlawSpin}) do become identical to Jacobi-Bianchi identities if we postulate that
\begin{eqnarray}
-\frac{1}{2}T^{\mu\nu}=G^{\mu\nu}
\label{energy}
\end{eqnarray}
and
\begin{eqnarray}
S^{\alpha\mu\nu}=F^{\alpha\mu\nu}.
\label{spin}
\end{eqnarray}

It is possible to see that the equation (\ref{energy}) can be split up into its antisymmetric and symmetric parts, and the last part can be separated into the spin and the metric dependent parts, as follow
\begin{eqnarray}
\nonumber
-\frac{1}{2}T^{\mu\nu}=E^{\mu\nu}
+\frac{1}{4}\left(\frac{1}{2}g_{\mu\nu}S_{\rho\sigma\theta}S^{\rho\sigma\theta}-
S_{\mu\sigma\theta}S_{\nu}^{\phantom{\nu}\sigma\theta}\right)
+\frac{1}{2}\nabla_{\rho}S^{\rho}_{\phantom{\rho}\mu\nu}
\end{eqnarray}
in which we can easily recognize the purely metric Einstein curvature tensor $E$ while all the other terms depend only on the spin, one of them being symmetric, and the other antisymmetric; this last antisymmetric spin dependent term does not contain any information that is not already contained into the equation (\ref{spin}), and so it can be removed; we can then re-write equations (\ref{energy}) taking into account only its symmetric part as 
\begin{eqnarray}
-\frac{1}{2}T^{\mu\nu}_{(\mathrm{symmetric})}=E^{\mu\nu}
+\frac{1}{4}\left(\frac{1}{2}g_{\mu\nu}S_{\rho\sigma\theta}S^{\rho\sigma\theta}-
S_{\mu\sigma\theta}S_{\nu}^{\phantom{\nu}\sigma\theta}\right)
\end{eqnarray}
and in particular we have the contraction
\begin{eqnarray}
\frac{1}{2}T=R-\frac{1}{4}S_{\rho\sigma\theta}S^{\rho\sigma\theta}.
\end{eqnarray}
On the other hand, equation (\ref{spin}) can not be contracted, nor decomposed, but we can re-write it in terms of the Cartan pseudo-vector as
\begin{eqnarray}
V^{\alpha}=\frac{\hbar}{4}\overline{\psi}\gamma^{\alpha}\gamma^{P}\psi
\end{eqnarray}
in which we can see that the property to be parity odd is showed from both sides of the equation, and so the whole equation is parity invariant even if each side is not; the existence of a parity violating term (whose presence can also be seen into the Dirac field equation) is a very interesting feature of this theory, because it allows us to consider the issue of CP violation in a spontaneous way, since the term that induces the CP violation is naturally contained into the original theory (see for example the work of Andrianov, Giacconi and Soldati \cite{a-g-s}).

Field equations (\ref{energy}) and (\ref{spin}) can also be obtained by a variation of the action of Dirac field with respect to the vierbein and to the spin connection, respectively.

This fact strengthen the validity of the Einstein-Cartan-Sciama-Kibble theory, because Dirac theory provides a fundamental physical field for the geometrical background.

General considerations about Dirac field in curved-twisted spacetime are taken into account by de Berredo-Peixoto, Helayel-Neto and Shapiro in \cite{bp-hn-s}, as well as by Hehl in \cite{h}, and see also the general review of Shapiro in \cite{s}; applications to Dirac field are considered in the work of Watanabe and Hayashi in \cite{w-h}.
%%%%%%%%%%%%%%%%%%%%%%%%%%%%%%%%%%%%%%%%%%%%%%%%%%%%%%%%%%%%%%%%%%%%%%%%%%%%%%%%%%%%%%%%%%%%%%%%%%%
%%%%%%%%%%%%%%%%%%%%%%%%%%%%%%%%%%%%%%%%%%%%%%%%%%%%%%%%%%%%%%%%%%%%%%%%%%%%%%%%%%%%%%%%%%%%%%%%%%%
\chapter{Gauge Theories}
The general idea that lies on the ground of gauge theory was first formulated by Weyl in the 1920s, and had the purpose to reduce electromagnetic potentials to a compensative field needed to establish the scale invariance of a given system: that's why \emph{gauge} symmetry. Although the attempt of Weyl failed, the basic concept has been kept, and further developed so to reach the formulation that nowadays we know to be one of the most fundamental theories we have at our disposal to describe nature.

Weyl's idea was to consider a dynamical system, and to require the gauge symmetry to be local; the introduction of a compensative field with well defined properties would have preserved the symmetry at a local level: this naturally made the room necessary to place a new field in; once this new field was considered, and its properties studied, he recognized it to be the field that represents the electromagnetic interaction we were used to introduce by hand, and which was at that point justified as the field required to have a symmetry in which we believe as a matter of principle.

The first interaction to be ``gauged'' has been electromagnetism, and it has been followed by the two nuclear forces, which present the idea of a further generalization, since the underlying symmetry is described by a non-abelian group.

For those gauge theories described by a non-commutative group, the symmetry represents the invariance of the laws of nature for a mixing of the physical fields, i.e. when the Lagrangian is written in terms of many fields, and we interchange them between each other, nothing, in the initial Lagrangian, is supposed to change.

These are the symmetries we will be interested in the following. 
%%%%%%%%%%%%%%%%%%%%%%%%%%%%%%%%%%%%%%%%%%%%%%%%%%%%%%%%%%%%%%%%%%%%%%%%%%%%%%%%%%%%%%%%%%%%%%%%%%%
\section{The Principle of Gauge Symmetry}
Gauge symmetry, as we said, is connected to the idea that we can mix a set of different physical fields leaving the physical situation unchanged; this means that, thinking to collect the fields in a row like a vector, and thinking the mixing as represented by a matrix shuffling its components, it does not matter which component of this vector we are considering, because the only physical meaning is that of the vector as a whole.
%%%%%%%%%%%%%%%%%%%%%%%%%%%%%%%%%%%%%%%%%%%%%%%%%%%%%%%%%%%%%%%%%%%%%%%%%%%%%%%%%%%%%%%%%%%%%%%%%%%
\subsection{Gauge Symmetric Fields}
We start by giving the following
\begin{dfn}
Let be a set of $k$ fields, which can be tensors of any rank, $\phi^{a}$ with $a=1,...,k$, and let be given a $k\times k$ matrix $T$ belonging to a certain Lie group $\mathbb{G}$; if they transform as
\begin{eqnarray}
\begin{tabular}{|c|}
\hline\\
$(\phi')^{a}=T^{a}_{\phantom{a}b}(\phi)^{b}$\\ \\
\hline
\end{tabular}
\end{eqnarray}
or equivalently
\begin{eqnarray}
\begin{tabular}{|c|}
\hline\\
$\phi'=T\phi$\\ \\
\hline
\end{tabular}
\end{eqnarray}
then they are a set of Gauge Invariant Fields, or Gauge Symmetric Fields.
\end{dfn}

As well as what we have done for tensors and spinors, we can write down the principle of Gauge Symmetry
\begin{pre}[Gauge Symmetry]
Properties of physical quantities have to be mathematically expressed in the form of vanishing gauge symmetric fields.
\end{pre}
%%%%%%%%%%%%%%%%%%%%%%%%%%%%%%%%%%%%%%%%%%%%%%%%%%%%%%%%%%%%%%%%%%%%%%%%%%%%%%%%%%%%%%%%%%%%%%%%%%%
\subsection{Covariant Differentiation\\ on Gauge Symmetric Fields}
As we did for tensors and spinors, we have to consider the fact that this matrix can also be local, in the sense that its parameters can depends on the point we consider, and so we have to introduce compensative fields, to get the gauge symmetric derivative.
\subsubsection{The Gauge Potentials}
We have that
\begin{dfn}
A set of fields transforming as
\begin{eqnarray}
\begin{tabular}{|c|}
\hline\\
$(A'_{\mu})^{a}_{\phantom{a}b}=
T^{a}_{\phantom{a}p}(A_{\mu})^{p}_{\phantom{p}q}(T^{-1})^{q}_{\phantom{q}b}
-\partial_{\mu}T^{a}_{\phantom{a}p}(T^{-1})^{p}_{\phantom{p}b}$\\ \\
\hline
\end{tabular}
\end{eqnarray}
or equivalently
\begin{eqnarray}
\begin{tabular}{|c|}
\hline\\
$A'_{\mu}=T(A_{\mu}-T^{-1}\partial_{\mu}T)T^{-1}$\\ \\
\hline
\end{tabular}
\end{eqnarray}
is called set of Gauge Potentials.
\end{dfn}

We stress that, unlike the case of tensor, but like the case of spinors, no Cartan tensor could possibly be defined for gauge potentials!

The analogous of Riemann tensor, is 
\begin{dt}
Given the gauge potentials, the quantity
\begin{eqnarray}
\begin{tabular}{|c|}
\hline\\
$(F_{\mu \nu})^{a}_{\phantom{a}b}=
\partial_{\mu}(A_{\nu})^{a}_{\phantom{a}b}-\partial_{\nu}(A_{\mu})^{a}_{\phantom{a}b}
+(A_{\mu})^{a}_{\phantom{a}p}(A_{\nu})^{p}_{\phantom{p}b}
-(A_{\nu})^{a}_{\phantom{a}p}(A_{\mu})^{p}_{\phantom{p}b}$\\ \\
\hline
\end{tabular}
\end{eqnarray}
or equivalently
\begin{eqnarray}
\begin{tabular}{|c|}
\hline\\
$F_{\mu \nu}=\partial_{[\mu}A_{\nu]}+[A_{\mu},A_{\nu}]$\\ \\
\hline
\end{tabular}
\end{eqnarray}
transforms as 
\begin{eqnarray}
\begin{tabular}{|c|}
\hline\\
$(F'_{\mu\nu})^{a}_{\phantom{a}b}=
T^{a}_{\phantom{a}p}(F_{\mu\nu})^{p}_{\phantom{p}q}(T^{-1})^{q}_{\phantom{q}b}$\\ \\
\hline
\end{tabular}
\end{eqnarray}
or equivalently
\begin{eqnarray}
\begin{tabular}{|c|}
\hline\\
$F'_{\mu \nu}=TF_{\mu \nu}T^{-1}$\\ \\
\hline
\end{tabular}
\end{eqnarray}
and it is antisymmetric in its indices, and it is called Gauge Strength.
\end{dt}

Now, the fact that all these quantities have two different types of indices allows us to think that we can split apart the dependence on the tensorial index from the one of the index that labels the components, as well as what we have done for spinors, by writing
\begin{eqnarray}
\nonumber
(F_{\mu \nu})^{a}_{\phantom{a}b}=F_{\mu \nu}^{(k)}(G_{(k)})^{a}_{\phantom{a}b}\\
\nonumber
(A_{\nu})^{a}_{\phantom{a}b}=A_{\nu}^{(k)}(G_{(k)})^{a}_{\phantom{a}b}
\end{eqnarray}
in which the quantities $F_{\mu \nu}^{(k)}$ and $A_{\nu}^{(k)}$ are tensorial fields not related to some known quantity, and so fundamental, while the matrices $G_{(k)}$ are constant matrices, about which we can have more informations considering this lemma
\begin{dt}
\label{lemma}
Given a matrix Lie group $\mathbb{G}$ and its corresponding matrix Lie Algebra $\mathbb{A}$ we have that: 
\begin{itemize}
\item[1)]
\begin{itemize}
\item[a)] if $T \in \mathbb{G}$ then $T^{-1}\partial_{\mu}T \in \mathbb{A}$
\item[b)] if $T \in \mathbb{G}$ and $G \in \mathbb{A}$ then $TGT^{-1} \in \mathbb{A}$
\end{itemize}
\item[2)] if $G_{(i)} \in \mathbb{A}$ then $[G_{(i)},G_{(j)}] \in \mathbb{A}$, hence it is possible to write it in the form 
					\begin{eqnarray}
					\begin{tabular}{|c|}
					\hline\\
					$[G_{(i)},G_{(j)}]=C_{ij}^{\phantom{ij} k}G_{(k)}$\\ \\
					\hline
					\end{tabular}
					\end{eqnarray}
for some coefficients $C_{ij}^{\phantom{ij} k}$ called Structure Constants
\item[3)] we can write 
					\begin{eqnarray}
					\begin{tabular}{|c|}
					\hline\\
					$\mathrm{tr}\it{(G_{(i)}G_{(j)})=L \delta_{ij}}$\\ \\
					\hline
					\end{tabular}
					\end{eqnarray}
for some coefficient $L$.
\end{itemize}
\end{dt}

In this way, we can see that the choice for which the matrices $G_{(k)} \in \mathbb{A}$ gives the fact that both gauge potentials and strengths are algebra valued fields, which we can write according to the decomposition above.

We can now give the following 
\begin{thm}
Given a set of matrices $G_{(k)}$ that are a basis of generators for the Lie algebra correspondent to the Lie group, we can decompose the gauge potentials as
\begin{eqnarray}
(A_{\nu})^{a}_{\phantom{a}b}=A_{\nu}^{(k)}(G_{(k)})^{a}_{\phantom{a}b}
\end{eqnarray}
and the gauge strengths as
\begin{eqnarray}
(F_{\mu \nu})^{a}_{\phantom{a}b}=F_{\mu \nu}^{(k)}(G_{(k)})^{a}_{\phantom{a}b}
\end{eqnarray}
in terms of the fields $A_{\nu}^{(k)}$ and $F_{\mu \nu}^{(k)}$ verifying the relationship
\begin{eqnarray}
F_{\mu \nu}^{(k)}=
\partial_{\mu}A_{\nu}^{(k)}-\partial_{\nu}A_{\mu}^{(k)}+C_{ij}^{\phantom{ij} k}A_{\mu}^{(i)}A_{\nu}^{(j)}
\label{gaugestrength}
\end{eqnarray}
and considered as fundamental fields.
\end{thm}
%%%%%%%%%%%%%%%%%%%%%%%%%%%%%%%%%%%%%%%%%%%%%%%%%%%%%%%%%%%%%%%%%%%%%%%%%%%%%%%%%%%%%%%%%%%%%%%%%%%
\subsubsection{The Covariant Derivation}
Given the gauge potentials, we can determine the gauge covariant derivation by giving the following
\begin{dt}
Given the gauge potentials, if $\phi$ is a gauge symmetric field then the quantity
\begin{eqnarray}
\begin{tabular}{|c|}
\hline\\
$\Delta_{\mu}\phi^{(a)}=\partial_{\mu}\phi^{(a)}+(A_{\mu})^{a}_{\phantom{a}b}\phi^{(b)}$\\ \\
\hline
\end{tabular}
\end{eqnarray}
or equivalently
\begin{eqnarray}
\begin{tabular}{|c|}
\hline\\
$\Delta_{\beta}\phi=\partial_{\beta}\phi+A_{\beta}\phi$\\ \\
\hline
\end{tabular}
\end{eqnarray}
transforms as
\begin{eqnarray}
\begin{tabular}{|c|}
\hline\\
$(\Delta_{\beta}\phi')^{a}=
T^{a}_{\phantom{a}p}(\Delta_{\beta}\phi)^{p}$\\ \\
\hline
\end{tabular}
\end{eqnarray}
or equivalently
\begin{eqnarray}
\begin{tabular}{|c|}
\hline\\
$(\Delta_{\beta}\phi)'=T\Delta_{\beta}\phi$\\ \\
\hline
\end{tabular}
\end{eqnarray}
and it is called Gauge Covariant Derivative of the gauge symmetric field $\phi$ with respect to the coordinate $\beta$.
\end{dt}

Having the lemma \ref{lemma}, we can also give this
\begin{thm}
Given a set of matrices $G_{(k)}$ that are a basis of generators for the Lie algebra correspondent to the Lie group, we can decompose the gauge covariant derivative as
\begin{eqnarray}
\Delta_{\mu}\phi^{(a)}=\partial_{\mu}\phi^{(a)}+A_{\mu}^{(k)}(G_{(k)})^{a}_{\phantom{a}b}\phi^{(b)}
\label{gaugecovderivative}
\end{eqnarray}
for the gauge symmetric field.
\end{thm}

We notice that in the expression for the strength (\ref{gaugestrength}) it does not matter which particular representation we choose (since the Cs do not depend on the representation we choose for the given algebra), while for the gauge covariant derivative (\ref{gaugecovderivative}) the representation plays a fundamental role.

In the environment created by the same formalism many different gauge theories can find place; what does distinguish one another is the only two things we did not specified yet: the particular form of the gauge group, so the specific structure of the Lie algebra, and its representations.

We can list some particular cases: 
\begin{itemize}
\item[1)] $SO(2)$ - This group is abelian, so we have vanishing commutation relations, and hence the structure constants are zero 
\begin{eqnarray}\nonumber
F_{\mu \nu}=\partial_{\mu}A_{\nu}-\partial_{\nu}A_{\mu},
\end{eqnarray}
and an explicit representation for the elements of the Lie algebra is trivially the identity matrix, so that there is no mixing, and we can write 
\begin{eqnarray}\nonumber
\Delta_{\mu}\phi=\partial_{\mu}\phi+A_{\mu}\phi.
\end{eqnarray}
\item[2)] $SO(3)$, $SU(2)$ - This group is not abelian and its structure constants are the completely antisymmetric $3$-dimensional coefficients of Ricci-Levi-Civita $\varepsilon^{abc}$, so 
\begin{eqnarray}\nonumber
F_{\mu \nu}^{(a)}=\partial_{\mu}A_{\nu}^{(a)}-\partial_{\nu}A_{\mu}^{(a)}
+\varepsilon^{a}_{\phantom{a}bc}A_{\mu}^{(b)}A_{\nu}^{(c)}.
\end{eqnarray}

We have that:
\begin{itemize}
\item[2.1)] $SO(3)$ - Georgi-Glashow model.
In this case the gauge group is $SO(3)$ and we have that one representation is given by using the structure constants themselves
\begin{eqnarray}\nonumber
(G_{(k)})^{i}_{\phantom{i}j}=-\varepsilon^{i}_{\phantom{i}jk}
\end{eqnarray}
and so 
\begin{eqnarray}\nonumber
\Delta_{\mu}\phi^{(a)}=
\partial_{\mu}\phi^{(a)}+\varepsilon^{a}_{\phantom{a}bc}A_{\mu}^{(b)}\phi^{(c)}.
\end{eqnarray}

We know that $3$-dimensional vectors with vector product are in fact a Lie Algebra; this suggest to use for this formulae the vector product, so to have a more compact notation; we have 
\begin{eqnarray}\nonumber
\vec{F}_{\mu \nu}=
\partial_{\mu}\vec{A}_{\nu}-\partial_{\nu}\vec{A}_{\mu}+\vec{A}_{\mu}\wedge\vec{A}_{\nu}
\end{eqnarray}
and the gauge invariant derivative for the triplet of fields is given by
\begin{eqnarray}\nonumber
\Delta_{\mu}\vec{\phi}=\partial_{\mu}\vec{\phi}+\vec{A}_{\mu}\wedge\vec{\phi}.
\end{eqnarray}
\item[2.2)] $SU(2)$ - Yang-Mills model.
In this model the gauge group is $SU(2)$ and we can use the irreducible complex representation given by Pauli matrices divided by two
\begin{eqnarray}\nonumber
G_{(k)}=\frac{1}{2}\sigma_{(k)}
\end{eqnarray}
and
\begin{eqnarray}\nonumber
\Delta_{\mu}\phi^{(a)}=
\partial_{\mu}\phi^{(a)}+A_{\mu}^{(k)}\frac{(\sigma_{(k)})^{a}_{\phantom{a}b}}{2}\phi^{(b)}.
\end{eqnarray}
\end{itemize}
\end{itemize}
%%%%%%%%%%%%%%%%%%%%%%%%%%%%%%%%%%%%%%%%%%%%%%%%%%%%%%%%%%%%%%%%%%%%%%%%%%%%%%%%%%%%%%%%%%%%%%%%%%%
\section{The Principle of Minimal Action:\\ Interactions}
As we did before for covariant theories, also for gauge symmetrical theories we will obtain field equations as Euler-Lagrange field equations, by varying the action. 

Since the photon is massless, field equations for the electromagnetic field have to be linear in the gauge strength, and so, the Lagrangian for the electromagnetic field must be quadratic in the gauge strength\footnote{Actually, there is the possibility to consider also a coupling between the gauge strength and the gauge potential: it is the so-called Chern-Simons modification of electrodynamics; but this coupling in $4$ dimensions is consistent only if we introduce an auxiliary, particular vector field, making the coupling non-minimal, and \emph{ad hoc}. For this reasons, and since a massive photon has never been observed up to now, we will not consider the Chern-Simons term in the action.}; in this way, we can build up only two inequivalent actions, namely
\begin{eqnarray}
\nonumber
S=S_{a}+S_{b}=a\int F_{\alpha \beta}F^{\alpha \beta}\sqrt{|g|}dV
+b\int F_{\alpha \beta}(\ast F)^{\alpha \beta}\sqrt{|g|}dV,
\end{eqnarray}
but we can also notice that 
\begin{eqnarray}
\nonumber
S_{b}=b\int F_{\alpha \beta}(\ast F)^{\alpha \beta}\sqrt{|g|}dV\equiv
\frac{b}{2}\int F_{\alpha\beta}F_{\mu\nu}\varepsilon^{\alpha\beta\mu\nu}\sqrt{|g|}dV=\\
\nonumber
=2b\int \nabla_{\alpha}A_{\beta}\nabla_{\mu}A_{\nu}\varepsilon^{\alpha\beta\mu\nu}\sqrt{|g|}dV=
2b\int \nabla_{\alpha}A_{\beta}\nabla_{\mu}(\varepsilon^{\alpha\beta\mu\nu}A_{\nu})\sqrt{|g|}dV=\\
\nonumber
=2b\int (\nabla_{\alpha}(A_{\beta}\nabla_{\mu}(\varepsilon^{\alpha\beta\mu\nu}A_{\nu}))
-A_{\beta}\nabla_{\alpha}\nabla_{\mu}(\varepsilon^{\alpha\beta\mu\nu}A_{\nu}))\sqrt{|g|}dV=\\
\nonumber
=2b\int (\nabla_{\alpha}(A_{\beta}\nabla_{\mu}(\varepsilon^{\alpha\beta\mu\nu}A_{\nu}))
-\frac{1}{2}A_{\beta}[\nabla_{\alpha},\nabla_{\mu}](\varepsilon^{\alpha\beta\mu\nu}A_{\nu}))\sqrt{|g|}dV=\\
\nonumber
=2b\int \nabla_{\alpha}(A_{\beta}\nabla_{\mu}(\varepsilon^{\alpha\beta\mu\nu}A_{\nu}))\sqrt{|g|}dV
\end{eqnarray}
and so $S_{b}$ is the volume integral of the divergence of a vector that can be neglected, and the only action we can write is then
\begin{eqnarray}
\nonumber
S=\int F_{\alpha \beta}F^{\alpha \beta}\sqrt{|g|}dV.
\end{eqnarray}

Even if we have no \emph{a priori} indications that the Lagrangian has to be written in such a form also for non-abelian cases, we will keep this form also for non-commutative gauge theories.

General assumptions for the Lagrangian require, beside its quadratic form, that it has to be gauge invariant.

Because of the transformation law of the gauge strength, we can calculate the product between two of them, and then take the trace, so to have
\begin{eqnarray}\nonumber
\mathrm{tr}\it{(F_{\mu \nu}F_{\alpha \beta})'}=
\mathrm{tr}\it{(TF_{\mu \nu}T^{-1}TF_{\alpha \beta}T^{-1})}=
\mathrm{tr}\it{(TF_{\mu \nu}F_{\alpha \beta}T^{-1})}=\\
\nonumber
=\mathrm{tr}\it{(F_{\mu \nu}F_{\alpha \beta}T^{-1}T)}=
\mathrm{tr}\it{(F_{\mu \nu}F_{\alpha \beta})}
\end{eqnarray}
and finally, by taking the contraction, we get that $\mathrm{tr}(F_{\mu \nu}F^{\mu \nu})$ is a quadratic gauge invariant scalar. 

In this way, the expression for the action is given as
\begin{dfn}
The action for gauge fields is
\begin{eqnarray}
\nonumber
S=\int \mathrm{tr}\it{(F_{\alpha \beta}F^{\alpha \beta})}\sqrt{|g|}dV.
\end{eqnarray}
\end{dfn}
And this action gives field equations as 
\begin{thm}
Gauge fields equations are given as 
\begin{eqnarray}
\nonumber
V^{\alpha}\equiv\nabla_{\rho}F^{\rho\alpha}+2[A_{\rho},F^{\rho\alpha}]=0
\end{eqnarray}
where the vector of matrices $V$ is divergenceless.
\end{thm}
%%%%%%%%%%%%%%%%%%%%%%%%%%%%%%%%%%%%%%%%%%%%%%%%%%%%%%%%%%%%%%%%%%%%%%%%%%%%%%%%%%%%%%%%%%%%%%%%%%%
%%%%%%%%%%%%%%%%%%%%%%%%%%%%%%%%%%%%%%%%%%%%%%%%%%%%%%%%%%%%%%%%%%%%%%%%%%%%%%%%%%%%%%%%%%%%%%%%%%%
%%%%%%%%%%%%%%%%%%%%%%%%%%%%%%%%%%%%%%%%%%%%%%%%%%%%%%%%%%%%%%%%%%%%%%%%%%%%%%%%%%%%%%%%%%%%%%%%%%%
%%%%%%%%%%%%%%%%%%%%%%%%%%%%%%%%%%%%%%%%%%%%%%%%%%%%%%%%%%%%%%%%%%%%%%%%%%%%%%%%%%%%%%%%%%%%%%%%%%%
\part{Developments}
\chapter{Conformal Theories}
In spite of the sense of necessity that characterizes its geometrical background, the fundamental theory lacks the same degree of necessity in the dynamics: while the principles of covariance for changes of system of reference, basis and gauge seem to be very simple, general and necessary principles, sufficient to define the structure of the geometry, the dynamics is shaped by the principle of minimal action, but all the actions we studied were defined through phenomenological considerations, or, as for the case of gravity, not defined at all.

A principle we could have used to fix the form of all the terms in the action could have been a sort of Occam's razor, confining the action to the least order (so to say, second order gravity, first order matter fields and second order interactions); but this principle is, again, a phenomenological principle, since we can not be sure, as a matter of principle, that this should really hold for any physical system.

In order to fix the form of the action in a unique way from a theoretical viewpoint we have to find a principle of symmetry which can allow one and only one action.

This principle can be the principle of Conformal Symmetry.
\section{The Principle of Conformal Symmetry}
\subsection{Conformal Gravity}
Conformal Symmetry is a symmetry that stretches lengths without changing angles amplitudes, and under such a transformation, the character of a Riemannian space changes drastically; nevertheless, such a change is a scale transformation, and, before a scale is fixed (by using quantum considerations), we can be allowed to think that nature is scale invariant: so, we can assume Nature to be conformally invariant, i.e. two Riemannian spaces related by a conformal transformation, although very different from a geometrical point of view, are, seen as physical spaces, equivalent.

Considering the line element defined by the First Fundamental Form $ds$, we can see that a transformation that stretches this element is $ds'=\Omega ds$; since the coordinates won't undergo any change, the conformal transformations is equivalent to the transformation of the metric only $g'_{\alpha \beta}=\Omega^{2}g_{\alpha \beta}$ for any $\Omega=\Omega(x)$; and so \begin{pre}[Conformal Symmetry]
Let be given a Riemann space, with metric $g$; let be given a function $\Omega=\Omega(x)$; the transformation for which
\begin{eqnarray}
\begin{tabular}{|c|}
\hline\\
$g'_{\alpha \beta}=\Omega^{2}g_{\alpha \beta}$\\ \\
\hline
\end{tabular}
\label{ctm}
\end{eqnarray}
is said to be a Conformal Transformation. Physical laws must be invariant under conformal transformations.
\end{pre}

After giving this definition of conformal transformations for the metric tensor, we can think to extend it to the other fundamental quantity, Cartan tensor; since Cartan tensor has no \emph{a priori} connections with the metric, we cannot know how we can implement conformal transformations for it, unless we consider other fields. It is possible, in fact, to find which transformation Cartan tensor must undergo, if we consider it as part of the spin connection, and we require conformal invariance for spinors. Anyway, the treatment of conformally invariant spinor fields, together with Cartan tensor, is far beyond the aim of this thesis; a comprehensive treatment is presented by Buchbinder and Shapiro in \cite{b-s} and Hammond in \cite{ha}.

In the following, we will consider the conformally invariant theories without Cartan tensor nor spin, or spinorial fields of matter.

For the Riemannian spacetime we are going to take into account hereafter, the only geometrical field is then the metric tensor, whose conformal transformation laws have already been established; and in order to implement the principle of conformal symmetry, no non-conformal tensors have to be used: a quite complete treatment of conformal symmetry is given by Fulton, Rohrlich and Witten in \cite{f-r-w}.

It is actually possible to get from the metric tensor alone another conformally covariant tensor: it is Weyl tensor, whose expression is given as follow
\begin{dt}
The tensor 
\begin{eqnarray}
C_{\alpha\beta\mu\nu}=R_{\alpha\beta\mu\nu}
+\frac{1}{2}(R_{\alpha[\nu}g_{\mu]\beta}+R_{\beta[\mu}g_{\nu]\alpha})
+\frac{1}{6}Rg_{\alpha[\mu}g_{\nu]\beta}
\end{eqnarray}
has the same symmetry properties Riemann tensor has, and it is irreducible; Weyl tensor verifies the property
\begin{eqnarray}
\begin{tabular}{|c|}
\hline\\
$C'_{\alpha \beta \mu \nu}=\Omega^{2}C_{\alpha \beta \mu \nu}$\\ \\
\hline
\end{tabular}
\end{eqnarray}
that is, it is conformally covariant, and it is called Irreducible Riemann Curvature Tensor, or Conformal Curvature Tensor, or Weyl Curvature Tensor.
\end{dt}

A particular property of Weyl tensor is expressed as 
\begin{eqnarray}
\nonumber
C_{\alpha \beta \mu \nu}C^{\alpha \beta \mu \nu}=
R_{\alpha \beta \mu \nu}R^{\alpha \beta \mu \nu}-2R_{\beta \nu}R^{\beta \nu}+\frac{1}{3}R^{2}
\end{eqnarray}
as a geometric identity.

It is possible to prove that
\begin{thm}
It is always possible to find a scale factor $\Omega$ for which the metric transforms into the constant metric if and only the conformal curvature tensor vanishes.
\end{thm}

For a comprehensive discussion about Weyl conformal tensor, see, for example, the textbook of Wald \cite{wa}.

Thus, the principle of conformal invariance can be translated into a mathematical statement by saying that all we have to use in conformal theory is Weyl conformal curvature tensor $C_{\alpha \beta \mu \nu}$.

For this reason, the action will not be written in terms of Riemann curvatures, but in terms of Weyl curvatures.

Moreover, in $4$ dimensions (and using Weyl tensor's symmetry properties and tracelessness), it is possible to prove that all the possible different conformally invariant actions reduce, up to equivalence, to one only! 

The action
\begin{eqnarray}
S=\int C_{\alpha \beta \mu \nu}C^{\alpha \beta \mu \nu}\sqrt{|g|}dV
\end{eqnarray}
is the sole conformally invariant action we can have.

We have that 
\begin{dfn}
The (only) conformally invariant action for the gravitational field is
\begin{eqnarray}
S=\int [C_{\alpha \beta \mu \nu}C^{\alpha \beta \mu \nu}+k\mathscr{L}_{p}]\sqrt{|g|}dV
\end{eqnarray}
in which the constant $k$ has to be determined.
\end{dfn}

This action gives the field equations 
\begin{thm}[Field Equations]
Conformal gravitational field equations are given as 
\begin{eqnarray}
\nonumber
\nabla^{2}R_{\alpha\beta}-\frac{1}{3}\nabla_{\alpha}\nabla_{\beta}R
-\frac{1}{6}g_{\alpha\beta}\nabla^{2}R+\\
\nonumber
+(R_{\alpha \rho \mu \nu}R_{\beta}^{\phantom{\beta} \rho \mu \nu}
-\frac{1}{4}g_{\alpha\beta}R_{\sigma \rho \mu \nu}R^{\sigma \rho \mu \nu})-\\
\nonumber
-2(R_{\alpha \rho}R_{\beta}^{\rho}-\frac{1}{4}g_{\alpha\beta}R_{\sigma \rho}R^{\sigma \rho})+\\
+\frac{1}{3}R(R_{\alpha\beta}-\frac{1}{4}g_{\alpha\beta}R)=-\frac{k}{4}T_{\alpha\beta};
\label{cfe}
\end{eqnarray}
equations (\ref{cfe}) are the fundamental field equations of conformal gravity.\\
$\square$
\footnotesize
\textit{Proof.} The variation of the action is
\begin{eqnarray}
\nonumber
\delta S=\int (2\nabla^{2}R_{\alpha\beta}-\frac{2}{3}\nabla_{\alpha}\nabla_{\beta}R
-\frac{1}{3}g_{\alpha\beta}\nabla^{2}R
-4R_{\alpha \rho}R_{\beta}^{\rho}+\frac{2}{3}RR_{\alpha\beta}
+2R_{\alpha \rho \mu \nu}R_{\beta}^{\phantom{\beta} \rho \mu \nu}-\\
\nonumber
-\frac{1}{2}g_{\alpha\beta}R_{\sigma \rho \mu \nu}R^{\sigma \rho \mu \nu}
+g_{\alpha\beta}R_{\sigma \rho}R^{\sigma \rho}
-\frac{1}{6}g_{\alpha\beta}R^{2})\delta g^{\alpha\beta}\sqrt{|g|}dV
\end{eqnarray}
so that the principle of minimal action $\delta S=0$ gives the final result.
\normalsize
$\blacksquare$
\end{thm}

We wish to remark that we can write field equations in the form 
\begin{eqnarray}
\nonumber
W_{\alpha\beta}=-\frac{k}{4}T_{\alpha\beta}
\end{eqnarray}
if we define the tensor
\begin{eqnarray}
\nonumber
W_{\alpha\beta}\equiv
\nabla^{2}R_{\alpha\beta}-\frac{1}{3}\nabla_{\alpha}\nabla_{\beta}R
-\frac{1}{6}g_{\alpha\beta}\nabla^{2}R+\\
\nonumber
+(R_{\alpha \rho \mu \nu}R_{\beta}^{\phantom{\beta} \rho \mu \nu}
-\frac{1}{4}g_{\alpha\beta}R_{\sigma \rho \mu \nu}R^{\sigma \rho \mu \nu})-\\
\nonumber
-2(R_{\alpha \rho}R_{\beta}^{\rho}-\frac{1}{4}g_{\alpha\beta}R_{\sigma \rho}R^{\sigma \rho})+\\
\nonumber
+\frac{1}{3}R(R_{\alpha\beta}-\frac{1}{4}g_{\alpha\beta}R)
\end{eqnarray}
which is symmetric, traceless and divergenceless: the fact that it is symmetric is a general property; the tracelessness comes directly from conformal invariance, as we can see for an infinitesimal conformal variation of the metric, which is
\begin{eqnarray}
\nonumber
g'_{\alpha\beta}=\Omega^{2}g_{\alpha\beta}
\approx(1+\delta \gamma)g_{\alpha\beta}=
g_{\alpha\beta}+\delta \gamma g_{\alpha\beta}
\end{eqnarray}
and from which
\begin{eqnarray}
\nonumber
\delta g_{\alpha\beta}\equiv g'_{\alpha\beta}-g_{\alpha\beta}=\delta \gamma g_{\alpha\beta}
\end{eqnarray}
so that in the action it will appear as 
\begin{eqnarray}
\nonumber
\delta S=\int 2W_{\alpha\beta}\delta g^{\alpha\beta}\sqrt{|g|}dV=
\int 2W_{\alpha\beta}g^{\alpha\beta}\delta \gamma\sqrt{|g|}dV=
\int 2W_{\alpha}^{\alpha}\delta \gamma\sqrt{|g|}dV
\end{eqnarray}
and so, after the assumption of minimal action, we get the tracelessness; finally the divergencelessness is, again, a general feature of such a tensor.

Further, we would like to stress that no cosmological term can appear, being $\Lambda g_{\alpha\beta}$ such that its trace is equal to $4\Lambda$, and so it is not traceless. 
%%%%%%%%%%%%%%%%%%%%%%%%%%%%%%%%%%%%%%%%%%%%%%%%%%%%%%%%%%%%%%%%%%%%%%%%%%%%%%%%%%%%%%%%%%%%%%%%%%%
\subsection{Conformal Gauge Fields}
For the conformal invariance of the gauge fields we note that, by definition of gauge strength, the gauge potential can not transform, and so we trivially have that  
\begin{dfn}
Let be given a function $\Omega=\Omega(x)$; the transformation for which
\begin{eqnarray}
\begin{tabular}{|c|}
\hline\\
$A_{\mu}'=A_{\mu}$\\ \\
\hline
\end{tabular}
\label{ctmg}
\end{eqnarray}
is said to be a Conformal Transformation for gauge fields.
\end{dfn}

Then, we see that the gauge strength does not transform $F_{\mu \nu}'=F_{\mu \nu}$.

This can tell us something more about the action we wrote for gauge fields; since we know that the completely covariant form of the gauge strength does not transform, which means that only $2$ field strength must be allowed into the action, then the quadratic action is not only the simplest, but also the only case of conformally invariant action we can consider for gauge fields!
%%%%%%%%%%%%%%%%%%%%%%%%%%%%%%%%%%%%%%%%%%%%%%%%%%%%%%%%%%%%%%%%%%%%%%%%%%%%%%%%%%%%%%%%%%%%%%%%%%%
\subsection{Conformal Scalar Fields}
In the part about fundamentals, we didn't talk about Scalar Fields: they do not appear naturally within the context of such theories; moreover, they have never been observed in Nature as fundamental fields.

Nonetheless, it could always be possible to consider composite scalar fields that appear in the theory, playing a central role, like Higgs fields.

The simplest model for scalar fields is given by the action
\begin{eqnarray}
\nonumber
S=\int (g^{\alpha\beta}D_{\alpha}\phi \ D_{\beta}\phi-m^{2}\phi^{2})\sqrt{|g|}dV
\end{eqnarray}
where $g_{\mu\nu}$ is the metric of the Riemannian spacetime, and where $D_{\alpha}\phi \equiv \partial_{\alpha}\phi+A_{\alpha}\phi$ with $A_{\alpha}$ the gauge fields; we see that this model can not be conformally invariant if the scalar fields are massive: so we will consider massless scalar fields 
\begin{eqnarray}
\nonumber
S=\int g^{\alpha\beta}D_{\alpha}\phi \ D_{\beta}\phi\sqrt{|g|}dV
\end{eqnarray}
for which the conformal invariance can actually be achieved by the transformation law $\phi'=\Omega^{-1}\phi$.

In the previous transformation law, $\Omega$ was considered constant; in general, however, we want to have the possibility to consider the function $\Omega$ as much general as possible.

The presence of the partial derivative operating on a generic function will give an extra term, breaking the conformal invariance; as well as for the previous cases, then, we have to introduce a compensative field able to restore the conformal invariance of the whole action.

In the case of conformal theories, interestingly enough, we do not need to introduce any non-conformally invariant compensative field, since such a field is already provided by the theory itself: it is Ricci curvature scalar $R$. 

We know, in fact, that its transformation law under a conformal transformation is given as
\begin{eqnarray}
\nonumber
R'=\Omega^{-2}(R-6\nabla^{\mu}\ln{\Omega}\nabla_{\mu}\ln{\Omega}-6\nabla^{2}\ln{\Omega})
\end{eqnarray}
while the transformation law for the gauge covariant derivative of the scalar fields is given as 
\begin{eqnarray}
\nonumber
(D_{\mu}\phi)'=\Omega^{-1}(D_{\mu}\phi-\phi\partial_{\mu}\ln{\Omega})
\end{eqnarray}
and so we can consider a new generic term in the action $\xi R \phi^{2}$, which transforms as
\begin{eqnarray}
\nonumber
(\xi R \phi^{2})'=\Omega^{-4}(\xi R\phi^{2}-
6\xi \phi^{2}\nabla^{\mu}\ln{\Omega}\nabla_{\mu}\ln{\Omega}-6\xi \phi^{2}\nabla^{2}\ln{\Omega})\equiv\\
\nonumber
\Omega^{-4}(\xi R\phi^{2}-
6\xi \phi^{2}D^{\mu}\ln{\Omega}D_{\mu}\ln{\Omega}-6\xi \phi^{2}D^{2}\ln{\Omega}),
\end{eqnarray}
and for which the action will be written as 
\begin{eqnarray}
\nonumber
S=\int ((D\phi)^{2}+\xi R \phi^{2})\sqrt{|g|}dV.
\end{eqnarray}

This action transforms as
\begin{eqnarray}
\nonumber
S'\!=S\!-\!\!\int [(6\xi-1)\phi^{2}((D\ln{\Omega})^{2}+D^{2}\ln{\Omega})
+D_{\mu}(\phi^{2}D^{\mu}\ln{\Omega})]\sqrt{|g|}dV
\end{eqnarray}
and taking the factor $\xi$ equal to $1/6$ gives
\begin{eqnarray}\nonumber
S'=S-\int D_{\mu}(\phi^{2}D^{\mu}\ln{\Omega})\sqrt{|g|}dV\equiv
S-\int \nabla_{\mu}(\phi^{2}D^{\mu}\ln{\Omega})\sqrt{|g|}dV
\end{eqnarray}
in which the last term is the integral of a divergence, which can be neglected into the action, and, up to equivalence, we get
\begin{eqnarray}\nonumber
S'=S.
\end{eqnarray}

Then, the action
\begin{eqnarray}
\nonumber
S=\int ((D\phi)^{2}+\frac{1}{6}R \phi^{2})\sqrt{|g|}dV
\end{eqnarray}
is the only conformally invariant action we can have for the scalar fields.

We can generalize it even more by adding a conformally invariant potential $V=V(\phi)$: in a $4$-dimensional geometry, renormalization is possible only if we consider the potential to be at most fourth order in the power of $\phi$, that is $V=\Lambda\phi^{4}$, and we have that the transformation law for scalars tells us that such a potential would actually be conformally invariant. 

We have that the action
\begin{eqnarray}
\nonumber
S=\int ([(D\phi)^{2}+\frac{1}{6}R \phi^{2}]-\Lambda\phi^{4})\sqrt{|g|}dV
\end{eqnarray}
is the most general conformally invariant action for scalar fields we can have.

So, we can give the following
\begin{dfn}
Let be given a function $\Omega=\Omega(x)$; the transformation for which
\begin{eqnarray}
\begin{tabular}{|c|}
\hline\\
$\phi'=\Omega^{-1}\phi$\\ \\
\hline
\end{tabular}
\label{cts}
\end{eqnarray}
is said to be a Conformal Transformation for scalar fields.
\end{dfn}

And thus, we have
\begin{dfn}
The conformally invariant action for scalar fields is
\begin{eqnarray}
S=\int ((D\phi)^{2}+\frac{1}{6}R \phi^{2}-\Lambda\phi^{4})\sqrt{|g|}dV
\end{eqnarray}
where $R$ is the Ricci curvature scalar, and $\Lambda$ is a constant.
\end{dfn}

This action gives the field equations 
\begin{thm}[Field Equations]
Conformal scalar fields equations are
\begin{eqnarray}
D^{2}\phi-\frac{1}{6}R \phi+2\Lambda\phi^{3}=0.
\label{csfe}
\end{eqnarray}
\end{thm}
%%%%%%%%%%%%%%%%%%%%%%%%%%%%%%%%%%%%%%%%%%%%%%%%%%%%%%%%%%%%%%%%%%%%%%%%%%%%%%%%%%%%%%%%%%%%%%%%%%%
\section{Conformal Symmetry and its Breakdown}
The action defined above can be re-written as
\begin{eqnarray}
\nonumber
S=\int ((D\phi)^{2}+\frac{1}{6}R \phi^{2}-\Lambda\phi^{4})\sqrt{|g|}dV=\\
\nonumber
=\int ((D\phi)^{2}-[-\frac{1}{6}R \phi^{2}+\Lambda\phi^{4}])\sqrt{|g|}dV=\\
\nonumber
=\int ((D\phi)^{2}-V(\phi))\sqrt{|g|}dV
\end{eqnarray}
in which we have moved the non-derivative dynamical term into the potential, which is now given as 
\begin{eqnarray}
\nonumber
V(\phi)=\Lambda\phi^{4}-\frac{1}{6}R \phi^{2}.
\end{eqnarray}

If we want that a conformal symmetry breaking occurs in a spontaneous way, we have to be in a situation in which the ground state is such that the conformal symmetry is still a property of the system when the fields are in an unstable stationary point of the potential; because of the instability, the configuration of fields will spontaneously tend to the stable stationary point of the potential, which is, however, non-conformally invariant.

The potential is such that 
\begin{eqnarray}
\nonumber
\frac{\partial V(\phi)}{\partial \phi}=4\Lambda\phi^{3}-\frac{1}{3}R \phi
\end{eqnarray}
and so
\begin{eqnarray}
\nonumber
\frac{\partial^{2} V(\phi)}{\partial \phi^{2}}=12\Lambda\phi^{2}-\frac{1}{3}R;
\end{eqnarray}
the condition to have a spontaneous conformal symmetry breaking is given when the stationary solution $\phi_{0}=0$ is unstable and the stationary solution $12\Lambda\phi_{0}^{2}=R_{0}$ is stable: this means that $R_{0}>0$.

The spontaneous breakdown of the conformal symmetry corresponds to the situation in which the vacuum expectation value of the fields is given as $R_{0}>0$, which means that the ground state has the geometry of an Anti-deSitter spacetime.

This condition is achieved by choosing the constant $\Lambda$ to be positive, i.e. $\Lambda=\lambda^{2}$.

The potential turns out to be
\begin{eqnarray}\nonumber
V(\phi)=\lambda^{2}\phi^{4}-\frac{1}{6}R\phi^{2}
\end{eqnarray}
and so the action will be 
\begin{eqnarray}
\nonumber
S=\int ((D\phi)^{2}-[\lambda^{2}\phi^{4}-\frac{1}{6}R\phi^{2}])\sqrt{|g|}dV,
\end{eqnarray}
with the condition $\phi_{0}^{2}=\frac{R_{0}}{12\lambda^{2}}$ for the vacuum expectation value.

The final step toward the construction of an action that is conformally invariant but at the same time able to produce the spontaneous breakdown of its conformal symmetry is then accomplished, and so we can write the action as 
\begin{dfn}
The conformally invariant action for scalar fields is
\begin{eqnarray}
S=\int ((D\phi)^{2}+\frac{1}{6}R \phi^{2}-\lambda^{2}\phi^{4})\sqrt{|g|}dV
\end{eqnarray}
where $\lambda$ is a constant, and where we have to assume the additional condition $12\lambda^{2}\phi_{0}^{2}=R_{0}$ for the VEV.
\end{dfn}

This action gives the field equations 
\begin{thm}[Field Equations]
Conformal scalar fields equations are
\begin{eqnarray}
(D^{2}-\frac{1}{6}R+2\lambda^{2}\phi^{2})\phi=0,
\label{ssbcsfe}
\end{eqnarray}
where the condition $\phi_{0}^{2}=\frac{1}{12\lambda^{2}}R_{0}$ is assumed for the VEV.
\end{thm}
%%%%%%%%%%%%%%%%%%%%%%%%%%%%%%%%%%%%%%%%%%%%%%%%%%%%%%%%%%%%%%%%%%%%%%%%%%%%%%%%%%%%%%%%%%%%%%%%%%%
\section{Broken Symmetry in Conformal Theories}
We can now put all the pieces together, to get the final action; it is important to remark that, because of the presence of the non-minimal dynamical coupling between gravity and scalar fields $R\phi^{2}$, we have to expect extra terms in the expression of the energy-momentum tensor that enters in the gravitational field equations, as discussed by Callan, Coleman and Jackiw in \cite{c-c-j}.

The final action will be then
\begin{dfn}
The action for conformally invariant scalar fields in interaction with gauge fields in a curved spacetime, with spontaneous conformal symmetry breaking is
\begin{eqnarray}
\nonumber
S=\int (C_{\alpha \beta \mu \nu}C^{\alpha \beta \mu \nu}
-\frac{1}{4L}\mathrm{tr}\it{(F_{\alpha \beta}F^{\alpha \beta})}+\\
+\phi\left(\frac{R}{6}-D^{2}\right)\phi-\lambda^{2}\phi^{4})
\sqrt{|g|}dV
\label{finalaction}
\end{eqnarray}
where $\lambda$ is a constant, and where we have to assume the additional condition $12\lambda^{2}\phi_{0}^{2}=R_{0}$ for the VEV.
\end{dfn}

This action gives the field equations 
\begin{thm}[Field Equations]
Conformal field equations for gravity are 
\begin{eqnarray}
\nonumber
\nabla^{2}R_{\alpha\beta}-\frac{1}{3}\nabla_{\alpha}\nabla_{\beta}R
-\frac{1}{6}g_{\alpha\beta}\nabla^{2}R+\\
\nonumber
+(R_{\alpha \rho \mu \nu}R_{\beta}^{\phantom{\beta} \rho \mu \nu}
-\frac{1}{4}g_{\alpha\beta}R_{\sigma \rho \mu \nu}R^{\sigma \rho \mu \nu})-\\
\nonumber
-2(R_{\alpha \rho}R_{\beta}^{\rho}-\frac{1}{4}g_{\alpha\beta}R_{\sigma \rho}R^{\sigma \rho})+\\
+\frac{1}{3}R(R_{\alpha\beta}-\frac{1}{4}g_{\alpha\beta}R)=-\frac{k}{4}T_{\alpha\beta}
\label{fieldequationsgravity}
\end{eqnarray}
where
\begin{eqnarray}
T_{\alpha\beta}=T^{(\mathrm{gauge})}_{\alpha\beta}+T^{(\mathrm{scalar})}_{\alpha\beta}
\end{eqnarray}
with
\begin{eqnarray}
T^{(\mathrm{gauge})}_{\alpha\beta}=
\frac{1}{4}g_{\alpha\beta}F^{2}-F^{(k)}_{\rho\alpha}F^{(k)\rho}_{\phantom{(k)\rho}\beta}
\end{eqnarray}
and
\begin{eqnarray}
\nonumber
T^{(\mathrm{scalar})}_{\alpha\beta}=
2(D_{\alpha}\phi D_{\beta}\phi-\frac{1}{2}g_{\alpha\beta}(D\phi)^{2})+\\
\nonumber
-\frac{1}{3}(D_{\alpha}D_{\beta}\phi^{2}-g_{\alpha\beta}D^{2}\phi^{2})+\\
+\frac{1}{3}E_{\alpha\beta}\phi^{2}+g_{\alpha\beta}\lambda^{2}\phi^{4};
\end{eqnarray}
conformal field equations for gauge fields are 
\begin{eqnarray}
\nabla_{\rho}F^{(k)\rho\alpha}+C^{k}_{\phantom{k}ij}A^{(i)}_{\rho}F^{(j)\rho\alpha}=2J^{(k)\alpha}
\end{eqnarray}
where
\begin{eqnarray}
J^{(k)\alpha}=-(D^{\alpha}\phi)^{\dagger} G^{(k)} \phi;
\end{eqnarray}
conformal field equations for scalars are
\begin{eqnarray}
(D^{2}-\frac{1}{6}R+2\lambda^{2}\phi^{2})\phi=0,
\end{eqnarray}
with the condition $\phi_{0}^{2}=\frac{1}{12\lambda^{2}}R_{0}$ assumed for the VEV.
\end{thm}

We have the properties
\begin{thm}
The energy-momentum tensor is conserved
\begin{eqnarray}
D_{\alpha}T^{\alpha\beta}=0
\end{eqnarray}
because
\begin{eqnarray}
D_{\alpha}T^{\alpha\beta}_{(\mathrm{gauge})}=0
\end{eqnarray}
and
\begin{eqnarray}
D_{\alpha}T^{\alpha\beta}_{(\mathrm{scalar})}=0,
\end{eqnarray}
and the current vector is also conserved
\begin{eqnarray}
D_{\alpha}J^{(k)\alpha}=0.
\end{eqnarray}
\end{thm}
%%%%%%%%%%%%%%%%%%%%%%%%%%%%%%%%%%%%%%%%%%%%%%%%%%%%%%%%%%%%%%%%%%%%%%%%%%%%%%%%%%%%%%%%%%%%%%%%%%%
\subparagraph{The Gauss-Bonnet Topological Term.} We would like to end this chapter with a consideration about one feature of the field equations for gravity.

In the second-order theory field equations come from just one piece, but in higher-order theories there are many pieces that have to appear with different coefficients: this allows different possible choices for the coefficients, and it could even be possible to have a choice for which we can give to the field equations many properties they may not have \emph{a priori}.

For example, in fourth-order theories of gravity, the contraction of the field equations gives
\begin{eqnarray}
\nonumber
(3a+b+c)\nabla^{2}R=-\frac{k}{4}T
\end{eqnarray}
where $T$ is the trace of the energy-momentum tensor; if we want to describe conformal theories we need $T=0$, and so
\begin{eqnarray}
\nonumber
(3a+b+c)\nabla^{2}R=0,
\end{eqnarray}
which means $3a+b+c=0$, and this condition is verified by the choice of the coefficients $a=\frac{1}{3}$, $b=2$ and $c=-1$ in the conformal action. 

The other choice we would like to talk about is quite interesting. The field equations for fourth-order gravity are 
\begin{eqnarray}
\nonumber
(b+4c)\nabla^{2}R_{\mu \nu}+(2a+\frac{b}{2})g_{\mu \nu}\nabla^{2}R
-(2a+b+2c)\nabla_{\mu}\nabla_{\nu}R-\\
\nonumber
-2c(\frac{1}{4}g_{\mu \nu}R_{\alpha \beta \rho \sigma}R^{\alpha \beta \rho \sigma}-
R_{\nu \beta \rho \sigma}R_{\mu}^{\phantom{\mu} \beta \rho \sigma})
+(4c+2b)R_{\nu \beta \mu \sigma}R^{\beta \sigma}-\\
\nonumber
-(\frac{b}{2}g_{\mu \nu}R_{\alpha \beta}R^{\alpha \beta}
+4cR_{\nu \beta}R_{\mu}^{\phantom{\mu} \beta})
+2aRR_{\nu \mu}-\frac{a}{2}g_{\nu \mu}R^{2}=-\frac{k}{2} T_{\mu \nu}
\end{eqnarray}
and we see that we can remove the dynamical terms from the geometrical left-hand-side requiring that
\begin{eqnarray}
\nonumber
(b+4c)\nabla^{2}R_{\mu \nu}=0\\
\nonumber
(2a+\frac{b}{2})g_{\mu \nu}\nabla^{2}R=0\\
\nonumber
(2a+b+2c)\nabla_{\mu}\nabla_{\nu}R=0
\end{eqnarray}
and so 
\begin{eqnarray}
\nonumber
b+4c=0\\
\nonumber
2a+\frac{b}{2}=0\\
\nonumber
2a+b+2c=0
\end{eqnarray}
which gives the condition
\begin{eqnarray}
\nonumber
4a=-b=4c.
\end{eqnarray}

If the condition $4a=-b=4c$ is satisfied, then field equations will not have the dynamical geometrical terms; the corresponding action (\ref{action2}) would read
\begin{eqnarray}
\nonumber
S_{(a,-4a,a)}=\int
[a(R^{2}-4R^{\alpha\beta}R_{\alpha\beta}+R^{\alpha\theta\mu\rho}R_{\alpha\theta\mu\rho})
+k\mathscr{L}_{p}]\sqrt{|g|}dV
\end{eqnarray}
or, in the vacuum 
\begin{eqnarray}
\nonumber
S_{(a,-4a,a)}=a\int
(R^{2}-4R^{\alpha\beta}R_{\alpha\beta}+R^{\alpha\theta\mu\rho}R_{\alpha\theta\mu\rho})\sqrt{|g|}dV.
\end{eqnarray}

This action turns out to be a very peculiar action, and Gauss and Bonnet were able to prove that it always vanishes; the reason for this fact is that its Lagrangian can be written as a quadridivergence of some vector, and the integral of a quadridivergence of any vector taken in a volume with a surface on which the vector vanishes is always zero.

Gauss-Bonnet theorem can be written as 
\begin{thm}[Gauss-Bonnet]
It is always possible to write
\begin{eqnarray}\nonumber
R_{\alpha \beta \mu \nu}R^{\alpha \beta \mu \nu}-4R_{\alpha \beta}R^{\alpha \beta}+R^{2}=\nabla_{\alpha}V^{\alpha}
\end{eqnarray}
for some vector $V^{\alpha}$.
\end{thm}

This theorem defines what we will call the Gauss-Bonnet topological term $R_{\alpha \beta \mu \nu}R^{\alpha \beta \mu \nu}-4R_{\alpha \beta}R^{\alpha \beta}+R^{2}$; this term is very important, because it shows that in a fourth-order action 
\begin{eqnarray}
\nonumber
S_{(a,b,c)}=\int
(aR^{2}+bR^{\alpha\beta}R_{\alpha\beta}+cR^{\alpha\theta\mu\rho}R_{\alpha\theta\mu\rho})\sqrt{|g|}dV
\end{eqnarray}
we can always replace the Riemann-squared piece with a sum of a Ricci-squared piece plus a scalar-squared plus a quadridivergence that we will always be able to remove, leaving 
\begin{eqnarray}
\nonumber
S_{(a,b,c)}=
\int (aR^{2}+bR^{\alpha\beta}R_{\alpha\beta}+4cR_{\alpha \beta}R^{\alpha \beta}-cR^{2})\sqrt{|g|}dV\\
\nonumber
=\int((a-c)R^{2}+(b+4c)R^{\alpha\beta}R_{\alpha\beta})\sqrt{|g|}dV
\end{eqnarray}
and so, we can rename their still undefined coefficients as
\begin{eqnarray}
\nonumber
S_{(a,b,c)}=
\int((a-c)R^{2}+(b+4c)R^{\alpha\beta}R_{\alpha\beta})\sqrt{|g|}dV\\
\nonumber
=\int(AR^{2}+BR^{\alpha\beta}R_{\alpha\beta})\sqrt{|g|}dV=S_{(A,B)}.
\end{eqnarray}

Gauss-Bonnet topological term allows us to assume a fourth-order action containing the Ricci-squared and the scalar-squared pieces only!

From this action we can calculate field equations, and require their contraction to be zero, to have a conformal theory; the constraint we have to take into account is $3A+B=0$; we will choose then $A=-\frac{2}{3}$ and $B=2$ to get the conformally invariant action for the theory.

In this way, we have that 
\begin{dfn}
The only conformally invariant Gauss-Bonnet modified action for the gravitational field is
\begin{eqnarray}
S=\int [(2R^{\alpha\beta}R_{\alpha\beta}-\frac{2}{3}R^{2})+k\mathscr{L}_{p}]\sqrt{|g|}dV
\end{eqnarray}
in which the constant $k$ has to be determined.
\end{dfn}

This action gives the field equations 
\begin{thm}[Field Equations]
Gauss-Bonnet modified conformal gravitational field equations are given as 
\begin{eqnarray}
\nonumber
\nabla^{2}R_{\mu \nu}-\frac{1}{6}g_{\mu \nu}\nabla^{2}R-\frac{1}{3}\nabla_{\mu}\nabla_{\nu}R+\\
\nonumber
+2R^{\alpha\beta}(R_{\nu \beta \mu \alpha}-\frac{1}{4}g_{\mu \nu}R_{\alpha \beta})-\\
-\frac{2}{3}R(R_{\nu \mu}-\frac{1}{4}g_{\nu \mu}R)=-\frac{k}{4} T_{\mu \nu};
\label{mcfe}
\end{eqnarray}
equations (\ref{mcfe}) are the fundamental Gauss-Bonnet modified field equations of conformal gravity.
\end{thm}
%%%%%%%%%%%%%%%%%%%%%%%%%%%%%%%%%%%%%%%%%%%%%%%%%%%%%%%%%%%%%%%%%%%%%%%%%%%%%%%%%%%%%%%%%%%%%%%%%%%
%%%%%%%%%%%%%%%%%%%%%%%%%%%%%%%%%%%%%%%%%%%%%%%%%%%%%%%%%%%%%%%%%%%%%%%%%%%%%%%%%%%%%%%%%%%%%%%%%%%
\chapter{Conformal Theory of $SO(3)$}
\section{The gauge symmetry group $SO(3)$:\\ the Georgi-Glashow Model}
Once that the action for conformal theory is give as in equation (\ref{finalaction}), eventually Gauss-Bonnet modified, the only thing that is still missing is the gauge symmetry group, and its representation.

We will consider here the simplest non-commutative example: the gauge symmetry group $SO(3)$, called the Georgi-Glashow model (see \cite{g-g}).

In the Georgi-Glashow model, as we said before, we have the explicit choice for the gauge covariant derivative and the gauge fields given as 
\begin{eqnarray}
D_{\mu}\phi^{(a)}=\partial_{\mu}\phi^{(a)}+\varepsilon^{a}_{\phantom{a}bc}A_{\mu}^{(b)}\phi^{(c)},
\end{eqnarray}
\begin{eqnarray}
F_{\mu \nu}^{(a)}=\partial_{\mu}A_{\nu}^{(a)}-\partial_{\nu}A_{\mu}^{(a)}
+\varepsilon^{a}_{\phantom{a}bc}A_{\mu}^{(b)}A_{\nu}^{(c)}\end{eqnarray}
and with the compact $3$-dimensional notation, we have
\begin{eqnarray}
D_{\mu}\vec{\phi}=\partial_{\mu}\vec{\phi}+\vec{A}_{\mu}\wedge\vec{\phi}
\end{eqnarray}
and
\begin{eqnarray}
D^{2}\vec{\phi}=D^{\mu}D_{\mu}\vec{\phi}=\nabla^{\mu}\nabla_{\mu}\vec{\phi}
+2\vec{A}^{\mu}\wedge\nabla_{\mu}\vec{\phi}+\vec{A}_{\mu}\wedge(\vec{A}^{\mu}\wedge\vec{\phi})+
\nabla_{\mu}\vec{A}^{\mu}\wedge\vec{\phi}
\end{eqnarray}
and so 
\begin{eqnarray}
[D_{\alpha},D_{\beta}]\vec{\phi}=\vec{F}_{\alpha \beta}\wedge\vec{\phi}
\end{eqnarray}
with
\begin{eqnarray}
\vec{F}_{\mu \nu}=\partial_{\mu}\vec{A}_{\nu}-\partial_{\nu}\vec{A}_{\mu}+\vec{A}_{\mu}\wedge\vec{A}_{\nu}
\end{eqnarray}
for which we have that
\begin{eqnarray}
D_{\mu}\vec{F}^{\mu \nu}=\nabla_{\mu}\vec{F}^{\mu \nu}+\vec{A}_{\mu}\wedge\vec{F}^{\mu \nu}
\end{eqnarray}
with
\begin{eqnarray}
D_{\mu}\vec{F}_{\alpha \nu}+D_{\alpha}\vec{F}_{\nu \mu}+D_{\nu}\vec{F}_{\mu \alpha}=0
\end{eqnarray}
if we want to stop at the second order derivative.

Given the gauge symmetry group, its representation and the threedimensional notation, it is possible to write the action in a completely explicit form as follow
\begin{dfn}
The action for conformally invariant scalar fields in interaction with $SO(3)$ symmetric gauge fields in a curved spacetime, with spontaneous conformal symmetry breaking is
\begin{eqnarray}
S=\int (C_{\alpha \beta \mu \nu}C^{\alpha \beta \mu \nu}
-\frac{1}{4}\vec{F}_{\mu \nu}\cdot\vec{F^{\mu \nu}}
+\vec{\phi}\cdot\left(\frac{R}{6}-D^{2}\right)\vec{\phi}-\lambda^{2}\phi^{4})\sqrt{|g|}d^{4}x
\label{veryfinalactionexplicit}
\end{eqnarray}
where $\lambda$ is a constant, and where we have to assume the additional condition $12\lambda^{2}\phi_{0}^{2}=R_{0}$ for the VEV.
\end{dfn}

This action gives the field equations 
\begin{thm}[Field Equations]
Conformal field equations for gravity are 
\begin{eqnarray}
\nonumber
\nabla^{2}R_{\alpha\beta}-\frac{1}{3}\nabla_{\alpha}\nabla_{\beta}R
-\frac{1}{6}g_{\alpha\beta}\nabla^{2}R+\\
\nonumber
+(R_{\alpha \rho \mu \nu}R_{\beta}^{\phantom{\beta} \rho \mu \nu}
-\frac{1}{4}g_{\alpha\beta}R_{\sigma \rho \mu \nu}R^{\sigma \rho \mu \nu})-\\
\nonumber
-2(R_{\alpha \rho}R_{\beta}^{\rho}-\frac{1}{4}g_{\alpha\beta}R_{\sigma \rho}R^{\sigma \rho})+\\
+\frac{1}{3}R(R_{\alpha\beta}-\frac{1}{4}g_{\alpha\beta}R)=-\frac{k}{4}T_{\alpha\beta}
\label{fieldequationsgravityexplicit}
\end{eqnarray}
where
\begin{eqnarray}
T_{\alpha\beta}=T^{(\mathrm{gauge})}_{\alpha\beta}+T^{(\mathrm{scalar})}_{\alpha\beta}
\end{eqnarray}
with
\begin{eqnarray}
T^{(\mathrm{gauge})}_{\alpha\beta}=
\frac{1}{4}g_{\alpha\beta}\vec{F}_{\mu\nu}\cdot\vec{F}^{\mu\nu}
-\vec{F}_{\alpha \mu}\cdot\vec{F}_{\beta}^{\phantom{\beta} \mu}
\end{eqnarray}
and 
\begin{eqnarray}
\nonumber
T^{(\mathrm{scalar})}_{\alpha\beta}=
2(D_{\alpha}\vec{\phi} \cdot D_{\beta}\vec{\phi}
-\frac{1}{2}g_{\alpha\beta}D_{\rho}\vec{\phi} \cdot D^{\rho}\vec{\phi})+\\
\nonumber
-\frac{1}{3}(D_{\alpha}D_{\beta}\phi^{2}-g_{\alpha\beta}D^{2}\phi^{2})+\\
+\frac{1}{3}E_{\alpha\beta}\phi^{2}+g_{\alpha\beta}\lambda^{2}\phi^{4};
\end{eqnarray}
conformal field equations for gauge fields are
\begin{eqnarray}
D_{\mu}\vec{F}^{\mu \nu}=2\vec{J}^{\nu}
\label{fieldequationsgaugeexplicit}
\end{eqnarray}
where
\begin{eqnarray}
\vec{J}_{\mu}=D_{\mu}\vec{\phi}\wedge\vec{\phi};
\end{eqnarray}
conformal field equations for scalars are
\begin{eqnarray}
(D^{2}-\frac{1}{6}R+2\lambda^{2}\phi^{2})\vec{\phi}=0,
\label{fieldequationsscalarexplicit}
\end{eqnarray}
where the condition of spontaneous conformal symmetry breaking is given with $\phi_{0}^{2}=\frac{1}{12\lambda^{2}}R_{0}$ for the VEV.
\end{thm}

In an explicit notation we have 
\begin{thm}[Field Equations]
Conformal field equations are given for the scalar fields as
\begin{eqnarray}
\nabla^{2}\vec{\phi}=(\vec{\phi}\wedge\nabla_{\mu}\vec{A}^{\mu}
-\vec{A}_{\mu}\wedge(\vec{A}^{\mu}\wedge\vec{\phi})
+2\nabla_{\mu}\vec{\phi}\wedge\vec{A}^{\mu})
+\frac{R}{6}\vec{\phi}
-2\lambda^{2}\phi^{2}\vec{\phi};
\end{eqnarray}
for the gauge fields as
\begin{eqnarray}
\nabla_{\mu}\vec{F}^{\mu \nu}=2\nabla^{\nu}\vec{\phi}\wedge\vec{\phi}+
+2(\vec{A}^{\nu}\wedge\vec{\phi})\wedge\vec{\phi}+
\vec{F}^{\mu \nu}\wedge\vec{A}_{\mu};
\end{eqnarray}
and for the gravitational field as
\begin{eqnarray}\nonumber
\frac{1}{6}g_{\mu \nu}\nabla^{2}R-\nabla^{2}R_{\mu \nu}+\frac{1}{3}\nabla_{\mu}\nabla_{\nu}R+\\
\nonumber
-2R^{\sigma \beta}(R_{\beta \nu \sigma \mu }-\frac{1}{4}g_{\mu \nu}R_{\sigma \beta})+
\frac{2}{3}R(R_{\nu \mu}-\frac{1}{4}g_{\nu \mu}R)=\\
\nonumber
=\frac{1}{4}\left[\right.
(\frac{1}{4}g_{\nu \mu}\vec{F}_{\alpha \beta}\cdot \vec{F}^{\alpha \beta}
-\vec{F}_{\mu \rho}\cdot \vec{F}_{\nu}^{\phantom{\nu} \rho})+\\
\nonumber
+(2(\vec{A}_{\mu}\cdot \vec{A}_{\nu}\phi^{2}-(\vec{A}_{\mu}\cdot \vec{\phi})(\vec{A}_{\nu}\cdot \vec{\phi}))
-g_{\nu \mu}(A^{2}\phi^{2}-(\vec{A}^{\rho}\cdot \vec{\phi})(\vec{A}_{\rho}\cdot \vec{\phi})))+\\
\nonumber
+(2(A_{\nu}\cdot \vec{\phi}\wedge\nabla_{\mu}\vec{\phi}+A_{\mu}\cdot \vec{\phi}\wedge\nabla_{\nu}\vec{\phi})
-2g_{\nu \mu}A^{\rho}\cdot \vec{\phi}\wedge\nabla_{\rho}\vec{\phi})+\\
\nonumber
+(\frac{1}{3}g_{\mu \nu}\nabla^{2}\phi^{2}-\frac{1}{3}\nabla_{\mu}\nabla_{\nu}\phi^{2}+
2\nabla_{\mu}\vec{\phi}\cdot \nabla_{\nu}\vec{\phi}-g_{\nu \mu}\nabla_{\rho}\vec{\phi}\cdot \nabla^{\rho}\vec{\phi})+\\
g_{\nu \mu}\lambda^{2}\phi^{4}+\frac{1}{3}E_{\nu \mu}\phi^{2}\left.\right].
\end{eqnarray}
The condition of spontaneous conformal symmetry breaking is given as $\phi_{0}^{2}=\frac{1}{12\lambda^{2}}R_{0}$ for the VEV.
\end{thm}

We have the properties
\begin{thm}
The energy-momentum tensor is conserved
\begin{eqnarray}
D_{\alpha}T^{\alpha\beta}=0
\end{eqnarray}
because
\begin{eqnarray}
D_{\alpha}T^{\alpha\beta}_{(\mathrm{gauge})}=0
\end{eqnarray}
and
\begin{eqnarray}
D_{\alpha}T^{\alpha\beta}_{(\mathrm{scalar})}=0,
\end{eqnarray}
and the current vector is also conserved
\begin{eqnarray}
D_{\alpha}\vec{J}^{\alpha}=0.
\end{eqnarray}
\end{thm}

It is important to remark that the conservation laws, as well as the irreducibility of gravitational field equations, are indeed obtained by using only the field equations themselves: for example, the irreducibility is given by the fact that
\begin{eqnarray}
\nonumber
0=-D_{\rho}\vec{\phi} \cdot D^{\rho}\vec{\phi}
+\frac{1}{2}\nabla^{2}\phi^{2}-\frac{1}{6}R\phi^{2}+2\lambda^{2}\phi^{4}
\end{eqnarray}
because of the irreducibility of the energy-momentum tensor of the gauge fields in $4$ dimensions, so that
\begin{eqnarray}\nonumber
-D_{\rho}\vec{\phi} \cdot D^{\rho}\vec{\phi}+\frac{1}{2}\nabla^{2}\phi^{2}-\frac{1}{6}R\phi^{2}+2\lambda^{2}\phi^{4}=0
\end{eqnarray}
which can be re-written as
\begin{eqnarray}\nonumber
-D_{\rho}\vec{\phi} \cdot D^{\rho}\vec{\phi}+\frac{1}{2}D^{2}\phi^{2}-\frac{1}{6}R\phi^{2}+2\lambda^{2}\phi^{4}=0
\end{eqnarray}
then
\begin{eqnarray}\nonumber
\vec{\phi} \cdot (D^{2}\vec{\phi}-\frac{1}{6}R\vec{\phi}+2\lambda^{2}\phi^{2}\vec{\phi})=0
\end{eqnarray}
which is automatically satisfied by taking into account the field equations for the scalar fields; the same field equations together with the field equations for the gauge fields provide these laws for the energy-momentum tensors
\begin{eqnarray}\nonumber
\nabla_{\nu}T_{(\mathrm{gauge})}^{\mu \nu}=-2\vec{F}^{\mu \rho}\cdot \vec{J}_{\rho}
\end{eqnarray}
and
\begin{eqnarray}\nonumber
\nabla_{\nu}T_{(\mathrm{scalar})}^{\mu \nu}=2(\vec{F}^{\mu \rho}\cdot \vec{J}_{\rho}
+\frac{1}{12}R\nabla^{\mu}\phi^{2}-\frac{1}{2}\lambda^{2}\nabla^{\mu}\phi^{4})
\end{eqnarray}
which gives
\begin{eqnarray}\nonumber
\nabla_{\nu}T^{\mu \nu}=0
\end{eqnarray}
expressing the conservation law for the energy-momentum of the total system; finally, if we consider the field equations for the gauge fields, we get that
\begin{eqnarray}\nonumber
2D_{\nu}\vec{J}^{\nu}=D_{\nu}D_{\mu}\vec{F}^{\mu \nu} \equiv 0
\end{eqnarray}
or considering the field equations for the scalar fields, we get
\begin{eqnarray}
\nonumber
0 \equiv D^{2}\vec{\phi}\wedge \vec{\phi}=D_{\nu}(D^{\nu}\vec{\phi}\wedge \vec{\phi})=D_{\nu}\vec{J}^{\nu}
\end{eqnarray}
so that, in any case, we get the conservation law for the current.

In this way, we have proven that we actually have conservation laws, but they are all already contained in the field equations.
%%%%%%%%%%%%%%%%%%%%%%%%%%%%%%%%%%%%%%%%%%%%%%%%%%%%%%%%%%%%%%%%%%%%%%%%%%%%%%%%%%%%%%%%%%%%%%%%%%%
\section{The spacetime symmetry group $\mathbb{R}\times SO(3)$:\\ stationary spherically symmetric spaces}
Among all the possible applications we can consider, the most important is certainly the case of spherical symmetry, which is also stationary, therefore the space is split into the time and the spherically symmetric ordinary space, i.e. the distribution of energy does not depend explicitly on time, neither on the angles, but only on the radial coordinate. Then, the space is topologically equivalent to $\mathbb{R}\times SO(3)$, where $SO(3)$ is the group of the spatial isometries.

This structure for spatial isometries allow us to consider $3$ linearly independent $4$-dimensional Killing vectors in spherical coordinates, as follow
\begin{eqnarray}
\xi_{(1)}=(0,0,\cos{\varphi},-\sin{\varphi}\cot{\theta})\\
\xi_{(2)}=(0,0,-\sin{\varphi},-\cos{\varphi}\cot{\theta})\\
\xi_{(3)}=(0,0,0,1)
\end{eqnarray}
for which it is obviously
\begin{eqnarray}
[\xi_{(i)},\xi_{(j)}]=\varepsilon_{ijk}\xi_{(k)}
\end{eqnarray}
so, the correct representation of the symmetry group for $SO(3)$.

The metric tensor in spherical coordinates has to be isometric, that is its Lie derivative must be zero; a straightforward calculation gives the fact that for a diagonal metric it is of the form
\begin{eqnarray}
\nonumber
g_{tt}=A(r)\\
\nonumber
g_{rr}=-B(r)\\
\nonumber
g_{\theta\theta}=-C(r)\\
\nonumber
g_{\varphi\varphi}=-C(r)\sin^{2}{\theta}
\end{eqnarray}
in which it can be proven that in a frame at rest with respect to the origin of the coordinates, we can choose $C(r)=r^{2}$, so
\begin{eqnarray}
\nonumber
g_{tt}=A(r)\\
\nonumber
g_{rr}=-B(r)\\
\nonumber
g_{\theta\theta}=-r^{2}\\
\nonumber
g_{\varphi\varphi}=-r^{2}\sin^{2}{\theta},
\end{eqnarray}
and Mannheim and Kazanas in \cite{m-k/conf} showed how we can always reduce it to a form in which $A=\frac{1}{B}=1+h(r)$ for a smooth function $h(r)$, so that
\begin{eqnarray}
g_{tt}=1+h(r)\\
g_{rr}=-\frac{1}{1+h(r)}\\
g_{\theta\theta}=-r^{2}\\
g_{\varphi\varphi}=-r^{2}\sin^{2}{\theta}.
\end{eqnarray}

Given this metric, we can compute the symmetric connection
\begin{eqnarray}
\Gamma^{\theta}_{r \theta}=\Gamma^{\varphi}_{r \varphi}=\frac{1}{r}\\
\Gamma^{\varphi}_{\varphi \theta}=\cot{\theta}\\
\Gamma^{\theta}_{\varphi \varphi}=-\sin{\theta}\cos{\theta}\\
\Gamma^{r}_{\varphi \varphi}=\sin^{2}{\theta}\Gamma^{r}_{\theta \theta}=-r(1+h)\sin^{2}{\theta}\\
\Gamma^{r}_{r r}=-\frac{h'}{2(1+h)}\\
\Gamma^{r}_{t t}=\frac{1}{2}h'(1+h)\\
\Gamma^{t}_{r t}=\frac{h'}{2(1+h)}.
\end{eqnarray}

Finally, Riemann curvature tensor has all the components vanishing except
\begin{eqnarray}
R_{trtr}=-\frac{h''}{2}\\
R_{t\theta t\theta}=-\frac{r(1+h)h'}{2}\\
R_{r\theta r\theta}=\frac{rh'}{2(1+h)}\\
R_{t\varphi t\varphi}=-\frac{r(1+h)h'}{2}\sin^{2}{\theta}\\
R_{r\varphi r\varphi}=\frac{rh'}{2(1+h)}\sin^{2}{\theta}\\
R_{\varphi \theta \varphi \theta}=r^{2}h\sin^{2}{\theta},
\end{eqnarray}
Ricci tensor has all the components vanishing except
\begin{eqnarray}
R_{tt}=(1+h)(\frac{h''}{2}+\frac{h'}{r})\\
R_{rr}=-\frac{1}{(1+h)}(\frac{h''}{2}+\frac{h'}{r})\\
R_{\theta \theta}=-(rh'+h)\\
R_{\varphi \varphi}=-(rh'+h)\sin^{2}{\theta}
\end{eqnarray}
and the scalar curvature is given by
\begin{eqnarray}
R=\frac{(r^{2}h)''}{r^{2}}.
\end{eqnarray}

With all these forms, it is possible to calculate all the products and the contractions, and all the physical quantities that enter in the field equations.

For example, we can give the expression of the derivatives of scalars as follow
\begin{eqnarray}\nonumber
\nabla_{t}\nabla_{t}A=-\frac{(1+h)h'}{2}A'\\
\nonumber
\nabla_{r}\nabla_{r}A=A''+\frac{h'}{2(1+h)}A'\\
\nonumber
\nabla_{\theta}\nabla_{\theta}A=r(1+h)A'\\
\nonumber
\nabla_{\varphi}\nabla_{\varphi}A=r(1+h)\sin^{2}{\theta}A'
\end{eqnarray}
so that 
\begin{eqnarray}\nonumber
\nabla^{2}A=-((1+h)A''+\left(\frac{2(1+h)+rh'}{r}\right)A')
\end{eqnarray}
and these equations hold for any scalar, included the curvature scalar $R$.

We also have specific expressions for the curvatures; the squared Riemann 
\begin{eqnarray}\nonumber
R^{\alpha \beta \mu \nu}R_{\alpha \beta \mu \nu}=
4\left(\left(\frac{h''}{2}\right)^{2}
+\left(\frac{h'}{r}\right)^{2}+\left(\frac{h}{r^{2}}\right)^{2}\right)
\end{eqnarray} 
which is always positive; the products of Riemann and Ricci are
\begin{eqnarray}\nonumber
R^{\alpha \beta}R_{\alpha t \beta t}=
\frac{(1+h)}{2}\left(\frac{h''^{2}}{2}+\frac{h'h''}{r}+\frac{2h'^{2}}{r^{2}}+\frac{2hh'}{r^{3}}\right)
\end{eqnarray}
\begin{eqnarray}\nonumber
R^{\alpha \beta}R_{\alpha r \beta r}=
-\frac{1}{2(1+h)}\left(\frac{h''^{2}}{2}+\frac{h'h''}{r}+\frac{2h'^{2}}{r^{2}}+\frac{2hh'}{r^{3}}\right)
\end{eqnarray}
\begin{eqnarray}\nonumber
R^{\alpha \beta}R_{\alpha \theta \beta \theta}=
-\frac{r^{2}}{2}\left(\frac{h'h''}{r}+\frac{2h'^{2}}{r^{2}}+\frac{2hh'}{r^{3}}+\frac{2h^{2}}{r^{4}}\right)
\end{eqnarray}
\begin{eqnarray}\nonumber
R^{\alpha \beta}R_{\alpha \varphi \beta \varphi}=
-\frac{r^{2}\sin^{2}{\theta}}{2}\left(\frac{h'h''}{r}+\frac{2h'^{2}}{r^{2}}+\frac{2hh'}{r^{3}}+\frac{2h^{2}}{r^{4}}\right)
\end{eqnarray}
from which we can get the squared of Ricci
\begin{eqnarray}\nonumber
R_{\alpha \beta}R^{\alpha \beta}=
2\left(\left(\frac{h''}{2}+\frac{h'}{r}\right)^{2}+\left(\frac{h'}{r}+\frac{h}{r^{2}}\right)^{2}\right)
\end{eqnarray}
always positive: so, the squared of the three main curvatures are all positive; the squared of Weyl conformal tensor is given as
\begin{eqnarray}\nonumber
C^{\alpha\beta\mu\nu}C_{\alpha\beta\mu\nu}
=\frac{4}{3}\left(\frac{h''}{2}-\frac{h'}{r}+\frac{h}{r^{2}}\right)^{2}
\end{eqnarray}
which is always positive.

For the derivatives of curvatures we have 
\begin{eqnarray}\nonumber
\nabla^{2}R_{t t}=
-(1+h)R''_{t t}-\left(\frac{2(1+h)-rh'}{r}\right)R'_{t t}+\\
\nonumber
+\left(h''+\frac{2h'}{r}-\frac{h'^{2}}{(1+h)}\right)R_{t t}
\end{eqnarray}
\begin{eqnarray}\nonumber
\nabla^{2}R_{r r}=
-(1+h)R''_{r r}-\left(\frac{2(1+h)+3rh'}{r}\right)R'_{r r}+\\
\nonumber
+\left(-h''-\frac{2h'}{r}+\frac{4(1+h)}{r^{2}}-\frac{h'^{2}}{(1+h)}\right)R_{r r}-\frac{4}{r^{4}}R_{\theta \theta}
\end{eqnarray}
\begin{eqnarray}\nonumber
\nabla^{2}R_{\theta \theta}=
-(1+h)R''_{\theta \theta}+\left(\frac{1+h-rh'}{r}\right)R'_{\theta \theta}
+\left(\frac{2h'}{r}\right)R_{\theta \theta}
\end{eqnarray}
\begin{eqnarray}\nonumber
\nabla^{2}R_{\varphi \varphi}=
-(1+h)R''_{\varphi \varphi}+\left(\frac{1+h-rh'}{r}\right)R'_{\varphi \varphi}
+\left(\frac{2h'}{r}\right)R_{\varphi \varphi}
\end{eqnarray}
from which
\begin{eqnarray}\nonumber
\nabla^{2}R=-((1+h)R''+\left(\frac{2(1+h)+rh'}{r}\right)R').
\end{eqnarray}
%%%%%%%%%%%%%%%%%%%%%%%%%%%%%%%%%%%%%%%%%%%%%%%%%%%%%%%%%%%%%%%%%%%%%%%%%%%%%%%%%%%%%%%%%%%%%%%%%%%
\section{Spatial and gauge mixed $SO(3)$ symmetry}
The following \emph{Ansatz} we will make is to consider the fact that the lowest energy solutions are those with the maximal symmetry (see \cite{c-l}), which is $SO(3)$, as for the spatial coordinates.

So, the explicit assumption on the components of the scalar fields is
\begin{eqnarray}
\phi^{(a)}=f(r^{2})\frac{r^{a}}{r}
\end{eqnarray}
where $f(r)$ is an undefined function.

For the gauge potentials we have to notice that, since they are vectors, they also have to be covariant under the space-time symmetry, which is described by a structure topologically equivalent to $\mathbb{R}\times SO(3)$, for which the general covariance has to be restricted to transformations that do not involve time, and for which the $3$-dimensional covariance is restricted to the symmetry group $SO(3)$; the temporal component will not transform, and the pure spatial ones will transform according to 
\begin{eqnarray}\nonumber
A'^{a}=\frac{\partial x'^{a}}{\partial x^{b}}A^{b}
\end{eqnarray}
where 
\begin{eqnarray}\nonumber
\frac{\partial x'^{a}}{\partial x^{b}}\in SO(3).
\end{eqnarray}

Once that the general covariance is broken, and the form for these vectors is fixed, it makes sense to assume them of the form 
\begin{eqnarray}
A^{0 (a)}=0
\end{eqnarray}
and
\begin{eqnarray}
A^{i (a)}=g(r^{2})\sum_{k=1}^{k=3}\varepsilon^{aik}r^{k}
\end{eqnarray}
where $g(r)$ is an undefined function.

In terms of $3$-dimensional notation they are
\begin{eqnarray}
\vec{A}_{0}=0\\
\vec{A}_{a}=g(r^{2})\vec{n}_{a}\wedge\vec{r}
\end{eqnarray}
where $\vec{n}_{a}$ is the unity vector along the $a$ axis, and
\begin{eqnarray}
\vec{\phi}=f(r^{2})\frac{\vec{r}}{r}.
\end{eqnarray}

Now, since the frame we have chosen is in spherical coordinates, we have to write all these fields with respect to the spherical coordinates $(r,\theta,\varphi)$. 

For the scalar fields there is no problem, and a simple calculation gives
\begin{eqnarray}
\phi^{(1)}=f(r^{2})\sin{\theta}\sin{\varphi}\\
\phi^{(2)}=f(r^{2})\sin{\theta}\cos{\varphi}\\
\phi^{(3)}=f(r^{2})\cos{\theta}.
\end{eqnarray}

For the gauge fields we can notice that the explicit expression for Killing vectors for $SO(3)$ in Cartesian coordinates is
\begin{eqnarray}
\xi_{(1)}=
\left(\begin{array}{c}
0 \\
0 \\
z \\ 
-y
\end{array}\right); \ \
\xi_{(2)}=
\left(\begin{array}{c}
0 \\
-z \\
0 \\ 
x
\end{array}\right); \ \
\xi_{(3)}=
\left(\begin{array}{c}
0 \\
y \\
-x \\ 
0
\end{array}\right)
\end{eqnarray}
which means that they are proportional to the explicit expression of the gauge potentials; so we can write the relationships
\begin{eqnarray}
g\vec{\xi}^{\mu}=\vec{A}^{\mu};
\end{eqnarray}
the expression for the gauge potentials in spherical coordinates is easily given knowing the expression in spherical coordinates of the Killing vectors above
\begin{eqnarray}
A_{(1)}=
g(r^{2})\left(\begin{array}{c}
0 \\
0 \\
\cos{\phi} \\ 
-\sin{\phi}\cot{\theta}
\end{array}\right)\\
A_{(2)}=
g(r^{2})\left(\begin{array}{c}
0 \\
0 \\
-\sin{\phi} \\ 
-\cos{\phi}\cot{\theta}
\end{array}\right)\\
A_{(3)}=
g(r^{2})\left(\begin{array}{c}
0 \\
0 \\
0 \\ 
1
\end{array}\right).
\end{eqnarray}

Now we can define 
\begin{eqnarray}
1+r^{2}g(r^{2})=a(r^{2})
\end{eqnarray}
which will allow a formal simplification of all the formulae we are going to write down.

Firstly, we get the following invariants
\begin{eqnarray}
\phi^{2}=\vec{\phi}\cdot\vec{\phi}=f^{2}
\end{eqnarray}
\begin{eqnarray}
\vec{A}^{\mu}\cdot\vec{\phi}=0
\end{eqnarray}
\begin{eqnarray}
A^{2}=\vec{A}^{\mu}\cdot\vec{A}_{\mu}=-2rg
\end{eqnarray}
for some particular expressions involving the fields.

Considering the fields and their derivatives, then, we get
\begin{eqnarray}\nonumber
\nabla_{\mu}\vec{A}^{\mu}=0
\end{eqnarray}
\begin{eqnarray}\nonumber
\vec{A}^{\mu}\wedge\partial_{\mu}\vec{\phi}=2g\vec{\phi}
\end{eqnarray}
\begin{eqnarray}\nonumber
\partial_{\mu}\vec{\phi}\cdot\partial^{\mu}\vec{\phi}=
-\left((1+h)f'^{2}+2\frac{f^{2}}{r^{2}}\right).
\end{eqnarray}

With these basic expressions, we will compute all the fundamental ones; we have that the gauge covariant derivative of the scalar fields has a particular structure  
\begin{eqnarray}
D_{t}\vec{\phi}=0\\
D_{r}\vec{\phi}=\partial_{r}\vec{\phi}\\
D_{\theta}\vec{\phi}=a\partial_{\theta}\vec{\phi}\\
D_{\varphi}\vec{\phi}=a\partial_{\varphi}\vec{\phi}
\end{eqnarray}
and the gauge strength has also a particular structure
\begin{eqnarray}
\vec{F}_{t \mu}=0\\
\vec{F}_{r \mu}=\frac{a'}{a-1}\vec{A}_{\mu}\\
\vec{F}_{\theta \varphi}=\frac{1-a^{2}}{f}\sin{\theta}\vec{\phi}
\end{eqnarray}
which allows us to split the temporal and spatial part, talking about non-commutative electric and non-commutative magnetic fields, and they are such that $\vec{E}_{a}=0$, and
\begin{eqnarray}
\vec{B}^{\theta}=-\frac{1}{r^{2}\sin{\theta}}\frac{a'}{(a-1)}\vec{A}_{\varphi}\\
\vec{B}^{\varphi}=\frac{1}{r^{2}\sin{\theta}}\frac{a'}{(a-1)}\vec{A}_{\theta}\\
\vec{B}^{r}=\frac{1-a^{2}}{r^{2}f}\vec{\phi}
\end{eqnarray}
in which the symbol of vector represents the vectors in the internal space. 

We can build up the most important invariants as
\begin{eqnarray}
F^{2}\equiv \vec{F}_{\alpha \beta}\cdot \vec{F}^{\alpha \beta}
=\frac{2}{r^{2}}(2(1+h)(a')^{2}+\frac{(a^{2}-1)^{2}}{r^{2}})
\end{eqnarray}
and 
\begin{eqnarray}
(D\phi)^{2} \equiv D_{\mu}\vec{\phi} \cdot D^{\mu}\vec{\phi}=-((1+h)(f')^{2}+2\frac{a^{2}f^{2}}{r^{2}})
\end{eqnarray}
which will be used in the action.

Finally, let's give the expressions for the conserved quantities, starting from the current
\begin{eqnarray}
\vec{J}_{\mu}=\frac{af^{2}}{(1-a)}\vec{A}_{\mu}
\end{eqnarray}
and then giving the expressions for the energy-momentum tensor of the scalar fields as 
\begin{eqnarray}
\nonumber
T^{(\mathrm{scalar})}_{tt}=
\frac{1}{3}(1+h)[-2(1+h)ff''+(1+h)f'^{2}-ff'\left(h'+4\frac{(1+h)}{r}\right)\\
-\left(\frac{h'}{r}+\frac{h}{r^{2}}\right)f^{2}+6\frac{a^{2}f^{2}}{r^{2}}+3\lambda^{2}f^{4}]\\
\nonumber
T^{(\mathrm{scalar})}_{rr}=
-\frac{1}{3(1+h)}[-3(1+h)f'^{2}-ff'\left(h'+4\frac{(1+h)}{r}\right)\\
-\left(\frac{h'}{r}+\frac{h}{r^{2}}\right)f^{2}+6\frac{a^{2}f^{2}}{r^{2}}+3\lambda^{2}f^{4}]\\
\nonumber
T^{(\mathrm{scalar})}_{\theta \theta}=
\frac{1}{3}[2r^{2}(1+h)ff''-r^{2}(1+h)f'^{2}+2r(1+h+rh')ff'\\
+\left(\frac{h''}{2}+\frac{h'}{r}\right)r^{2}f^{2}-3\lambda^{2}r^{2}f^{4}]\\
\nonumber
T^{(\mathrm{scalar})}_{\varphi \varphi}=
\frac{1}{3}[2r^{2}(1+h)ff''-r^{2}(1+h)f'^{2}+2r(1+h+rh')ff'\\
+\left(\frac{h''}{2}+\frac{h'}{r}\right)r^{2}f^{2}-3\lambda^{2}r^{2}f^{4}]\sin^{2}{\theta}
\end{eqnarray} 
and for the energy-momentum tensor of the gauge fields as
\begin{eqnarray}
T^{(\mathrm{gauge})}_{t t}=
\frac{(1+h)}{e^{2}r^{2}}\left((1+h)a'^{2}+\frac{(a^{2}-1)^{2}}{2r^{2}}\right)\\
T^{(\mathrm{gauge})}_{r r}=
\frac{1}{e^{2}r^{2}(1+h)}\left((1+h)a'^{2}-\frac{(a^{2}-1)^{2}}{2r^{2}}\right)\\
T^{(\mathrm{gauge})}_{\theta \theta}=\frac{(a^{2}-1)^{2}}{2e^{2}r^{2}}\\
T^{(\mathrm{gauge})}_{\varphi \varphi}=\sin^{2}{\theta}\frac{(a^{2}-1)^{2}}{2e^{2}r^{2}};
\end{eqnarray}
the total energy-momentum tensor is 
\begin{eqnarray}
\nonumber
T_{tt}=
\frac{1}{3}(1+h)\left[\right.-2(1+h)ff''+(1+h)f'^{2}-ff'\left(h'+4\frac{(1+h)}{r}\right)+\\
-\left(\frac{h'}{r}+\frac{h}{r^{2}}\right)f^{2}+6\frac{a^{2}f^{2}}{r^{2}}+3\lambda^{2}f^{4}
+\frac{3}{e^{2}r^{2}}\left((1+h)a'^{2}+\frac{(a^{2}-1)^{2}}{2r^{2}}\right)\left.\right]\\
\nonumber
T_{rr}=
-\frac{1}{3(1+h)}\left[\right.-3(1+h)f'^{2}-ff'\left(h'+4\frac{(1+h)}{r}\right)-\\
-\left(\frac{h'}{r}+\frac{h}{r^{2}}\right)f^{2}+6\frac{a^{2}f^{2}}{r^{2}}+3\lambda^{2}f^{4}
-\frac{3}{e^{2}r^{2}}\left((1+h)a'^{2}-\frac{(a^{2}-1)^{2}}{2r^{2}}\right)\left.\right]\\
\nonumber
T_{\theta \theta}=
\frac{1}{3}[2r^{2}(1+h)ff''-r^{2}(1+h)f'^{2}+2r(1+h+rh')ff'\\
+\left(\frac{h''}{2}+\frac{h'}{r}\right)r^{2}f^{2}-3\lambda^{2}r^{2}f^{4}
+\frac{3(a^{2}-1)^{2}}{2e^{2}r^{2}}]\\
\nonumber
T_{\varphi \varphi}=
\frac{1}{3}[2r^{2}(1+h)ff''-r^{2}(1+h)f'^{2}+2r(1+h+rh')ff'\\
+\left(\frac{h''}{2}+\frac{h'}{r}\right)r^{2}f^{2}-3\lambda^{2}r^{2}f^{4}
+\frac{3(a^{2}-1)^{2}}{2e^{2}r^{2}}]\sin^{2}{\theta}
\end{eqnarray}
and this conclude the list of the main physical quantities we will need.

The substitution of all those quantities into the field equations obtained above leads to the explicit expression of the field equations of the Georgi-Glashow spherically symmetric Model; in those equations, there is the dependence on one independent variable only, the radial coordinate $r$, and we will refer to the (total) derivative with respect to $r$ with a prime, i.e. $\frac{df(r)}{dr}=f'$.

If we consider the equations for the gauge fields we see that all the components are actually proportional, and the only independent equation is given as
\begin{eqnarray}
\nonumber
((1+h)a')'=a(2f^{2}+\frac{a^{2}-1}{r^{2}}).
\end{eqnarray}

We have then
\begin{eqnarray}
\nonumber
((1+h)r^{2}f')'=f(2a^{2}+r^{2}(2\lambda^{2}f^{2}-\frac{(r^{2}h)''}{6r^{2}}))
\end{eqnarray}
as the only independent equation for the scalar field.

If we consider, instead, the equations for the gravitational field, we have that, because of their diagonalization, they reduce to four equations only, and, because of the proportionality between the two angular equations $W_{\varphi\varphi}=W_{\theta\theta}\sin^{2}{\theta}$ and $T_{\varphi\varphi}=T_{\theta\theta}\sin^{2}{\theta}$, we can take into account only one of them; then, the condition of tracelessness allows us to remove another equation, let's say the other angular one: only the temporal and the radial equations are left; finally, we can consider (see the work of Mannheim \cite{m/conf}) the linear combination of the radial and temporal components $T^{t}_{\phantom{t}t}-T^{r}_{\phantom{r}r}$, which gives
\begin{eqnarray}
\nonumber
(rh)''''=\frac{r}{2}(2f'^{2}-ff'')+\frac{3a'^{2}}{2r}
\end{eqnarray}
as the only independent equation for the gravitational field.

In general we do not know whether those field equations admit a solutions or not, and if they do, the solutions are extremely difficult to be found, since we have no general methods at our disposal that help us in order to look for them.

We can then try to search for a solution in a simplified situation.

Let us consider a situation in which no gauge field is present; the scalar fields now have only a global symmetry under the $SO(3)$ transformation of the Georgi-Glashow model. On the other side, however, the richness of such a physical system comes from the presence of the non-minimal coupling $R\phi^{2}$ between the scalar fields and the conformally curved spacetime; thus, for our purposes, the scalar-gravitational coupling will be an interesting enough situation. 

In the case in which the gauge fields are set to zero, the set of field equations is indeed remarkably simplified, and it consists of the equations
\begin{eqnarray}
((1+h)r^{2}f')'=fr^{2}(2\lambda^{2}f^{2}-\frac{(r^{2}h)''}{6r^{2}})\\
(rh)''''=\frac{r}{2}(2f'^{2}-ff'')
\end{eqnarray}
for the scalar and the gravitational field.

Now, one particular solution for this system of equations can easily be obtained as $f=v$; with it we have that the gravitational field behaves as $h(r)=-\frac{2m}{r}+\lambda^{2}v^{2}r^{2}$, for any value of the free parameter $m$. 

The solution
\begin{eqnarray}
f(r)=v\\
h(r)=-\frac{2m}{r}+\lambda^{2}v^{2}r^{2}
\label{solution}
\end{eqnarray} 
is very interesting because it contains a scalar field that turns out to be a vector in the $3$-dimensional internal space, with a (positive) constant norm and radial direction, therefore a topological soliton; the gravitational field behaves as a Schwarzschild gravitational field for short distances, while for large values of $r$ it behaves a quadratically in the radial coordinate, which could explain the problem of missing matter (see Mannheim and Kazanas \cite{m-k/conf}) for galactic rotational curves.

A problem this solution has is that the condition on the VEV is showed to be assumed for any excitations above the ground state, since $R=12\lambda^{2}v^{2}$ everywhere.

This problem, however, can be solved by considering that such a ground state can indeed be obtained far from the matter distribution that creates the gravitational field, where the metric of the spacetime tends to the value of the underlying Anti-deSitter structure of the ground state; considering a spacetime that has the asymptotic Anti-deSitter structure, we should be able to obtain the solution in equation (\ref{solution}) as the asymptotic approximation of a more general, physical solution. 

The theory here presented has been considered also in the works of Demir \cite{de}, Odintsov \cite{od} and, mainly, Mannheim and Kazanas in \cite{m-k}: Mannheim and Kazanas considered the scalar field in a gravitational background and found that, for a negative scalar field self-coupling, a mass scale is generated in the theory, and the metric corresponds to a conformal deSitter space-time; the case in which the behaviour of the scalar field is consistent with the geometry of a conformal Anti-deSitter spacetime, and so with a positive scalar self-coupling, has been considered here, and it can be found in the discussion by Edery, Fabbri and Paranjape in \cite{e-f-p}.

This is interesting since conformal Anti-deSitter spacetimes have risen in importance over the last few years in the context of the correspondence between string theory and conformal field theory, as discussed by Aharony, Gubser, Maldacena, Ooguri and Oz in \cite{a-g-m-o-o}.
%%%%%%%%%%%%%%%%%%%%%%%%%%%%%%%%%%%%%%%%%%%%%%%%%%%%%%%%%%%%%%%%%%%%%%%%%%%%%%%%%%%%%%%%%%%%%%%%%%%
%%%%%%%%%%%%%%%%%%%%%%%%%%%%%%%%%%%%%%%%%%%%%%%%%%%%%%%%%%%%%%%%%%%%%%%%%%%%%%%%%%%%%%%%%%%%%%%%%%%
%%%%%%%%%%%%%%%%%%%%%%%%%%%%%%%%%%%%%%%%%%%%%%%%%%%%%%%%%%%%%%%%%%%%%%%%%%%%%%%%%%%%%%%%%%%%%%%%%%%
%%%%%%%%%%%%%%%%%%%%%%%%%%%%%%%%%%%%%%%%%%%%%%%%%%%%%%%%%%%%%%%%%%%%%%%%%%%%%%%%%%%%%%%%%%%%%%%%%%%
\part{Epilogue}
\chapter{Geometry,\\ Theoretical Physics,\\ Phenomenology}
Arrived at this point of the discussion, we have material enough to make a quite comprehensive status of the situation.

The solution we discussed at the end of the last part was obtained as a particular solution of the model of a triplet of real scalar fields in a conformally Anti-deSitter spacetime; of course, there can be many more general solutions than the one we decided to look for. 

And of course, as we saw, it is possible, if we do not want to constrain ourselves to this global solution, to have a more general, local model including gauge fields.

However, the model we chose to consider was an $SO(3)$ gauge symmetry group, which has some problems in the electroweak unification framework of the Standard Model: in fact, this gauge symmetry (although it possesses an invariant subgroup $SO(2)$ that would correspond, after the spontaneous symmetry breaking mechanism, to a massless photon) gives rise to $3$ gauge fields only, and it would not fit into the framework of the Standard Model, which describes $4$ gauge fields (as Georgi and Glashow explains in \cite{g-g}).

Another generalization of the model presented here could be considering the gauge symmetry group as $SU(2)$, and then $U(1) \times SU(2)$.

Massless spinorial fields can be accommodated beside scalars and gauge fields into the action, and the straightest way to include them is to consider the action 
\begin{eqnarray}
\nonumber
S=\int (C_{\alpha \beta \mu \nu}C^{\alpha \beta \mu \nu}
-\frac{1}{4L}\mathrm{tr}\it{(F_{\alpha \beta}F^{\alpha \beta})}+\\
\nonumber
+\phi\left(\frac{R}{6}-D^{2}\right)\phi
+\frac{i\hbar}{2}(\overline{\psi}\gamma^{\mu}\mathbf{D}_{\mu}\psi-
\overline{\mathbf{D}_{\mu}\psi}\gamma^{\mu}\psi)-\\
\nonumber
-a\phi\overline{\psi}\psi-\lambda^{2}\phi^{4})\sqrt{|g|}dV
\end{eqnarray}
in which $\mathbf{D}_{\mu}=D_{\mu}+\omega_{\mu}$ is the spinorial derivative, where $D_{\mu}=\nabla_{\mu}+A_{\mu}$ is the gauge derivative, and $a$ is a generic constant (this action can be found in the works of Mannheim \cite{m} and Flanagan \cite{fl}); however, the term $-a(\phi\psi^{2})$ has only one scalar field, and so it cannot have gauge indices, and thus, we cannot have non-abelian gauge fields, but only electrodynamics: this tells us that we can actually have an action with spinors, but only paying the price of an even worse problem regarding electroweak unification.

An action that would not consider the Standard Model's electroweak unification is given without gauge fields as
\begin{eqnarray}
\nonumber
S=\int (C_{\alpha \beta \mu \nu}C^{\alpha \beta \mu \nu}
+\phi\left(\frac{R}{6}-\nabla^{2}\right)\phi
+\frac{i\hbar}{2}(\overline{\psi}\gamma^{\mu}\mathbf{D}_{\mu}\psi-
\overline{\mathbf{D}_{\mu}\psi}\gamma^{\mu}\psi)-\\
\nonumber
-a\phi\overline{\psi}\psi-\lambda^{2}\phi^{4})\sqrt{|g|}dV
\end{eqnarray}
in which $\mathbf{D}_{\mu}=\nabla_{\mu}+\omega_{\mu}$ is the spinorial derivative, and it could be compatible with the solitonic solution $\phi=v$ for massive fermionic fields.

Through this action it is also possible to justify the presence of scalar fields in terms of fermions: in fact, we can suppose that fermions enter in the action first, and then a scalar field is required to achieve the conformal symmetry. 

Then, it could be possible to think that some other extra field can be introduced in a non-minimally coupled way, so to achieve a natural inclusion of gauge fields.

Conformal symmetry is the key principle to get a unique action; if we require that all the pieces present in the action are conformally invariant \emph{and} that all the conformally invariant pieces are present in the action, we get an action which is the only action, up to equivalence, that we can define -- and it is this sense of unicity that makes Conformal Symmetry so intriguing a principle. 

Nonetheless, conformal symmetry is \emph{not} a symmetry of Nature. It is true that this can be seen as a consequence of the fact that an original conformal symmetry can be spontaneously broken; and it is even more amazing that the spontaneous symmetry breaking mechanism is required by the conformal symmetry itself! But we have no \emph{a priori} reason to think that, at some scale, the Universe should be conformally invariant, also because, if this was the case, there wouldn't even be any scale to talk about -- at this very point, the discussion slides into Epistemology, and the problem whether conformal symmetry is a good symmetry principle or not would belong to the domain of Philosophical disciplines.

If we do not want to consider conformal symmetry as a basic principle for theoretical physics, then we will not have a fixed form for the action: some pieces in the action would be fixed by phenomenological considerations, others would not be fixed at all.

One principle we could take into account is a principle of simplicity, stating that the action has to be taken at the least order derivative, that is we have to consider a least-order action.

This ``principle of least-order action'' gives, indeed, the action that produces the field equations of the ECSK theory, which we know to be the right ones at the present state of the observations.

This is, in fact, a good task accomplished; but from the point of view of theoretical physics, the principle of least-order action is not much more justified than the principle of conformal invariance, and again we are sliding into the domain of Philosophy.

And as long as we are not able to state a Principle of Symmetry that guides Geometry up to Theoretical Physics, we have to rule over the dynamics according to the principles of Phenomenology.
%%%%%%%%%%%%%%%%%%%%%%%%%%%%%%%%%%%%%%%%%%%%%%%%%%%%%%%%%%%%%%%%%%%%%%%%%%%%%%%%%%%%%%%%%%%%%%%%%%%
%%%%%%%%%%%%%%%%%%%%%%%%%%%%%%%%%%%%%%%%%%%%%%%%%%%%%%%%%%%%%%%%%%%%%%%%%%%%%%%%%%%%%%%%%%%%%%%%%%%
%%%%%%%%%%%%%%%%%%%%%%%%%%%%%%%%%%%%%%%%%%%%%%%%%%%%%%%%%%%%%%%%%%%%%%%%%%%%%%%%%%%%%%%%%%%%%%%%%%%
%%%%%%%%%%%%%%%%%%%%%%%%%%%%%%%%%%%%%%%%%%%%%%%%%%%%%%%%%%%%%%%%%%%%%%%%%%%%%%%%%%%%%%%%%%%%%%%%%%%
\appendix
\addcontentsline{toc}{part}{Appendices}
\chapter{The historical theory:\\ Weyl's Least-Order Gravity}
In this appendix, we would like to present the historical theory of Weyl gravity and electrodynamics, showing how Weyl tensor, used paying no attention to its property of conformal invariance, but only to its irreducibility, gives rise to a least order derivative approach\footnote{We would like to stress that least-order here means least order derivative of the curvature, not of the metric tensor; in fact, considering the metric tensor, we would get this theory as a third-order theory, which is the least order among all those theories for which we have derivatives of curvatures in the field equations, but which is not the least-order in general, since Einstein gravity is second-order.}; we will see that this theory has fundamental correlations with Maxwell theory of Electrodynamics.

Let us consider the background of a spacetime with a metric tensor; we will take Cartan tensor to be zero, in order to follow the original idea of Weyl. Then, we define Weyl tensor as usual, but using the symbol $W$.

We can calculate the quadridivergence of the both sides of the definition of Weyl tensor; using Jacobi-Bianchi identities, the identity we get becomes
\begin{eqnarray}
\nonumber
\nabla_{\alpha}W^{\alpha \beta \mu \nu}=\nabla_{\alpha}R^{\alpha \beta \mu \nu}
+\frac{1}{2}\nabla_{\alpha}R^{\alpha \nu}g^{\beta \mu}
-\frac{1}{2}\nabla_{\alpha}R^{\alpha \mu}g^{\beta \nu}+\\
\nonumber
+\frac{1}{2}\nabla^{\nu}R^{\beta \mu}-\frac{1}{2}\nabla^{\mu}R^{\beta \nu}
+\frac{1}{6}\nabla^{\mu}Rg^{\beta \nu}-\frac{1}{6}\nabla^{\nu}Rg^{\beta \mu}\equiv\\
\nonumber
\equiv \nabla^{\mu}R^{\beta \nu}-\nabla^{\nu}R^{\beta \mu}
+\frac{1}{4}\nabla^{\nu}Rg^{\beta \mu}
-\frac{1}{4}\nabla^{\mu}Rg^{\beta \nu}+\\
\nonumber
+\frac{1}{2}\nabla^{\nu}R^{\beta \mu}-\frac{1}{2}\nabla^{\mu}R^{\beta \nu}
+\frac{1}{6}\nabla^{\mu}Rg^{\beta \nu}-\frac{1}{6}\nabla^{\nu}Rg^{\beta \mu}=\\
\nonumber
=\frac{1}{2}\nabla^{\mu}R^{\beta \nu}-\frac{1}{2}\nabla^{\nu}R^{\beta \mu}
+\frac{1}{12}\nabla^{\nu}Rg^{\beta \mu}-\frac{1}{12}\nabla^{\mu}Rg^{\beta \nu}.
\end{eqnarray}

We see that in the expression of Weyl tensor, the only curvatures that enters are the Ricci curvature tensor and the Ricci curvature scalar, and we can write both of them in terms of the energy-momentum tensor and scalar, once Einstein's field equations are given; we have that
\begin{eqnarray}
\nonumber
R^{\sigma \mu}-\frac{1}{2}g^{\sigma \mu}R=kT^{\sigma \mu}
\end{eqnarray}
in which $k$ is a constant, and in which $T^{\sigma \mu}$ is the energy-momentum tensor; their contraction is 
\begin{eqnarray}
\nonumber
-R=kT
\end{eqnarray}
where $T$ is the trace of the energy-momentum tensor: with these expressions, we can invert the field equations as 
\begin{eqnarray}
\nonumber
R^{\sigma \mu}=k\left(T^{\sigma \mu}-\frac{1}{2}g^{\sigma \mu}T\right)\\
\nonumber
R=-kT.
\end{eqnarray}

The following substitution into the identity for Weyl tensor gives finally
\begin{eqnarray}
\nonumber
\nabla_{\alpha}W^{\alpha \beta \mu \nu}=\frac{k}{2}\left(
\nabla^{\mu}[T^{\beta \nu}-\frac{1}{3}g^{\beta \nu}T]
-\nabla^{\nu}[T^{\beta \mu}-\frac{1}{3}g^{\beta \mu}T]\right).
\end{eqnarray}

It is now justified the definition of the quantity
\begin{eqnarray}\nonumber
J^{\beta \mu \nu}=\frac{1}{2}\left(
\nabla^{\mu}[T^{\beta \nu}-\frac{1}{3}g^{\beta \nu}T]
-\nabla^{\nu}[T^{\beta \mu}-\frac{1}{3}g^{\beta \mu}T]\right)
\end{eqnarray}
which is irreducible, and which is such that 
\begin{eqnarray}
\nonumber
\nabla_{\sigma}J^{\sigma \mu \nu}=0.
\end{eqnarray}

Now, we have that
\begin{eqnarray}
\nonumber
\nabla_{\sigma}W^{\sigma \beta \mu \nu}=kJ^{\beta \mu \nu}.
\end{eqnarray}

So, the tensor $J$ is a conserved quantity
\begin{eqnarray}
\nabla_{\sigma}J^{\sigma \mu \nu}=0
\end{eqnarray}
which is the source of the least order derivative field equations
\begin{eqnarray}
\nabla_{\sigma}W^{\sigma \beta \mu \nu}=kJ^{\beta \mu \nu}.
\label{wem}
\end{eqnarray}

Those field equations are a way to re-write Einstein's field equations in the form of a quadridivergence of a geometrical irreducible curvature equal to a source of energy-momentum.

This form is analogous to the form of Maxwell equations for Electrodynamics, for we have that it exists a tensor $J$ such that
\begin{eqnarray}
\nabla_{\sigma}J^{\sigma}=0
\end{eqnarray}
which is the source of the first order derivative field equations
\begin{eqnarray}
\nabla_{\sigma}F^{\sigma \beta}=4\pi J^{\beta}.
\label{mem}
\end{eqnarray}

In an empty spacetime we would have the equations written as 
\begin{eqnarray}
\nonumber
\nabla_{\sigma}W^{\sigma \beta \mu \nu}=0
\end{eqnarray}
that is in a form of a conservation law; the vacuum form for Maxwell equations is also given as a conservation law as
\begin{eqnarray}
\nonumber
\nabla_{\sigma}F^{\sigma \beta}=0,
\end{eqnarray}
and both are conformally invariant.

The Weyl tensor $W$ and the Faraday-Maxwell field tensor $F$ are both irreducible quantities; also, they can be written in terms of a potential (see Lanczos in \cite{l}).

Maxwell's first order derivative field equations are formally analogous to Weyl's first order derivative field equations: they both are written as the source of the field equal to the quadridivergence of some irreducible geometrical quantity that admits a potential.

And this is the formal analogy between Maxwell and Weyl field equations that represents the core of Weyl's attempt of a unified theory of electromagnetism and gravity.
%%%%%%%%%%%%%%%%%%%%%%%%%%%%%%%%%%%%%%%%%%%%%%%%%%%%%%%%%%%%%%%%%%%%%%%%%%%%%%%%%%%%%%%%%%%%%%%%%%%
\chapter{Higher-Order Gravity\\ and Beyond}
As we concluded the part about Fundamentals, we presented some of the Higher-Order actions we can have.

In a Riemannian spacetime, when two derivatives of the metric were considered, we obtained that only one curvature had to be taken into account, while with four derivatives we could have also taken into account products of two curvatures; the situation became much more difficult starting from the sixth order and higher, since, beside products of curvatures, we had to consider also products of covariant derivatives of curvatures.

Considering then products of curvatures and products of covariant derivatives of curvatures in all their possible combinations, for the sixth-order action we got a total amount of $8$ terms for the triplet of curvatures, and $3$ terms coming from the doublet of covariant derivatives of curvatures (plus some more terms coming as products of one curvature with two covariant derivatives applied to one curvature, which could be written as a doublet of covariant derivatives of curvatures, up to a quadridivergence in the action), and all these terms gave together the action as
\begin{eqnarray}
\nonumber
S_{(\Phi,\Psi,\Theta,A,B,C,D,E,F,G,H)}=\int [(\Phi\nabla_{\rho}R\nabla^{\rho}R+
\Psi\nabla_{\rho}R_{\alpha \beta}\nabla^{\rho}R^{\alpha \beta}+\\
\nonumber
\Theta\nabla_{\rho}R_{\alpha \beta \mu \nu}\nabla^{\rho}R^{\alpha \beta \mu \nu}+AR^{3}+\\
\nonumber
BRR_{\alpha \beta}R^{\alpha \beta}+
CRR_{\alpha \beta \mu \nu}R^{\alpha \beta \mu \nu}+\\
\nonumber
DR_{\alpha \beta}R^{\beta \gamma}R^{\alpha}_{\gamma}+
ER^{\alpha \beta}R^{\gamma \delta}R_{\alpha \gamma \beta \delta}+\\
\nonumber
FR^{\alpha \beta}R_{\alpha \gamma \delta \mu}R_{\beta}^{\phantom{\beta}\gamma\delta\mu}+
GR^{\alpha\beta\gamma\delta}R_{\alpha\beta\mu\nu}R_{\gamma\delta}^{\phantom{\gamma\delta}\mu\nu}+\\
\nonumber
HR_{\alpha\beta\gamma\delta}R^{\beta\mu\delta\nu}
R^{\alpha}_{\phantom{\alpha}\mu}{}^{\gamma}_{\phantom{\gamma}\nu})+k\mathscr{L}_{p}]\sqrt{|g|}dV
\end{eqnarray}
that gave field equations as
\begin{eqnarray}
\nonumber
3HR_{\rho\theta\gamma\delta}R^{\theta\mu\delta\alpha}R^{\rho}_{\phantom{\rho}\mu}{}^{\gamma\beta}
-2\Phi g^{\alpha\beta}\nabla^{4}R-\Psi\nabla^{4}R^{\alpha\beta}
+\Psi\nabla_{\rho}\nabla^{\alpha}\nabla^{2}R^{\beta\rho}\\
\nonumber
+F\nabla_{\rho}\nabla_{\mu}(R^{\rho\sigma}R_{\sigma}^{\phantom{\sigma}\beta\mu\alpha}+
R^{\rho\sigma}R_{\sigma}^{\phantom{\sigma}\alpha\mu\beta})
+Eg^{\alpha\beta}\nabla_{\rho}\nabla_{\mu}(R_{\sigma\theta}R^{\sigma\mu\theta\rho})\\
\nonumber
+\Psi\nabla_{\rho}\nabla^{\beta}\nabla^{2}R^{\alpha\rho}
-\Psi g^{\alpha\beta}\nabla_{\rho}\nabla_{\sigma}\nabla^{2}R^{\sigma\rho}
-4\Theta\nabla_{\rho}\nabla_{\mu}\nabla^{2}R^{\rho\alpha\mu\beta}\\
\nonumber
-E\nabla_{\rho}\nabla^{\beta}(R_{\sigma\theta}R^{\sigma\alpha\theta\rho})-
E\nabla_{\rho}\nabla^{\alpha}(R_{\sigma\theta}R^{\sigma\beta\theta\rho})
-\frac{\Phi}{2}g^{\alpha\beta}\nabla_{\rho}R\nabla^{\rho}R\\
\nonumber
-\frac{3D}{2}\nabla_{\rho}\nabla^{\beta}(R^{\rho\nu}R^{\alpha}_{\nu})
+Bg^{\alpha\beta}\nabla_{\rho}\nabla_{\mu}(RR^{\mu\rho})
+Cg^{\alpha\beta}\nabla^{2}(R^{\rho\sigma\gamma\delta}R_{\rho\sigma\gamma\delta})\\
\nonumber
+E\nabla^{2}(R_{\sigma\theta}R^{\sigma\alpha\theta\beta})
+\frac{F}{2}\nabla^{2}(R^{\alpha\sigma\gamma\delta}R^{\beta}_{\phantom{\beta}\sigma\gamma\delta})
+\frac{F}{2}g^{\alpha\beta}
\nabla_{\rho}\nabla_{\sigma}(R^{\rho}_{\phantom{\rho}\mu\nu\theta}R^{\sigma\mu\nu\theta})\\
\nonumber
+B\nabla^{2}(RR^{\alpha\beta})
-B\nabla_{\rho}\nabla^{\beta}(RR^{\rho\alpha})
-F\nabla_{\rho}\nabla_{\mu}(R^{\beta\sigma}R_{\sigma}^{\phantom{\sigma}\rho\mu\alpha}+
R^{\alpha\sigma}R_{\sigma}^{\phantom{\sigma}\rho\mu\beta})\\
\nonumber
-C\nabla^{\alpha}\nabla^{\beta}(R^{\rho\sigma\gamma\delta}R_{\rho\sigma\gamma\delta})
+4C\nabla_{\sigma}\nabla_{\mu}(RR^{\beta\sigma\alpha\mu})
+Bg^{\alpha\beta}\nabla^{2}(R_{\mu\nu}R^{\mu\nu})\\
\nonumber
-B\nabla_{\rho}\nabla^{\alpha}(RR^{\rho\beta})
-B\nabla^{\alpha}\nabla^{\beta}(R_{\mu\nu}R^{\mu\nu})
+6G\nabla_{\rho}\nabla_{\mu}(R^{\alpha\mu\gamma\delta}R^{\beta\rho}_{\phantom{\beta\rho}\gamma\delta})\\
\nonumber
+3Ag^{\alpha\beta}\nabla^{2}R^{2}
-2\Theta \nabla_{\zeta}(\nabla^{\alpha}R^{\rho\mu\nu\zeta}R^{\beta}_{\phantom{\beta}\nu\mu\rho})
-3H\nabla_{\mu}\nabla_{\nu}(R^{\mu\theta\gamma\alpha}R^{\nu}_{\phantom{\nu}\gamma\theta}{}^{\beta})\\
\nonumber
-3A\nabla^{\alpha}\nabla^{\beta}R^{2}
+3H\nabla_{\rho}\nabla_{\nu}(R^{\theta\rho\gamma\nu}R{}_{\theta}{}^{\beta}{}_{\gamma}{}^{\alpha})
-2\Theta \nabla_{\zeta}(\nabla^{\beta}R^{\rho\mu\nu\zeta}R^{\alpha}_{\phantom{\alpha}\nu\mu\rho})\\
\nonumber
-\frac{1}{2}F\nabla_{\rho}\nabla^{\alpha}(R^{\rho\gamma\delta\sigma}R^{\beta}_{\phantom{\beta}\gamma\delta\sigma})
-\frac{1}{2}F\nabla_{\rho}\nabla^{\beta}(R^{\rho\gamma\delta\sigma}R^{\alpha}_{\phantom{\alpha}\gamma\delta\sigma})
-\Theta \nabla^{2}R^{\alpha\mu\nu\rho}R^{\beta}_{\phantom{\beta}\mu\nu\rho}\\
\nonumber
+2\Theta \nabla_{\kappa}(\nabla^{\beta}R^{\rho\mu\nu\alpha}R^{\kappa}_{\phantom{\kappa}\nu\mu\rho})
-E\nabla_{\rho}\nabla_{\mu}(R^{\mu\beta}R^{\rho\alpha}-R^{\rho\mu}R^{\alpha\beta})
-2\Phi R^{\alpha\beta}\nabla^{2}R\\
\nonumber
-\frac{\Psi}{2}g^{\alpha\beta}\nabla_{\rho}R_{\theta\sigma}\nabla^{\rho}R^{\theta\sigma}
-\frac{\Theta}{2}g^{\alpha\beta}\nabla^{\theta}R^{\sigma\mu\nu\rho}\nabla_{\theta}R_{\sigma\mu\nu\rho}
+\Theta \nabla^{\alpha}R^{\sigma\mu\nu\rho}\nabla^{\beta}R_{\sigma\mu\nu\rho}\\
\nonumber
-\Psi \nabla^{2}R^{\theta\beta}R^{\alpha}_{\theta}
-\Psi \nabla^{2}R^{\theta\alpha}R^{\beta}_{\theta}
+\Psi\nabla_{\rho}(\nabla^{\alpha}R^{\beta\mu}R^{\rho}_{\mu})
+\Psi\nabla_{\rho}(\nabla^{\beta}R^{\alpha\mu}R^{\rho}_{\mu})\\
\nonumber
+2\Phi\nabla^{\alpha}\nabla^{\beta}\nabla^{2}R
+\Psi\nabla^{\alpha}R^{\rho\sigma}\nabla^{\beta}R_{\rho\sigma}
-\Psi\nabla_{\rho}(\nabla^{\alpha}R^{\rho\mu}R^{\beta}_{\mu}+\nabla^{\beta}R^{\rho\mu}R^{\alpha}_{\mu})\\
\nonumber
+\frac{E}{2}g^{\alpha\beta}R^{\sigma\rho}R^{\gamma\delta}R_{\sigma\gamma\rho\delta}
-\Theta \nabla^{2}R^{\beta\mu\nu\rho}R^{\alpha}_{\phantom{\alpha}\mu\nu\rho}
+2\Theta \nabla_{\kappa}(\nabla^{\alpha}R^{\rho\mu\nu\beta}R^{\kappa}_{\phantom{\kappa}\nu\mu\rho})\\
\nonumber
-\frac{3D}{2}\nabla_{\rho}\nabla^{\alpha}(R^{\rho\nu}R^{\beta}_{\nu})
+\frac{3D}{2}g^{\alpha\beta}\nabla_{\rho}\nabla_{\mu}(R^{\rho\nu}R^{\mu}_{\nu})
+\frac{3D}{2}\nabla^{2}(R^{\alpha\nu}R^{\beta}_{\nu})\\
\nonumber
-\frac{C}{2}g^{\alpha\beta}RR^{\rho\sigma\gamma\delta}R_{\rho\sigma\gamma\delta}
+CR^{\alpha\beta}R^{\rho\sigma\gamma\delta}R_{\rho\sigma\gamma\delta}
+2CRR^{\alpha\sigma\gamma\delta}R^{\beta}_{\phantom{\beta}\sigma\gamma\delta}\\
\nonumber
-\frac{F}{2}g^{\alpha\beta}R^{\rho\sigma}R_{\rho\mu\nu\theta}R_{\sigma}^{\phantom{\sigma}\mu\nu\theta}
-\frac{3}{2}ER^{\sigma\rho}R^{\beta\delta}R^{\alpha}_{\phantom{\alpha}\sigma\rho\delta}
-\frac{D}{2}g^{\alpha\beta}R_{\rho\sigma}R^{\rho\gamma}R^{\sigma}_{\gamma}\\
\nonumber
+2FR_{\rho\sigma}R^{\alpha\mu\nu\rho}R^{\beta}_{\phantom{\beta}\mu\nu}{}^{\sigma}
+3DR_{\rho\sigma}R^{\rho\alpha}R^{\sigma\beta}
-\frac{H}{2}R_{\rho\theta\gamma\delta}R^{\theta\mu\delta\nu}
R^{\rho}_{\phantom{\rho}\mu}{}^{\gamma}_{\phantom{\gamma}\nu}g^{\alpha\beta}\\
\nonumber
+2BRR^{\alpha}_{\nu}R^{\beta\nu}
+\frac{1}{2}FR^{\alpha\theta}R^{\beta\gamma\delta\sigma}R_{\theta\gamma\delta\sigma}
+\Phi \nabla^{\alpha}R\nabla^{\beta}R
-\frac{3}{2}ER^{\sigma\rho}R^{\alpha\delta}R^{\beta}_{\phantom{\beta}\sigma\rho\delta}\\
\nonumber
+\frac{1}{2}FR^{\beta\theta}R^{\alpha\gamma\delta\sigma}R_{\theta\gamma\delta\sigma}
+\frac{G}{2}R^{\rho\sigma\gamma\delta}R_{\rho\sigma\mu\nu}
R_{\gamma\delta}^{\phantom{\gamma\delta}\mu\nu}g^{\alpha\beta}
+3GR^{\alpha\sigma\gamma\delta}R^{\beta}_{\phantom{\beta}\sigma\mu\nu}
R_{\gamma\delta}^{\phantom{\gamma\delta}\mu\nu}\\
\nonumber
-\frac{1}{2}BR_{\mu\nu}R^{\mu\nu}\left(g^{\alpha\beta}R-2R^{\alpha\beta}\right)
-\frac{1}{2}AR^{2}\left(g^{\alpha\beta}R-6R^{\alpha\beta}\right)=-\frac{k}{2}T^{\alpha \beta}
\end{eqnarray}
for the sixth-order gravity; of course, if we go higher with the order of derivatives of the metric tensor, we get more and more terms, written as products of four curvatures plus products of one curvature with a doublet of covariant derivatives of curvatures plus a doublet of two covariant derivatives applied to one curvature for the eighth-order action; and of course, we can go even higher, considering as many orders as we want (for general discussions about higher-order actions, see, for example, Lovelock in \cite{lo}). 

Anyway, it is not in the action, but in the field equations that derivatives of curvatures play a fundamental role, and the derivatives of curvatures appear in the field equations starting from the fourth-order; for these theories, it is in fact possible to get vacuum equations that allow all the curvatures to be different from zero, and so the gravitational field will be contained in all the curvatures as well. 

This is different from the case of Einstein gravity, in which the second-order action provided field equations without derivatives of curvature, whose vacuum form reduced to the simple expression of the vanishing of Ricci curvature tensor, confining all the physical gravitational effects into the Riemann curvature tensor.

We can then say that the gravitational field contained in Riemann curvature tensor is constrained in it in Einstein's least-order gravity, while in higher-order gravity it can ``propagate'' also in Ricci curvature tensor, and even in Ricci curvature scalar.

Moreover, Einstein gravity is the limit case of a zero Cartan tensor of the Einstein-Cartan-Sciama-Kibble gravity, in which the presence of Cartan tensor allows a natural description of the spin of matter fields (the Dirac field); if we want to include Cartan tensor and spinor fields into this picture then the extension to higher-order theories would have actions with terms containing products of Cartan tensors plus one term containing the product of one Cartan tensor with the spin tensor, giving rise to field equations relating the spin to derivatives of Cartan tensor, with the consequence that the vacuum field equations would allow a non-vanishing Cartan tensor solution: this solution would then describe a Cartan tensor propagating out of the spin distribution, which is an effect that has never been observed up to now. So, it seems that only in the scheme of the least-order ECSK gravity the inclusion of matter fields can be simply achieved.

Another, last feature that drastically distinguishes the least-order theory from higher-order theories is that in absence of a suitable principle (such as for the conformal symmetry principle), higher-order theories depend on many parameters that can not be \emph{a priori} chosen, while the least-order theory is (up to equivalence) unique.
%%%%%%%%%%%%%%%%%%%%%%%%%%%%%%%%%%%%%%%%%%%%%%%%%%%%%%%%%%%%%%%%%%%%%%%%%%%%%%%%%%%%%%%%%%%%%%%%%%%
\chapter{Electrodynamics}
A fundamental physical field we never talked about is the Electromagnetic field. 

The reason for which we kept it out of the previous discussion is that the main point this thesis wanted to focus on did not involve the electromagnetic field, and so there was no reason to introduce a concept that we were not going to use.

However, electrodynamics has very interesting features, especially under the viewpoint of the conformal invariance.

Considering all the physical gauge fields we have observed so far, the Strong field, mediated by gluons, is confined inside hadrons, while the Weak field is mediated by the massive $W^{\pm}$ and $Z_{0}$; for both of them, the short range of their interaction does provide a scale in the theory, while the electromagnetic field is the only physical field we know to be \emph{a priori and a posteriori}, always conformally invariant.

Let us consider then a charge distribution (verifying the conservation law $\nabla_{\mu}J^{\mu}=0$) as the source of the electromagnetic field's dynamics, governed by the fundamental Maxwell equations
\begin{eqnarray}
\nabla_{\nu}F^{\nu \mu}=J^{\mu};
\label{em}
\end{eqnarray}
these equations come with the structural Maxwell equations
\begin{eqnarray}
\partial_{\rho}F_{\mu \alpha}+\partial_{\alpha}F_{\rho \mu}+\partial_{\mu}F_{\alpha \rho}=0.
\label{sem}
\end{eqnarray}
We see that \emph{even in the general case} equations (\ref{em}) and (\ref{sem}) do not contain any completely antisymmetric Cartan tensor; also, equations (\ref{sem}) do not contain the metric tensor neither: considering equations (\ref{sem}) as the equations that define its structure and equations (\ref{em}) as the equations that determine its dynamics, we can say that the structure of the electromagnetic field is defined independently on the metric of the spacetime, and that both its structure and its dynamics do not have interactions with Cartan tensor!

In the following, then, we will consider Cartan tensor to be zero without losing generality, as long as we deal only within the framework of electrodynamics.

Equation (\ref{sem}) can be solved to give the solution 
\begin{eqnarray}
F_{\alpha \rho}=\partial_{\alpha}A_{\rho}-\partial_{\rho}A_{\alpha}
\label{structuralsolution}
\end{eqnarray}
for some vector $A_{\mu}$ defined up to a gauge transformation as 
\begin{eqnarray}
A'_{\mu}=A_{\mu}-\partial_{\mu}f
\label{gauge}
\end{eqnarray}
for some function $f$ undefined.

Having this solution, field equations (\ref{em}) can be written as 
\begin{eqnarray}  
\nabla^{2}A^{\mu}-R^{\mu}_{\phantom{\mu} \alpha}A^{\alpha}
-g^{\beta \mu}\partial_{\beta}(\nabla_{\nu}A^{\nu})=J^{\mu}
\end{eqnarray}
and, because of the freedom ensured by the equation (\ref{gauge}), we can choose the function $f$ to give rise to the condition $\nabla_{\nu}A^{\nu}=0$, which leaves 
\begin{eqnarray}  
\nabla^{2}A^{\mu}-R^{\mu}_{\phantom{\mu} \alpha}A^{\alpha}=J^{\mu},
\label{edpotentials}
\end{eqnarray}
which is the expression of the Laplacian of the potentials $A$, and in which we can see that the Ricci curvature tensor takes part in it.

On the other hand, we can also consider the two sets of Maxwell equations (\ref{em}) and (\ref{sem}) to see that 
\begin{eqnarray}  
\nabla^{2}F_{\alpha \beta}-\frac{1}{3}RF_{\alpha \beta}+F^{\mu \nu}C_{\alpha \beta \mu \nu}
=\nabla_{\alpha}J_{\beta}-\nabla_{\beta}J_{\alpha},
\label{edstrengths}
\end{eqnarray}
which is the expression of the Laplacian of the strength $F$, and in which both Ricci curvature scalar and Weyl conformal curvature tensor take part in the dynamics of the electromagnetic field.

These are the basic expressions that describe electrodynamics.

Now, let us consider the expression that relates the potential to the strength
\begin{eqnarray}
\nonumber
F_{\alpha \rho}=\partial_{\alpha}A_{\rho}-\partial_{\rho}A_{\alpha}
\end{eqnarray}
and the analogous expression we saw for the general case of gauge fields 
\begin{eqnarray}
\nonumber
F_{\alpha \rho}=\partial_{\alpha}A_{\rho}-\partial_{\rho}A_{\alpha}+[A_{\alpha},A_{\rho}];
\end{eqnarray}
for a conformal transformation, the coordinates do not transform, and neither do the partial derivatives, which are then conformally invariant, and so, supposing that $A$ transforms as $A'=\Omega^{p}A$, we see that
\begin{eqnarray}
\nonumber
F'_{\alpha \rho}=\partial_{\alpha}A'_{\rho}-\partial_{\rho}A'_{\alpha}=
\Omega^{p}(\partial_{\alpha}A_{\rho}-\partial_{\rho}A_{\alpha})=\Omega^{p}F_{\alpha \rho}
\end{eqnarray}
but
\begin{eqnarray}
\nonumber
F'_{\alpha \rho}=\partial_{\alpha}A'_{\rho}-\partial_{\rho}A'_{\alpha}+[A'_{\alpha},A'_{\rho}]=
\Omega^{p}(\partial_{\alpha}A_{\rho}-\partial_{\rho}A_{\alpha}+\Omega^{p}[A_{\alpha},A_{\rho}])
\end{eqnarray}
and so we do not get the correct form for the gauge fields, unless we suppose $p=0$; this shows that in the general case, the non-linearity of the expression that relates potentials and strengths forces the conformal transformation to be trivially $A'=A$, while for the particular case of the Maxwell electrodynamics no constraint actually comes to dictate the specific conformal transformation law the field $A$ has to undergo.

Hence, it is only due to a reason of analogy that we will consider the transformation law of the field $A$ to be given as 
\begin{eqnarray}
A'_{\alpha}=A_{\alpha};
\end{eqnarray}
with this conformal transformation law, it is easy to see that the structural Maxwell field equations
\begin{eqnarray}
\nonumber
\partial_{\rho}F_{\mu \alpha}+\partial_{\alpha}F_{\rho \mu}+\partial_{\mu}F_{\alpha \rho}=0,
\end{eqnarray}
and, in the vacuum, Maxwell field equations 
\begin{eqnarray}
\nonumber
\nabla_{\nu}F^{\nu \mu}=0
\end{eqnarray}
are indeed conformally invariant.
%%%%%%%%%%%%%%%%%%%%%%%%%%%%%%%%%%%%%%%%%%%%%%%%%%%%%%%%%%%%%%%%%%%%%%%%%%%%%%%%%%%%%%%%%%%%%%%%%%%
\tableofcontents%INDICE%%%%%%%%%%%%%%%%%%%%%%%%%%%%%%%%%%%%%%%%%%%%%%%%%%%%%%%%%%%%%%%%%%%%%%%%%%%%
%%%%%%%%%%%%%%%%%%%%%%%%%%%%%%%%%%%%%%%%%%%%%%%%%%%%%%%%%%%%%%%%%%%%%%%%%%%%%%%%%%%%%%%%%%%%%%%%%%%
\newpage
%%%%%%%%%%%%%%%%%%%%%%%%%%%%%%%%%%%%%%%%%%%%%%%%%%%%%%%%%%%%%%%%%%%%%%%%%%%%%%%%%%%%%%%%%%%%%%%%%%%
\noindent \textbf{Acknowledgments}. I would like to thank Prof.~Silvio Bergia since he is the one who taught me the Foundations of Relativity, and Prof.~Manu B.~Paranjape because this thesis was born from the collaboration we had together; I would also like to thank Christian B\"ohmer, the group of Prof.~Alan Guth (Max Tegmark, Yi Mao and Serkan Cabi), Prof.~Ilya Shapiro and Prof.~Friedrich W.~Hehl for the interesting discussions about Cartan (torsion) tensor, and Prof.~Valter Moretti and Prof.~Massimo Ferri for the discussions we have had on Mathematics, in general. A special thanks goes to Prof.~Lee Smolin.

And a final, most important thanks goes to Marie-H\'el\`ene Genest, who shares with me the pleasure of knowing things.
%%%%%%%%%%%%%%%%%%%%%%%%%%%%%%%%%%%%%%%%%%%%%%%%%%%%%%%%%%%%%%%%%%%%%%%%%%%%%%%%%%%%%%%%%%%%%%%%%%%
%%%%%%%%%%%%%%%%%%%%%%%%%%%%%%%%%%%%%%%%%%%%%%%%%%%%%%%%%%%%%%%%%%%%%%%%%%%%%%%%%%%%%%%%%%%%%%%%%%%
%%%%%%%%%%%%%%%%%%%%%%%%%%%%%%%%%%%%%%%%%%%%%%%%%%%%%%%%%%%%%%%%%%%%%%%%%%%%%%%%%%%%%%%%%%%%%%%%%%%
%%%%%%%%%%%%%%%%%%%%%%%%%%%%%%%%%%%%%%%%%%%%%%%%%%%%%%%%%%%%%%%%%%%%%%%%%%%%%%%%%%%%%%%%%%%%%%%%%%%

%%%%%%%%%%%%%%%%%%%%%%%%%%%%%%%%%%%%%%%%%%%%%%%%%%%%%%%%%%%%%%%%%%%%%%%%%%%%%%%%%%%%%%%%%%%%%%%%%%%
\end{document}